\documentclass[a4paper,11pt]{JHEP3}
\usepackage[OT4]{fontenc}
\usepackage{amsmath}
\usepackage[active]{srcltx}

\newcommand{\chiral}[3]{\!\big[\hspace*{-2pt}\begin{array}{c} \\[-19pt]{\scriptscriptstyle #1}\\[-9pt]{\scriptscriptstyle #2 #3}\end{array}\hspace*{-2pt}\big]}

\def\bra#1{\left\langle#1\right|}
\def\ket#1{\left|#1\right\rangle}

\newcommand{\V}[4]{V^{\,#2}_{#1\,#3}(#4)}

\author{
      Damian Chor\k{a}\.zkiewicz\footnote{e-mail: damian.ch@ift.uni.wroc.pl}\\
	 Institute of Theoretical Physics,
	 University of Wroc{\l}aw,\\
	 pl. M. Borna 1, 95-204~Wroc{\l}aw, Poland

}
\author{
      Leszek Hadasz\footnote{e-mail: hadasz@th.if.uj.edu.pl;\ corresponding author} \\
	 M. Smoluchowski Institute of Physics,
	 Jagiellonian University,\\
	 W.~Reymonta 4,
	 30-059~Krak\'ow, Poland

}
\author{
      Zbigniew Jask\'{o}lski \footnote{e-mail: jask@ift.uni.wroc.pl}\\
	 Institute of Theoretical Physics,
	 University of Wroc{\l}aw,\\
	 pl. M. Borna 1, 95-204~Wroc{\l}aw, Poland
}

\abstract{
Using a super scalar field  representation of the chiral vertex
operators we develop a general method of calculating braiding matrices for all types of
$N=1$ super-conformal 4-point blocks involving Ramond external weights. We give explicit
analytic formulae in a number of cases.
\vspace*{1cm}}

\title{ Braiding  properties
of the $N=1$ super-conformal blocks
(Ramond sector)}

\preprint{}
\keywords{Ramond algebra, chiral vertex operator, conformal blocks, supersymmetric Liouville field theory}

\begin{document}

\section{Introduction}

One of the fundamental principles of the 2-dimensional conformal field theory (CFT) is the convergence of
the operator product expansion \cite{Belavin:1984vu}. It in particular implies that
any 4-point function  factorizes in three different ways corresponding to
the scattering channels $s$, $t$, $u$. Equivalence of these decompositions is one of the basic
consistency condition of the theory usually referred to as the crossing symmetry or the bootstrap equation.

Using the factorization and the conformal properties of 3-point functions one can express any 4-point correlator
in terms of structure constants and holomorphic and anti-holomorphic 4-point conformal blocks \cite{Belavin:1984vu}.
In the rational CFT the crossing symmetry implies  monodromy relations between
conformal blocks in different channels \cite{Moore:1988qv,Felder:1989wr}. Monodromy matrices between $s-t$, and $t-u$ channels,
are called the fusion and the braiding matrices, respectively.
As the spectrum of a rational CFT is finite they are finite-dimensional.

A well known example of the CFT with a continuous spectrum is the Liouville theory. In this case
the fusion and braiding matrices were first calculated by Ponsot and Teschner using
representations of U${}_q({\rm sl}(2,{\mathbb R}))$ \cite{Ponsot:1999uf,Ponsot:2000mt}.
These matrices can be also obtained calculating the exchange relations for the chiral operators
in the scalar field representation \cite{Teschner:2001rv,Teschner:2003en}.
The explicite form of the integral kernels of the fusion and the braiding matrices
was used in the analytic proof that the Liouville structure constants
\cite{Dorn:1994xn,Zamolodchikov:1995aa} satisfy the bootstrap equation
 \cite{Ponsot:1999uf}. Although derived in the context of the Liouville field theory
 the fusion and braiding integral kernels are universal for all CFT.
 For instance  for the
 central charge and the spectrum of a minimal model the
 continuous integral over intermediate weights reduces to a finite sum by the ``contour
 pinching'' mechanizm \cite{Ponsot:2003ju,Hadasz:2004cm}.

 The structure of conformal blocks in the $N=1$ superconformal theory is considerably more complicated.
 It has been recently analyzed in the
 context of recursion relations in a number of papers
 \cite{Hadasz:2006sb,Belavin:2006,Belavin:2007gz,Belavin:2007eq,RR,Hadasz:2007nt,Hadasz:2008dt,Suchanek:2010kq}.
The form of the fusion matrix for the Neveu-Schwarz (NS) superconformal blocks
was first proposed in \cite{Hadasz:2007wi} on the basis of the properties
of super-symmetric extensions of $b$-hypergeometric functions. This result has been
confirmed in \cite{Chorazkiewicz:2008es} where the fusion matrix was derived from the
exchange relations of the chiral operators in the super scalar free field representation.
As in the case of the bosonic Liouville theory the explicit form of the fusion matrix
was used to
check the bootstrap equation in the NS sector of   the $N = 1$ super-symmetric Liouville theory
with the structure constants proposed in \cite{Rashkov:1996jx,Poghosian:1996dw}.

The main aim of the present paper is to derive the integral kernels of the braiding  matrices
in the Ramond (R) sector of $N = 1$ SCFT by calculating the exchange relation
for chiral vertex operators in the free super field representation.
The extension of the theory by the Ramond sector leads to four types of chiral
vertex operators. Their different compositions  correspond
 to different 4-point blocks of the Neveu-Schwarz  and Ramond fields
\cite{Suchanek:2010kq}.
We derive all technical ingredients necessary for
calculating the braiding matrices for all
types of $N = 1$ superconformal blocks involving external Ramond weights.
Due to a proliferation of types of superconformal blocks we
present detailed calculations only in a few cases. They were chosen to
illustrate all the technicalities involved.
The methods developed are  general and can be easily applied
to all other blocks.

Our first motivation was to complete the proof of the  bootstrap
equation in both sectors of the (GSO projected) $N = 1$ super Liouville theory.
The matrices we are to calculate are however
universal and
can be used to check the bootstrap equation and to calculate 4-point correlation functions in any $N = 1$ SCFT.
The second interesting problem is to find out if there is a supersymmetric counterpart
of the  relation between the fusion and the modular matrices
recently found in the standard CFT \cite{Hadasz:2009sw}.
Due to the technical complexity both problems are postponed to subsequent papers.

The paper is organized as follows. Following \cite{Hadasz:2006sb,Hadasz:2008dt,Suchanek:2010kq} we define
in  Section \ref{Chiral:ops}
the Neveu-Schwarz (NS) and
the Ramond (R) chiral vertex operators
and analyze some of their properties.  In Section \ref{section:chiral:scalar}
the construction of chiral superscalar field space representation of the NS and the R
 vertex operators
 is described.
In  Subsection 3.1 we derive braiding relations
for the chiral fermion fields.
Our method is to decompose the chiral fermion Fock space
into  Virasoro Verma modules and then
use  known braiding properties of the Ising model
chiral vertices.
In Subsection 3.2 we introduce the chiral fields and clarify their relation to
the chiral vertex operators of Section 2. In Subsection 3.3 the matrix elements
of the chiral fields necessary for their normalization are calculated.

    In   Section \ref{section:braid:rel}
we calculate the braiding matrices.
In Subsection 4.1 we derive the braiding kernel for the compositions of ordered exponentials and screening charges.
This result along with the results of Section 3 are used in Subsection 4.2 to calculated the braiding kernel
in several cases including pairs of NS-NS, R-NS and R-R chiral vertex operators.

The paper is supplemented by a number of appendixes.
Appendix \ref{Appendix:Ising} collects the properties of the  chiral vertex operators of the Ising model we need in
our construction of the chiral fermion fields. In Appendix \ref{Appendix:CWI} we derive in some specific case
the Ward identity for the fermionic current $S$ in the presence of Ramond  fields.
Appendix \ref{Appendix:Barnes} contains some relevant properties of the Barnes double gamma function.
In Appendix \ref{Appendix:Orthogonality} we derive the orthogonality relations we use in Subsection 4.1.

The paper is rather technical and some  remarks concerning conventions and notations can be helpful.
Let us first emphasize that the choice of chiral vertex operators in the Ramond sector is
determined  by the fact that the full theory is based on ``small representations'',  i.e.\ irreducible
representations of the left and the right $N = 1$ Ramond algebras
extended only by the common parity operator \cite{Hadasz:2008dt,Suchanek:2010kq}.

We shall adopt the symmetric form of the OPE  of the fermionic current with the  Ramond fields
\cite{Zamolodchikov:1988nm}:
\begin{eqnarray}
\label{def:R:fields}
S(z) R_{\beta}^{\pm}(w,\bar w)
& \sim &
\frac{i \beta {\rm e}^{\mp i\frac{\pi}{4}}}{(z-w)^{\frac{3}{2}}} R_{\beta}^{\mp}(w, \bar w)+ \dots,
\end{eqnarray}
and the standard normalization of the two-point function
\begin{eqnarray}
\label{normalization:Rplus}
\left\langle R_{\beta_2}^+(w_2,\bar w_2)R_{\beta_1}^+(w_1,\bar w_1)\right\rangle
& = &
\frac{\delta_{\beta_2+\beta_1,0}}{|w_2-w_1|^{2\Delta_{\beta_1}}}.
\end{eqnarray}
Formulae (\ref{def:R:fields}), (\ref{normalization:Rplus})
determine the braiding relations of the fermionic current with the Ramond fields up to a sign.
This is in order  related to the
normalization of the two-point function of the $R_{\beta}^-$ fields as can be easily verified
analyzing analytic properties of the
three-point functions\footnote{See Appendix B for a similar analysis of chiral correlators.}
\[
\left\langle S(z) R_{\beta_2}^\pm(w_2,\bar w_2)R_{\beta_1}^\mp(w_1,\bar w_1)\right\rangle.
\]
In the present paper we chose
\begin{equation}
\label{normalization:Rminus:I}
\begin{array}{rcl}
\left\langle R_{\beta_2}^-(w_2,\bar w_2)R_{\beta_1}^-(w_1,\bar w_1)\right\rangle
& = &
\frac{\delta_{\beta_2+\beta_1,0}}{|w_2-w_1|^{2\Delta_{\beta_1}}},
\\[6pt]
\sqrt{z-w}\,S(z)R^{\pm}_\beta(w,\bar w) & = &  \mp \,i\,\sqrt{w-z}\,R^{\pm}_\beta(w,\bar w)S(z)  .
\end{array}
\end{equation}
The opposite convention
\begin{equation}
\label{normalization:Rminus:II}
\begin{array}{rcl}
\left\langle R_{\beta_2}^-(w_2,\bar w_2)R_{\beta_1}^-(w_1,\bar w_1)\right\rangle
& = &
- \frac{\delta_{\beta_2+\beta_1,0}}{|w_2-w_1|^{2\Delta_{\beta_1}}},
\\[6pt]
\sqrt{z-w}\,S(z)R^{\pm}_\beta(w,\bar w) & = & \pm\, i\, \sqrt{w-z}\,R^{\pm}_\beta(w,\bar w)S(z),
\end{array}
\end{equation}
is also possible. It is used for instance in \cite{Hadasz:2008dt,Suchanek:2010kq}.

For the chiral Ramond fields we assume:
\begin{eqnarray}
\label{def:W:fields}
S(z) W^\pm_{{\scriptscriptstyle \rm \bf f}\,\beta}(w)
& \sim &
\frac{i \beta {\rm e}^{\mp i\frac{\pi}{4}}}{(z-w)^{\frac{3}{2}}}
W^\mp_{\bar{\scriptscriptstyle \rm \bf f}\,\beta}(w)
+ \dots,
\end{eqnarray}
where ${\scriptstyle \rm \bf f}= \rm e, o$ is the parity index and $\bar {\scriptstyle \rm \bf f}$ denotes the parity opposed to
${\scriptstyle \rm \bf f}$. Our convention   for braiding (\ref{normalization:Rminus:I})
takes the form\footnote{
The consistency of  (\ref{def:W:fields}) and (\ref{normalization:Rminus:I})
can be easily checked by explicit calculation in the representation we develop in Section 3.}
\begin{eqnarray}
\label{braiding:W:fields:}
\begin{array}{rcll}
\sqrt{z-w}\,S (z)W^\pm_{{\rm e}\,\beta}(w)
&=&
- \,i\,\sqrt{w-z}\,
W^\pm_{{\rm e}\,\beta}(w)
S(z),
\\[6pt]
\sqrt{z-w}\, S (z)W^\pm_{{\rm o}\,\beta}(w)
&=&
+ \,i\,\sqrt{w-z}\,
W^\pm_{{\rm o}\,\beta}(w)
S(z).
\end{array}
\end{eqnarray}
Choosing the principal argument of a complex number $\xi$ in the range $-\pi \leq {\rm Arg}\,\xi < \pi$
one can write (\ref{braiding:W:fields:}) as
\begin{eqnarray}
\label{braiding:W:fields}
\begin{array}{rcll}
S (z)W^\pm_{{\rm e}\,\beta}(w)
&=&
- \epsilon\,
W^\pm_{{\rm e}\,\beta}(w)
S(z)
\\
S (z)W^\pm_{{\rm o}\,\beta}(w)
&=&
+ \epsilon\,
W^\pm_{{\rm o}\,\beta}(w)
S(z)
\end{array}
,\;\;\; \;\;\epsilon =
 \Big\{
\begin{array}[c]{lllllllll}
\scriptstyle  +1&\scriptstyle {\rm for } &  \scriptstyle {\rm Arg}(z-w) >0   \\[-4pt]
\scriptstyle  -1&\scriptstyle  {\rm for } &\scriptstyle   {\rm Arg}(z-w) <0
\end{array}.
\end{eqnarray}
while (\ref{normalization:Rminus:I}) reads
\begin{eqnarray}
\label{braiding:R:fields:I}
S (z)R^\pm_{\beta}(w,\bar w)
&=&
\mp\epsilon\,
R^\pm_{\beta}(w,\bar w)
S(z)
,\;\;\; \;\;\epsilon =
 \Big\{
\begin{array}[c]{lllllllll}
\scriptstyle  +1&\scriptstyle {\rm for } &  \scriptstyle {\rm Arg}(z-w) >0   \\[-4pt]
\scriptstyle  -1&\scriptstyle  {\rm for } &\scriptstyle   {\rm Arg}(z-w) <0
\end{array}.
\end{eqnarray}
Braiding properties (\ref{braiding:W:fields:}), (\ref{braiding:W:fields}) are crucial for most of the calculations in the present paper.


Our notation for the chiral vertex operators and conformal blocks is organized as follows.
In the NS sector the chiral vertex operators are denoted by
$$
V_{\scriptscriptstyle \rm \bf f}\chiral{\Delta_2}{\Delta_3}{\Delta_1}(z)\;:\;{\cal V}_{\Delta_1} \;\to\;{\cal V}_{\Delta_3}
$$
where ${\scriptscriptstyle \rm \bf f}= \rm e, o$ is the parity index and the weights in the square brackets denote:
$\Delta_2$ -- the weight of the vertex itself, $\Delta_1$ -- the weight of the source and $\Delta_3$ -- the weight of the target
NS Verma module. The ``star'' vertices are defined by
$$
V_{\rm e}\chiral{*\Delta_2}{\Delta_3}{\Delta_1}(z)
=
\left\{S_{-{1\over 2}},V_{\rm o}\chiral{\Delta_2}{\Delta_3}{\Delta_1}(z)\right\}
,\;\;\;\;\;\;
V_{\rm o}\chiral{*\Delta_2}{\Delta_3}{\Delta_1}(z)
=
\left[S_{-{1\over 2}},V_{\rm e}\chiral{\Delta_2}{\Delta_3}{\Delta_1}(z)\right].
$$
In the other three sectors the rules are similar but the vertices acquire an additional $\pm$ index:
$$
V_{\scriptscriptstyle \rm \bf f}^\pm\chiral{\Delta_2}{\beta_3}{\beta_1}(z)\,:\,{\cal W}_{\beta_1} \;\to\;{\cal W}_{\beta_3}
\,,\;\;
V_{\scriptscriptstyle \rm \bf f}^\pm\chiral{\beta_2}{\Delta_3}{\beta_1}(z)\,:\,{\cal W}_{\beta_1} \;\to\;{\cal V}_{\Delta_3}
\,,\;\;
V_{\scriptscriptstyle \rm \bf f}^\pm\chiral{\beta_2}{\beta_3}{\Delta_1}(z)\,:\,{\cal V}_{\Delta_1} \;\to\;{\cal W}_{\beta_3}\,.
$$
This is related to the structure of Ward identities in these sectors. In contrast to the NS sector
 the 3-point conformal blocks  are determined up to four rather then two structure constants.
The $\pm$ values of the additional index correspond the choice of a basis of 3-point blocks required by
the ``small'' representation mentioned above.
The conformal weights of the Ramond modules are denoted by parameter $\beta$ which  emphasizes
the sign dependence but also encodes information about the sectors. The notation of vertices is consistent with the
notation of conformal blocks introduced in  \cite{Hadasz:2008dt,Suchanek:2010kq}.\footnote{The blocks
themselves are different as in the present paper our conventions for braiding and hence the Ward identities are different.} One has for instance
\begin{eqnarray*}
\mathcal{F}^{\scriptscriptstyle \rm \bf f}_{\Delta}\!
\left[^{\Delta_3 \; \pm \beta_2}_{\Delta_4 \; \;\;\;\beta_1} \right](z)
 & =&
 \bra{\nu_4}
V_{\scriptscriptstyle \rm \bf f}\chiral{\Delta_3}{\Delta_4}{\Delta}(1)
V_{\scriptscriptstyle \rm \bf f}^\pm\chiral{\beta_2}{\Delta}{\beta_1}(z)
\ket{w^+_1},
 \\
\mathcal{F}^{\scriptscriptstyle \rm \bf f}_{\Delta}\!
\left[^{\pm \beta_3 \; \pm \beta_2}_{\;\;\;\beta_4 \,\;\;\;\beta_1} \right] (z)
 & =&
 \bra{w^+_4}
V_{\scriptscriptstyle \rm \bf f}^\pm\chiral{\beta_3}{\beta_4}{\Delta}(1)
V_{\scriptscriptstyle \rm \bf f}^\pm\chiral{\beta_2}{\Delta}{\beta_1}(z)
\ket{w^+_1},
 \\
\mathcal{F}^{\scriptscriptstyle \rm \bf f}_{\beta}
\left[_{\hspace{4pt}\Delta_4 \; \hspace{4pt}\Delta_1}^{\pm\beta_3 \;\pm\beta_2} \right](z)
&=&
\bra{\nu_4}
V_{\scriptscriptstyle \rm \bf f}^\pm\chiral{\beta_3}{\Delta_4}{\beta}(1)
V_{\scriptscriptstyle \rm \bf f}^\pm\chiral{\beta_2}{\beta}{\Delta_1}(z)
\ket{\nu_1},
\\
\mathcal{F}^{\scriptscriptstyle \rm \bf f}_{\beta}
\left[^{\pm \beta_3 \;  \hspace{4pt}\Delta_2}_{ \hspace{4pt}\Delta_4 \; \pm \beta_1} \right](z)
&=&
\bra{\nu_4}
V_{\scriptscriptstyle \rm \bf f}^\pm\chiral{\beta_3}{\Delta_4}{\beta}(1)
V_{\scriptscriptstyle \rm \bf f}^\pm\chiral{\Delta_2}{\beta}{\beta_1}(z)
\ket{w^+_1}.
\end{eqnarray*}
Let us note  that the $\pm$ in front of $\beta$-s in the symbol
of a conformal block is related to the $\pm$ index of the corresponding vertex operator rather then
an actual sign of this parameter. (When there are two $\beta$-s in a column we write the signs in front of
the upper one.)
According to these notational rules
all braiding relations for the chiral vertex operators can be
easily translated into analytic continuation formulae for corresponding 4-point blocks.

Although very economic for denoting vertices and blocks
the $\Delta, \beta$ notation is not well suited for
the analytic expressions for the braiding matrices. For this purposes we
use in both sectors the  $\alpha$ parametrization of conformal weights:
$$
\Delta_{\rm NS} = {\alpha(Q-\alpha)\over 2},\;\;\;\;\;\;\Delta_{\rm R}={1\over 16} +{\alpha(Q-\alpha)\over 2}.
$$
The relation  $\alpha$ to $\beta$ in the Ramond sector is  straightforward $\alpha = {Q\over 2}-\sqrt{2}\beta $.

\section{Chiral vertex operators}
\label{Chiral:ops}
The relations of $N=1$ superconformal algebra
extended by
the fermion parity
operator $(-1)^F$ read
\begin{eqnarray}
\label{NS}
\nonumber
\left[L_m,L_n\right] & = & (m-n)L_{m+n} +\frac{c}{12}m\left(m^2-1\right)\delta_{m+n},
\\
\left[L_m,S_k\right] & = &\frac{m-2k}{2}S_{m+k},
\\
\nonumber
\left\{S_k,S_l\right\} & = & 2L_{k+l} + \frac{c}{3}\left(k^2 -\frac14\right)\delta_{k+l},
\\
\nonumber
[(-1)^F, L_m]&=&\{(-1)^F,S_k\}=0,
\end{eqnarray}
where $m,n \in \mathbb{Z}$ and   $k,l \in \mathbb{Z}+{1\over 2}$ in  the Neveu-Schwarz algebra sector
and $k,l \in \mathbb{Z}$ in the Ramond algebra one.

The NS supermodule ${\cal V}_\Delta$
of the highest weight $\Delta$ and the central charge $c$
is defined as a  free vector space  generated by all vectors
of the form
\begin{equation}
\label{basis}
\nu_{\Delta,MK}
\; = \;
L_{-M}S_{-K} \nu_{\Delta}
\; \equiv \;
L_{-m_j}\ldots L_{-m_1}S_{-k_i}\ldots S_{-k_1}\nu_{\Delta}\,,
\end{equation}
where
 $K = \{k_1,k_2,\ldots,k_i\}$ and
 $M = \{m_1,m_2,\ldots,m_j\}$ are
arbitrary ordered sets of  indices
$$
k_i > \ldots > k_2 > k_1 ,
\hspace{10pt} k_s\in \mathbb{N}-{1\over 2}, \hspace{20pt}
m_j \geqslant \ldots \geqslant m_2 \geqslant m_1,\hspace{10pt} m_r\in \mathbb{N}
$$
and $\nu_{\Delta}$ is the highest weight state with respect to the extended NS
algebra:
\begin{equation}
\label{highestNS}
L_0\nu_{\Delta}=\Delta\nu_{\Delta},
\hskip 5mm
(-1)^F\nu_{\Delta}=\nu_{\Delta},
\hskip 5mm
L_m\nu_{\Delta}= S_k\nu_{\Delta} =0,
\hskip 5mm
m\in \mathbb{N},\;\
k\in \mathbb{N}-{1\over 2}.
\end{equation}
In the Ramond sector the highest weight state is defined in a similar way
\begin{equation}
\label{highestR}
L_0w^+_{\beta}=\Delta w^+_{\beta},
\hskip 5mm
(-1)^Fw^+_{\beta}=w^+_{\beta},
\hskip 5mm
L_mw^+_{\beta}= S_kw^+_{\beta} =0,
\hskip 5mm
m\in \mathbb{N},\;\
k\in \mathbb{N}.
\end{equation}
A novel property is that the zero level subspace of the R supermodule ${\cal W}_\beta$ over
$w^+_\beta$ is 2-dimensional
\begin{equation}
\label{action:of:S0}
S_0  w^\pm_{\beta} =i {\rm e}^{\mp i{\pi\over 4}} \beta  w^\mp_\beta\;\;\;{\rm for}\;\;\;\Delta=
\textstyle {c\over 24}-\beta^2\neq 0.
\end{equation}
Hermitian forms $\langle
.\,,.\rangle_{c,\Delta}$ on ${\cal V}_\Delta$ and $\langle
.\,,.\rangle_{c,\beta}$ on
${\cal W}_{\beta}$ are uniquely determined  by the relations
\begin{equation}
\label{scalar:products}
(L_{m})^{\dag}=L_{-m},\;\;\;(S_{k})^{\dag}=S_{-k},\;\;\;
\langle \nu_{\Delta}, \nu_{\Delta}\rangle =1,\;\;\;
\langle w^+_{\beta}, w^+_{\beta}\rangle =1,\;\;\;
\langle w^+_{\beta},S_0\, w^+_{\beta}\rangle =0.
\end{equation}
They are  block-diagonal with respect to the $L_0$- and $(-1)^F$-gradings.

Following \cite{Suchanek:2010kq} we introduce 3-point blocks as
chiral 3-forms
(anti-linear in the left argument and
linear in the central and the right ones):
\begin{eqnarray*}
\mathcal{V}_{\Delta_3} \times \mathcal{V}_{\Delta_2} \times \mathcal{V}_{\Delta_1}
\ni(\xi_3, \xi_2, \xi_1)\;
& \longrightarrow &\;\;
\varrho_{\scriptscriptstyle \rm NN}(\xi_3, \xi_2, \xi_1|z)\in \mathbb{C},
\\
\mathcal{W}_{\beta_3} \times \mathcal{V}_{\Delta_2} \times \mathcal{W}_{\beta_1}
\ni(\eta_3, \xi_2, \eta_1)\;
& \longrightarrow &\;\;
\varrho_{\scriptscriptstyle \rm RR}(\eta_3, \xi_2, \eta_1|z)\in \mathbb{C},
\\
\mathcal{V}_{\Delta_3} \times \mathcal{W}_{\beta_2} \times \mathcal{W}_{\beta_1}
\ni(\xi_3, \eta_2, \eta_1)\;
& \longrightarrow &\;\;
\varrho_{\scriptscriptstyle \rm NR}(\xi_3, \eta_2, \eta_1|z)\in \mathbb{C},
\\
\mathcal{W}_{\beta_3} \times \mathcal{W}_{\beta_2} \times \mathcal{V}_{\Delta_1}
\ni(\eta_3, \eta_2, \xi_1)\;
& \longrightarrow &\;\;
\varrho_{\scriptscriptstyle \rm RN}(\eta_3, \eta_2, \xi_1|z)\in \mathbb{C},
\end{eqnarray*}
satisfying the ``bosonic'' (with respect to $L_n$) and the ``fermionic''
(with respect to $S_k$) Ward identities.
The ``bosonic'' identities are the same for 3-point blocks of all types.
We shall not use them in the present discussion (see
 \cite{Suchanek:2010kq} for their explicit form).

The ``fermionic'' Ward identities for the NN type of 3-point block take the form
\cite{Hadasz:2006sb}:
\begin{eqnarray}
\nonumber
\varrho_{\scriptscriptstyle \rm{NN}}( \xi_3,S_{k}\xi_2,\xi_1|z) &=&
 \sum\limits_{m=0}^{k+{1\over 2}}
 \left(\!
\begin{array}{c}
\scriptstyle k+{1\over 2}\\[-6pt]
\scriptstyle m
\end{array}
\!\right)
 (-z)^{m}
\Big(\varrho_{\scriptscriptstyle \rm{NN}}( S_{m-k}\xi_3,\xi_2,\xi_1|z)
\\
\nonumber
 &&\hspace{20pt} -\ (-1)^{|\xi_1|+|\xi_3|}\;
\varrho_{\scriptscriptstyle \rm{NN}}(\xi_3,\xi_2,S_{k-m}\xi_1|z) \Big),
\hskip 5mm k\geqslant-\scriptstyle {1\over 2},
\\[10pt]
\label{ward:NN}
\varrho_{\scriptscriptstyle \rm{NN}}(\xi_3,S_{-k}\xi_2,\xi_1|z) &=&
 \sum\limits_{m=0}^{\infty}
 \left(\!
\begin{array}{c}
\scriptstyle k-{3\over 2}+m\\[-6pt]
\scriptstyle m
\end{array}
\!\right)
z^{m}
\varrho_{\scriptscriptstyle \rm{NN}}(S_{k+m} \xi_3,\xi_2,\xi_1|z)
\\
\nonumber
 && \hspace{-95pt} -\;
(-1)^{|\xi_1|+|\xi_3|+k+{1\over 2}}
\sum\limits_{m=0}^{\infty}
 \left(\!
\begin{array}{c}
\scriptstyle k-{3\over 2}+m\\[-6pt]
\scriptstyle m
\end{array}
\!\right)
 z^{-k-m+{1\over 2}}
\varrho_{\scriptscriptstyle \rm{NN}}(\xi_3,\xi_2, S_{m-{1\over 2}}\xi_1|z), \hskip 5mm k>\scriptstyle {1\over 2},
\\ [10pt]
\nonumber
\varrho_{\scriptscriptstyle \rm{NN}}(S_{-k} \xi_3,\xi_2,\xi_1|z)
&=&
(-1)^{|\xi_1|+|\xi_3|+1}
\varrho_{\scriptscriptstyle \rm{NN}}( \xi_3,\xi_2,S_k\xi_1|z)
\\
\nonumber
 &+&
\sum\limits_{m=-1}^{l(k-{1\over 2})}
 \left(\!
\begin{array}{c}
\scriptstyle k+{1\over 2}\\[-6pt]
\scriptstyle m+1
\end{array}
\!\right)
  z^{k-{1\over 2} -m}
  \varrho_{\scriptscriptstyle \rm{NN}}(\xi_3,S_{m+{1\over 2}}\xi_2,\xi_1|z).
\end{eqnarray}
The form $\varrho_{\scriptscriptstyle \rm{NN}}$ is determined by the Ward identities up to two independent
constants
\begin{eqnarray*}
\varrho_{\scriptscriptstyle \rm NN}(\xi_3,\xi_2,\xi_1|z)
& =&
 \rho^{}_{\scriptscriptstyle \rm NN}(\xi_3,\xi_2,\xi_1|z)
 \varrho_{\scriptscriptstyle \rm NN}(\nu_3,\nu_2,\nu_1|1)
 \\
&+&  \rho^{*}_{\scriptscriptstyle \rm NN}(\xi_3,\xi_2,\xi_1|z)
 \varrho_{\scriptscriptstyle \rm NN}(\nu_3,*\nu_2,\nu_1|1)
\end{eqnarray*}
where $*\nu_i
\equiv
S_{-{1\over 2}}\nu_i$. For
$L_0$-eingenstates, $ L_0\,\xi_i  = \Delta_i(\xi_i)\xi_i$
\begin{eqnarray*}
\rho^{}_{\scriptscriptstyle \rm NN}(\xi_3,\xi_2,\xi_1|z)
&=&
z^{\Delta_3(\xi_3)- \Delta_2(\xi_2)-\Delta_1(\xi_1)}
\rho^{}_{\scriptscriptstyle \rm NN}(\xi_3,\xi_2,\xi_1|1),
\\
\rho^{*}_{\scriptscriptstyle \rm NN}(\xi_3,\xi_2,\xi_1|z)
&=&
z^{\Delta_3(\xi_3)- \Delta_2(\xi_2)-\Delta_1(\xi_1)}
\rho^{*}_{\scriptscriptstyle \rm NN}(\xi_3,\xi_2,\xi_1|1).
\end{eqnarray*}
Since the parity of the total number of fermionic excitations
is preserved in identities (\ref{ward:NN}):
\begin{eqnarray*}
\rho_{\scriptscriptstyle \rm NN }(S_{I}\nu_3,\nu_2,S_{J}\nu_1)
=
\rho^*_{\scriptscriptstyle \rm NN  }(S_{I}\nu_3,*\nu_2,S_{J}\nu_1)
= 0
\quad
&\mathrm{if}&
 \quad
|I| + |J | \in  \mathbb{N} -{1\over 2} ,\\
\rho^*_{\scriptscriptstyle \rm NN }(S_{I}\nu_3,\nu_2,S_{J}\nu_1)
=
\rho_{\scriptscriptstyle \rm NN  }(S_{I}\nu_3,*\nu_2,S_{J}\nu_1)
= 0
\quad
&\mathrm{if}&
 \quad
|I| + |J | \in  \mathbb{N}.
\end{eqnarray*}
The chiral vertex operators are defined by  their matrix elements
\begin{equation}
\label{chiral:vertex:NN}
\begin{array}{rcl}
\langle \xi_3|V_{\rm e}\chiral{\Delta_2}{\Delta_3}{\Delta_1}
(z)|\xi_1\rangle
&=&
\rho_{\scriptscriptstyle \rm  NN}(\xi_3,\nu_2,\xi_1|z),
\\   [8pt]
\langle \xi_3|
V_{\rm o}\chiral{\Delta_2}{\Delta_3}{\Delta_1}
(z)|\xi_1\rangle
&=&
\rho^*_{\scriptscriptstyle \rm  NN}(\xi_3,\nu_2,\xi_1|z),
\\        [8pt]
\langle \xi_3|
V_{\rm e}\chiral{*\Delta_2}{\Delta_3}{\Delta_1}
(z)|\xi_1\rangle
&=&
\rho^*_{\scriptscriptstyle \rm  NN}(\xi_3,*\nu_2,\xi_1|z),
\\   [8pt]
\langle \xi_3|
V_{\rm o}\chiral{*\Delta_2}{\Delta_3}{\Delta_1}
(z)|\xi_1\rangle
&=&
\rho_{\scriptscriptstyle \rm  NN}(\xi_3,*\nu_2,\xi_1|z).
\end{array}
\end{equation}
By the construction
$$
V_{\rm e}\chiral{*\Delta_2}{\Delta_3}{\Delta_1}
=\left\{
S_{-{1\over 2}},
V_{\rm o}\chiral{\Delta_2}{\Delta_3}{\Delta_1}
\right\}\;,\;\;\;\;
V_{\rm o}\chiral{*\Delta_2}{\Delta_3}{\Delta_1}
=\left[
S_{-{1\over 2}},
V_{\rm e}\chiral{\Delta_2}{\Delta_3}{\Delta_1}
\right].
$$
The ``fermionic'' Ward identities for the RR  3-point block read \cite{Suchanek:2010kq}:
\begin{eqnarray}
\nonumber
\varrho_{\scriptscriptstyle \rm RR}
(S_{-n} \eta_3,\xi_2,\eta_1|z)
&=&
(-1)^{|\eta_1|+|\eta_3|+1}
\varrho_{\scriptscriptstyle \rm RR}
( \eta_3,\xi_2,S_n\eta_1|z)
\\
\label{ward:RR}
&+&
\sum\limits_{k=-\frac12}^{\infty}
 \left(\!
\begin{array}[c]{c}
\scriptstyle n+{1\over 2}\\[-4pt]
\scriptstyle k+\frac12
\end{array}
\!\right)
  z^{n-k}
 \varrho_{\scriptscriptstyle \rm RR}(\eta_3,S_{k}\xi_2,\eta_1|z),
\\[4pt]
\nonumber
\sum_{p=0}^{\infty}
\left(\!
\begin{array}[c]{c}
\scriptstyle {1\over 2}\\[-4pt]
\scriptstyle p
\end{array}
\!\right)
 \ z^{\frac12 -p} \
\varrho_{\scriptscriptstyle \rm RR}(\eta_3,S_{p-k} \xi_2,\eta_1|z )
&=&
 \sum_{p=0}^{\infty}
\left(\!
\begin{array}[c]{c}
\scriptstyle {1\over 2}-k\\[-4pt]
\scriptstyle 2
\end{array}
\!\right)
 (-z)^p \
\varrho_{\scriptscriptstyle \rm RR}(S_{p+k - \frac12} \eta_3, \xi_2,\eta_1|z )
 \\
 \nonumber
 &&\hspace{-70pt} - (-1)^{|\eta_3|+|\eta_1|+1 } \sum_{p=0}^{\infty}
\left(\!
\begin{array}[c]{c}
\scriptstyle {1\over 2}-k\\[-4pt]
\scriptstyle 2
\end{array}
\!\right)
 (-z)^{ \frac12 -k-p}
 \varrho_{\scriptscriptstyle \rm RR}(\eta_3,\xi_2,S_{p}\eta_1 |z).
\end{eqnarray}
The 3-form
 $ \varrho_{\scriptscriptstyle \rm RR}(\eta_3, \xi_2, \eta_1|z)$ is determined
up to four rather then two constants:
\begin{equation*}
\begin{array}{rcl}
\varrho_{\scriptscriptstyle \rm RR}(\eta_3,\xi_2,\eta_1|z)
& =& 
 \rho^{++}_{\scriptscriptstyle \rm RR}(\eta_3,\xi_2,\eta_1|z)
 \varrho_{\scriptscriptstyle \rm RR}(w^+_3, \nu_2, w^+_1|1) \\
&+&
 \ \rho^{+-}_{\scriptscriptstyle \rm RR}(\eta_3,\xi_2,\eta_1|z)
 \varrho_{\scriptscriptstyle \rm RR}( w^+_3, \nu_2, w^-_1|1)\\
&+&
\ \rho^{-+}_{\scriptscriptstyle \rm RR}(\eta_3,\xi_2,\eta_1|z)
 \varrho_{\scriptscriptstyle \rm RR}( w^-_3, \nu_2, w^+_1|1)\\
&+& \ \rho^{--}_{\scriptscriptstyle \rm RR}(\eta_3,\xi_2,\eta_1|z)
 \varrho_{\scriptscriptstyle \rm RR}(w^-_3,\nu_2, w^-_1|1)\, . 
\end{array}
\end{equation*}
For $L_0$ eigenstates
$
 \rho^{\imath\jmath}_{\scriptscriptstyle \rm RR}(\eta_3,\xi_2,\eta_1|z)
 = z^{\Delta_3(\eta_3)- \Delta_2(\xi_2)-\Delta_1(\eta_1)}
 \rho^{\imath\jmath}_{\scriptscriptstyle \rm RR}(\eta_3,\xi_2,\eta_1|1).
$
As before the parity of the total number of fermionic excitations
is preserved  and therefore
\begin{eqnarray*}
\begin{array}{rclllll}
\rho^{++}_{\scriptscriptstyle \rm RR }(S_{M}\eta_3,\nu_2,S_{N}\eta_1)
=
\rho^{--}_{\scriptscriptstyle \rm RR  }(S_{M}\eta_3,\nu_2,S_{N}\eta_1)
&=& 0
\\
\rho^{+-}_{\scriptscriptstyle \rm RR }(S_{M}\eta_3,*\nu_2,S_{N}\eta_1)
=
\rho^{-+}_{\scriptscriptstyle \rm RR  }(S_{M}\eta_3,*\nu_2,S_{N}\eta_1)
&=& 0
\end{array}
\,
&\mathrm{if}&
 \,
\#N + \#M \in  2\mathbb{N} +1 ,\\
\begin{array}{rclllll}
\rho^{+-}_{\scriptscriptstyle \rm RR }(S_{M}\eta_3,\nu_2,S_{N}\eta_1)
=
\rho^{-+}_{\scriptscriptstyle \rm RR  }(S_{M}\eta_3,\nu_2,S_{N}\eta_1)
&=& 0
\\
\rho^{++}_{\scriptscriptstyle \rm RR }(S_{M}\eta_3,*\nu_2,S_{N}\eta_1)
=
\rho^{--}_{\scriptscriptstyle \rm RR  }(S_{M}\eta_3,*\nu_2,S_{N}\eta_1)
&=& 0
\end{array}
\,
&\mathrm{if}&
 \,
\#N + \#M \in  2\mathbb{N}.
\end{eqnarray*}
Using Ward identities (\ref{ward:RR}) one can derive the relations
\begin{equation}
\label{e-e,o-o:p}
\begin{array}{rcrrrrrrrrr}
\rho^{++}_{\scriptscriptstyle \rm RR}(S_{M}w_3^+, \nu_2, S_{N} w_1^+)
	&=& \rho^{--}_{\scriptscriptstyle \rm RR}(S_{M}w_3^-, \nu_2, S_{N} w_1^-),
\\
\rho^{--}_{\scriptscriptstyle \rm RR}(S_{M}w_3^+, \nu_2, S_{N} w_1^+)
	&=&\rho^{++}_{\scriptscriptstyle \rm RR}(S_{M}w_3^-, \nu_2, S_{N} w_1^-),
\\
 \rho^{+-}_{\scriptscriptstyle \rm RR}(S_{M}w_3^+, \nu_2, S_{N} w_1^-)
	&=& \rho^{-+}_{\scriptscriptstyle \rm RR}(S_{M}w_3^-, \nu_2, S_{N} w_1^+),
\\
 \rho^{-+}_{\scriptscriptstyle \rm RR}(S_{M}w_3^+, \nu_2,S_{N} w_1^-)
	&=& - \rho^{+-}_{\scriptscriptstyle \rm RR}(S_{M}w_3^-, \nu_2, S_{N} w_1^+),
\end{array}
\end{equation}
for the even number of fermionic operators  $\# M +\, \# N \in 2 \mathbb{N}$
 and
\begin{equation}
\label{e-e,o-o:np}
\begin{array}{rcrrrrr}
\rho^{+-}_{\scriptscriptstyle \rm RR}(S_{M}w_3^-, \nu_2,S_{N} w_1^-)
	&=& - i \rho^{-+}_{\scriptscriptstyle \rm RR}(S_{M}w_3^+, \nu_2, S_{N}w_1^+),
\\
\rho^{-+}_{\scriptscriptstyle \rm RR}(S_{M}w_3^-, \nu_2, S_{N}w_1^-)
	 &=& i \rho^{+-}_{\scriptscriptstyle \rm RR}(S_{M}w_3^+, \nu_2,S_{N} w_1^+),
\\
\rho^{++}_{\scriptscriptstyle \rm RR}(S_{M}w_3^-, \nu_2, S_{N}w_1^+)
	&=& -i  \rho^{--}_{\scriptscriptstyle \rm RR}(S_{M}w_3^+, \nu_2, S_{N} w_1^-),
\\
\rho^{--}_{\scriptscriptstyle \rm RR}(S_{M}w_3^-, \nu_2,S_{N} w_1^+)
	&=& -i \rho^{++}_{\scriptscriptstyle \rm RR}(S_{M}w_3^+, \nu_2, S_{N}w_1^-),
\end{array}
\end{equation}
for  $\# M +\, \# N \in 2 \mathbb{N}+1$.
One also has
\begin{equation}
\label{e-o}
\begin{array}{rcrrrrr}
\rho^{++}_{\scriptscriptstyle \rm RR}(S_{I}w_3^+, \nu_2, S_{J} w_1^+)
	&=& (-1)^{\# J} \rho^{+-}_{\scriptscriptstyle \rm RR}(S_{I}w_3^+, \nu_2, S_{J} w_1^-),
\\
\rho^{--}_{\scriptscriptstyle \rm RR}(S_{I}w_3^+, \nu_2, S_{J} w_1^+)
	&=& \,  i \,(-1)^{\# J} \rho^{-+}_{\scriptscriptstyle \rm RR}(S_{I}w_3^+, \nu_2, S_{J} w_1^-),
\\
\rho^{-+}_{\scriptscriptstyle \rm RR}(S_{I}w_3^+, \nu_2, S_{J} w_1^+)
	&=& (-1)^{\# J} \rho^{--}_{\scriptscriptstyle \rm RR}(S_{I}w_3^+, \nu_2, S_{J} w_1^-),
\\
\rho^{+-}_{\scriptscriptstyle \rm RR}(S_{I}w_3^+, \nu_2, S_{J} w_1^+)
&=&  i \, (-1)^{\# J} \rho^{++}_{\scriptscriptstyle \rm RR}(S_{I}w_3^+, \nu_2, S_{J} w_1^-).&
\end{array}
\end{equation}
Identical relations hold for $\nu$ replaced by $*\nu$.

An appropriate basis for the 3-point blocks takes the form:
\begin{equation}
\label{rhoRR}
 \begin{array}{rclllllll}
 \nonumber
 \rho^{(\pm)}_{\scriptscriptstyle \rm RR, e}
	&=& \rho^{++}_{\scriptscriptstyle RR} \pm \rho^{--}_{\scriptscriptstyle RR}\,,&&
 \rho^{(\pm)}_{\scriptscriptstyle \rm RR, o}
	&=& \rho^{+-}_{\scriptscriptstyle RR} \pm i\, \rho^{-+}_{\scriptscriptstyle RR}\,.&&
\end{array}
\end{equation}
The corresponding chiral vertex operators are given by
\begin{equation}
\label{chiral:vertex:RNR}
\begin{array}{rcl}
 \bra{\eta_3}
 V_{\scriptscriptstyle \rm \bf f}^\pm\chiral{\Delta_2}{\beta_3}{\beta_1}
(z)
 \ket{\eta_1}
&=& \rho^{(\pm)}_{\scriptscriptstyle \rm RR, f}(\eta_3,\nu,\eta_1|z),
\\
 \bra{\eta_3}
  V_{\scriptscriptstyle \rm \bf f}^\pm\chiral{*\Delta_2}{\beta_3}{\beta_1}
(z)
 \ket{\eta_1}
&=& \rho^{(\pm)}_{\scriptscriptstyle \rm RR, \bar{f}}(\eta_3,*\nu,\eta_1|z).
\end{array}
\end{equation}
In the mixed sectors the Ward identities read:
\begin{eqnarray}
\label{Ward_rhoNR}
\nonumber
&&\hspace{-70pt} \sum_{p=0}^{\infty}
\left(\!
\begin{array}[c]{c}
\scriptstyle n+{1\over 2}\\[-6pt]
\scriptstyle p
\end{array}
\!\right)
 \ z^{n+
\frac12 -p} \
\varrho_{\scriptscriptstyle \rm NR}
(\xi_3,S_{p} \eta_2,\eta_1 |z)
= \sum_{p=0}^{\infty}
\left(\!
\begin{array}[c]{c}
\scriptstyle {1\over 2}\\[-6pt]
\scriptstyle p
\end{array}
\!\right)
 (-z)^p \
\varrho_{\scriptscriptstyle \rm NR}
(S_{p-n - \frac12} \xi_3, \eta_2,\eta_1 |z)
 \\
&& + i (-1)^{|\xi_3|+|\eta_1|+1}  \sum_{p=0}^{\infty}
\left(\!
\begin{array}[c]{c}
\scriptstyle {1\over 2}\\[-6pt]
\scriptstyle p
\end{array}
\!\right)
 (-1)^p \ z^{\frac12 -p}
\varrho_{\scriptscriptstyle \rm NR}
(\xi_3, \eta_2,S_{n+p}\eta_1|z ) ,
\\  \nonumber
&& \hspace{-70pt} \sum_{p=0}^{\infty}
\left(\!
\begin{array}[c]{c}
\scriptstyle {1\over 2}\\[-6pt]
\scriptstyle p
\end{array}
\!\right)
 \ z^{\frac12 -p} \
\varrho_{\scriptscriptstyle \rm NR}
(\xi_3,S_{p-n} \eta_2,\eta_1 |z)
=
 \sum_{p=0}^{\infty}
\left(\!
\begin{array}[c]{c}
\scriptstyle {1\over 2}-n\\[-6pt]
\scriptstyle p
\end{array}
\!\right)
 (-z)^p \
 \varrho_{\scriptscriptstyle \rm NR}
(S_{p+n - \frac12} \xi_3, \eta_2,\eta_1 |z)
 \\  \nonumber
 &&+ i (-1)^{|\xi_3|+|\eta_1|+1 } \sum_{p=0}^{\infty}
\left(\!
\begin{array}[c]{c}
\scriptstyle {1\over 2}-n\\[-6pt]
\scriptstyle p
\end{array}
\!\right)
(-1)^{n+p}  z^{ \frac12-n -p}
 \varrho_{\scriptscriptstyle \rm NR}
(\xi_3,\eta_2,S_{p}\eta_1 |z)
 ,
\end{eqnarray}
\begin{eqnarray}\label{Ward_rhoRN}
 \nonumber
&&\hspace{-70pt} \sum_{p=0}^{\infty}
\left(\!
\begin{array}[c]{c}
\scriptstyle -n\\[-6pt]
\scriptstyle p
\end{array}
\!\right)
 \ z^{-p-n} \
\varrho_{\scriptscriptstyle \rm RN}
(\eta_3,S_{p} \eta_2,\xi_1|z )
=\sum_{p=0}^{\infty}
\left(\!
\begin{array}[c]{c}
\scriptstyle {1\over 2}\\[-6pt]
\scriptstyle p
\end{array}
\!\right)
 (-z)^p \
\varrho_{\scriptscriptstyle \rm RN}
(S_{n+p} \eta_3, \eta_2,\xi_1|z )
 \\
&& + i (-1)^{|\eta_3|+|\xi_1|+1}  \sum_{p=0}^{\infty}
\left(\!
\begin{array}[c]{c}
\scriptstyle {1\over 2}\\[-6pt]
\scriptstyle p
\end{array}
\!\right)
 (-1)^p \ z^{\frac12 -p}
\varrho_{\scriptscriptstyle \rm RN}
(\eta_3, \eta_2,S_{p-n-{1\over 2}}\xi_1 |z) ,
\\  \nonumber
&& \hspace{-70pt}
\varrho_{\scriptscriptstyle \rm RN}
(\eta_3,S_{-n} \eta_2,\xi_1|z)
= \sum_{p=0}^{\infty}
\left(\!
\begin{array}[c]{c}
\scriptstyle {1\over 2}-n\\[-6pt]
\scriptstyle p
\end{array}
\!\right)
 (-z)^p \
\varrho_{\scriptscriptstyle \rm RN}
(S_{p+n } \eta_3, \eta_2,\xi_1|z)
 \\  \nonumber
 && +i (-1)^{|\eta_3|+|\xi_1|+1 } \sum_{p=0}^{\infty}
\left(\!
\begin{array}[c]{c}
\scriptstyle {1\over 2}-n\\[-6pt]
\scriptstyle p
\end{array}
\!\right)
 (-1)^{n+p}\,z^{ \frac12 -n-p}
 \varrho_{\scriptscriptstyle \rm RN}
(\eta_3,\eta_2,S_{p-\frac12 }\xi_1|z ),
\end{eqnarray}
where $|\xi|,|\eta|$ denote parities of states $\xi\in {\cal V}_\Delta, \eta\in{\cal W}_\Delta$
\footnote{
Ward identities (\ref{Ward_rhoNR}), (\ref{Ward_rhoRN}) differ from the corresponding
ones   in  \cite{Hadasz:2008dt,Suchanek:2010kq} by the sign in front of  $i.$ This comes from
 the opposite convention (\ref{normalization:Rminus:II}) which was (implicitly) assumed in
\cite{Hadasz:2008dt,Suchanek:2010kq}.
}.
These  relations along with the ``bosonic'' Ward
identities
  determine each 3-point block  up to four constants:
\begin{eqnarray*}
\varrho_{\scriptscriptstyle \rm NR}(\xi_3,\eta_2,\eta_1|z)
& =&
 \rho^{++}_{\scriptscriptstyle \rm NR}(\xi_3,\eta_2,\eta_1|z)
 \varrho_{\scriptscriptstyle \rm NR}(\nu_3, w^+_2, w^+_1|1) \\
&+&  \rho^{+-}_{\scriptscriptstyle \rm NR}(\xi_3,\eta_2,\eta_1|z)
 \varrho_{\scriptscriptstyle \rm NR}(\nu_3, w^+_2, w^-_1|1)\\
&+& \rho^{-+}_{\scriptscriptstyle \rm NR}(\xi_3,\eta_2,\eta_1|z)
 \varrho_{\scriptscriptstyle \rm NR}(\nu_3, w^-_2, w^+_1|1)\\
&+&  \rho^{--}_{\scriptscriptstyle \rm NR}(\xi_3,\eta_2,\eta_1|z)
 \varrho_{\scriptscriptstyle \rm NR}(\nu_3, w^-_2, w^-_1|1),
\\[6pt]
\varrho_{\scriptscriptstyle \rm RN}(\eta_3,\eta_2,\xi_1|z)
&=& 
 \rho^{++}_{\scriptscriptstyle \rm RN}(\eta_3,\eta_2,\xi_1|z)
 \varrho_{\scriptscriptstyle \rm RN}(w^+_3, w^+_2, \nu_1|1)\\
&+& \rho^{+-}_{\scriptscriptstyle \rm RN}(\eta_3,\eta_2,\xi_1|z)
 \varrho_{\scriptscriptstyle \rm RN}(w^+_3, w^-_2, \nu_1|1 )
 \\
&+& \rho^{-+}_{\scriptscriptstyle \rm RN}(\eta_3,\eta_2,\xi_1|z)
 \varrho_{\scriptscriptstyle \rm RN}(w^-_3, w^+_2, \nu_1|1)\\
&+& \rho^{--}_{\scriptscriptstyle \rm RN}(\eta_3,\eta_2,\xi_1|z)
\varrho_{\scriptscriptstyle \rm RN}(w^-_3, w^-_2, \nu_1|1).
\end{eqnarray*}
For
$L_0$-eingenstates, $ L_0\,\xi_i  = \Delta_i(\xi_i)\xi_i$,
$ L_0\,\eta_j  = \Delta_j(\eta_j)\,\eta_j$ one has:
\begin{eqnarray*}
\rho^{\imath\jmath}_{\scriptscriptstyle \rm NR }
(\xi_3,\eta_2,\eta_1|z)&=& z^{\Delta_3(\xi_3)- \Delta_2(\eta_2)-\Delta_1(\eta_1)}\
\rho^{\imath\jmath}_{\scriptscriptstyle \rm NR }
(\xi_3,\eta_2,\eta_1|1),\\
\rho^{\imath\jmath}_{\scriptscriptstyle \rm RN }
(\eta_3,\eta_2,\xi_1|z)&=& z^{\Delta_3(\eta_3)- \Delta_2(\eta_2)-\Delta_1(\xi_1)}\
\rho^{\imath\jmath}_{\scriptscriptstyle \rm RN }
(\eta_3,\eta_2,\xi_1|1),\;\;\;\;{\imath,\jmath}=\pm\ .
\end{eqnarray*}
Using Ward identities (\ref{Ward_rhoNR}), (\ref{Ward_rhoRN}) one can derive the  relations
\begin{equation}
\label{tNR1}
\begin{array}{rcr}
\rho^{+-}_{\scriptscriptstyle \rm NR  }(S_{I}\nu, w_2^-, S_{J}w_1^+) &=&
i\, \rho^{--}_{\scriptscriptstyle \rm NR }(S_{I}\nu, w_2^+, S_{J}w_1^+),
\\
\rho^{-+}_{\scriptscriptstyle \rm NR }(S_{I}\nu, w_2^-, S_{J} w_1^+) &=&
 \rho^{++}_{\scriptscriptstyle \rm NR }(S_{I}\nu, w_2^+, S_{J}w_1^+),
\\
\rho^{++}_{\scriptscriptstyle \rm NR }(S_{I}\nu, w_2^-, S_{J}w_1^+)
&=& i\, \rho^{-+}_{\scriptscriptstyle \rm NR }(S_{I}\nu, w_2^+, S_{J}w_1^+),
\\
\rho^{--}_{\scriptscriptstyle \rm NR }(S_{I}\nu, w_2^-, S_{J} w_1^+) &=&
 \rho^{+-}_{\scriptscriptstyle \rm NR }(S_{I}\nu, w_2^+, S_{J}w_1^+),
\end{array}
\end{equation}
\begin{equation}
\label{tRN1}
\begin{array}{rcr}
\rho^{+-}_{\scriptscriptstyle \rm RN  }(S_{J}w_3^+, w_2^-, S_{I}\nu) &=&
 \rho^{++}_{\scriptscriptstyle \rm RN }( S_{J}w_3^+, w_2^+,S_{I}\nu)  ,
\\
\rho^{-+}_{\scriptscriptstyle \rm RN  }(S_{J}w_3^+, w_2^-, S_{I}\nu) &=&
i\, \rho^{--}_{\scriptscriptstyle \rm RN }( S_{J}w_3^+, w_2^+,S_{I}\nu) ,
\\
\rho^{++}_{\scriptscriptstyle \rm RN  }(S_{J}w_3^+, w_2^-, S_{I}\nu) &=&
i\,\rho^{+-}_{\scriptscriptstyle \rm RN }( S_{J}w_3^+, w_2^+,S_{I}\nu)   ,
\\
\rho^{--}_{\scriptscriptstyle \rm RN  }(S_{J}w_3^+, w_2^-, S_{I}\nu) &=&
 \rho^{-+}_{\scriptscriptstyle \rm RN }( S_{J}w_3^+, w_2^+,S_{I}\nu)  ,
\end{array}
\end{equation}
and
\begin{equation}
\label{tNR2}
\begin{array}{rcr}
\rho^{+-}_{\scriptscriptstyle \rm NR }(S_{I}\nu, w_2^+, S_{J}w_1^-) &=&
\;\;(-1)^{\# J}\, \rho^{++}_{\scriptscriptstyle \rm NR }(S_{I}\nu, w_2^+, S_{J}w_1^+),
\\
\rho^{-+}_{\scriptscriptstyle \rm NR }(S_{I}\nu, w_2^+, S_{J} w_1^-) &=&
- i\, (-1)^{\# J}\, \rho^{--}_{\scriptscriptstyle \rm NR }(S_{I}\nu, w_2^+, S_{J}w_1^+),
\\
\rho^{++}_{\scriptscriptstyle \rm NR }(S_{I}\nu, w_2^+, S_{J}w_1^-) &=&
- i\, (-1)^{\# J}\, \rho^{+-}_{\scriptscriptstyle \rm NR }(S_{I}\nu, w_2^+, S_{J}w_1^+),
\\
\rho^{--}_{\scriptscriptstyle \rm NR }(S_{I}\nu, w_2^+, S_{J} w_1^-) &=&
\;\;(-1)^{\# J}\, \rho^{-+}_{\scriptscriptstyle \rm NR }(S_{I}\nu, w_2^+, S_{J}w_1^+),
\end{array}
\end{equation}
\begin{equation}
\label{tRN2}
\begin{array}{rcr}
\rho^{+-}_{\scriptscriptstyle \rm RN  }(S_{J}w_3^-, w_2^+, S_{I}\nu) &=&
- i\,(-1)^{\# I}\,
\rho^{--}_{\scriptscriptstyle \rm RN }( S_{J}w_3^+, w_2^+,S_{I}\nu)    ,
\\
\rho^{-+}_{\scriptscriptstyle \rm RN  }(S_{J}w_3^-, w_2^+, S_{I}\nu) &=&
(-1)^{\# I}\,
\rho^{++}_{\scriptscriptstyle \rm RN }( S_{J}w_3^+, w_2^+,S_{I}\nu)  ,
\\
\rho^{++}_{\scriptscriptstyle \rm RN  }(S_{J}w_3^-, w_2^+, S_{I}\nu) &=&
-i\,(-1)^{\# I}\,
\rho^{-+}_{\scriptscriptstyle \rm RN }( S_{J}w_3^+, w_2^+,S_{I}\nu) ,
\\
\rho^{--}_{\scriptscriptstyle \rm RN  }(S_{J}w_3^-, w_2^+, S_{I}\nu) &=&
(-1)^{\# I}\,
 \rho^{+-}_{\scriptscriptstyle \rm RN }( S_{J}w_3^+, w_2^+,S_{I}\nu)  .  \\[5pt]
\end{array}
\end{equation}
For states with definite parities some forms identically vanish:
\begin{eqnarray*}
\rho^{\pm \pm}_{\scriptscriptstyle \rm NR }(S_{I}\nu,w^{+}_2,S_{J}w_1^+)
=
\rho^{\pm \pm}_{\scriptscriptstyle \rm RN  }(S_{J}w_3^+, w_2^-, S_{I}\nu)
= 0
\quad
&\mathrm{if}&
 \quad
(2|I| + \# J ) \in 2 \mathbb{N} +1 ,\\
\rho^{\pm \mp}_{\scriptscriptstyle \rm NR }(S_{I}\nu,w^{+}_2,S_{J}w_1^+)
=
\rho^{\pm \mp}_{\scriptscriptstyle \rm RN  }(S_{J}w_3^+, w_2^-, S_{I}\nu)
= 0
\quad
&\mathrm{if}&
 \quad
(2|I| + \# J ) \in 2 \mathbb{N}.
\end{eqnarray*}

The decomposition of the 4-point functions of Ramond fields into conformal blocks
suggests the following  convenient choice of a basis of
the 3-point blocks \cite{Hadasz:2008dt,Suchanek:2010kq}
\begin{equation}
\label{3ptBNR}
\begin{array}{rclcrclll}
\rho^{(\pm)}_{\scriptscriptstyle \rm NR, e}
&=&
\rho^{++}_{\scriptscriptstyle \rm NR} \pm
        \rho^{--}_{\scriptscriptstyle \rm NR}\,,   \;\;\;
&&
\;\;\;
\rho^{(\pm)}_{\scriptscriptstyle \rm RN, e}
&=&
\rho^{++}_{\scriptscriptstyle \rm RN} \pm
        \rho^{--}_{\scriptscriptstyle \rm RN}\,,
\\
\rho^{(\pm)}_{\scriptscriptstyle \rm NR, o}
&=&
\rho^{+-}_{\scriptscriptstyle \rm NR} \pm i
        \rho^{-+}_{\scriptscriptstyle \rm NR}\,,  \;\;\;
&&
\;\;\;
\rho^{(\pm)}_{\scriptscriptstyle \rm RN, o}
&=&
 \rho^{-+}_{\scriptscriptstyle \rm RN} \pm i
        \rho^{+-}_{\scriptscriptstyle \rm RN}\,.
\end{array}
\end{equation}
The chiral vertex operators are then defined by  their matrix elements as follows
\begin{equation}
\label{chiral:vertex:definition}
\begin{array}{rcl}
\langle \xi_3|
 V_{\scriptscriptstyle \rm \bf f}^\pm\chiral{\beta_2}{\beta_3}{\Delta_1}
(z)|\eta_1\rangle
&=&
\rho^{(\pm)}_{\scriptscriptstyle \rm  NR, \bf f}(\xi_3,w_2^+,\eta_1|z)\, ,
\\   [8pt]
\langle \eta_3|
 V_{\scriptscriptstyle \rm \bf f}^\pm\chiral{\beta_2}{\Delta_3}{\beta_1}
(z)|\xi_1\rangle
&=&
\rho^{(\pm)}_{\scriptscriptstyle \rm  RN, \bf f}(\eta_3,w_2^+,\xi_1|z) \,, \;\;\;{\scriptstyle \rm \bf f} = { \rm e,o  }.
\end{array}
\end{equation}

\section{Chiral superscalar}
\label{section:chiral:scalar}

\subsection{Chiral fermion}

In the NS sector the chiral fermion field decomposes into half-integer modes:
\begin{eqnarray}
\label{chiral:fermion:NS}
\psi(z)
& = &
\sum\limits_{r \in {\mathbb Z}+\frac12} \psi_r\, z^{-r-{1\over 2}},\;\;
\{\psi_r , \psi_s\}=\delta_{r+s},\;\;\;
	      \{(-1)^{F} , \psi_s\}=0
,\;\;\;
	      \psi_{r}^{\dagger}=\psi_{-r}.
\end{eqnarray}
The algebra of modes
is realized
in the Fock space ${\cal F}_{\rm NS}$ generated out of the
vacuum $|\Omega_{\rm F} \rangle$
satisfying  $
\psi_r|\Omega_{\rm F}\rangle = 0,\; r> 0,\;(-1)^{F}| \Omega_{\rm F} \rangle=| \Omega_{\rm F} \rangle,\; \langle\Omega_{\rm F}| \Omega_{\rm F} \rangle=1.
$

\noindent In the R sector $\psi(z)$ has the integer mode decomposition:
\begin{eqnarray}
\label{chiral:fermion:R}
\psi(z)
& = &
\sum\limits_{m \in {\mathbb Z}} \psi_m\, z^{-m-{1\over 2}},\;\;
\{\psi_m , \psi_n\}=\delta_{m+n},\;\;\;
	      \{(-1)^{F} , \psi_m\}=0
,\;\;\;
	      \psi_{m}^{\dagger}=\psi_{-m},
\end{eqnarray}
and the vacuum state of
the Fock space ${\cal F}_{\rm R}$ is
doubly degenerated
$$
\psi_0|\Omega^\pm\rangle \propto |\Omega^\mp\rangle,\;\;
(-1)^F|\Omega^\pm\rangle = \pm|\Omega^\pm\rangle,\;\;\langle\Omega^\pm| \Omega^\pm \rangle=1,\;\;
\langle\Omega^\pm| \Omega^\mp \rangle=0.
$$
Both Fock spaces carry the $c={1\over 2}$ Virasoro algebra representation
\begin{eqnarray*}
L_0&=&
\sum\limits_{k>0}\left(k+\textstyle{1\over 2}\right)\psi_{-k}\psi_k + \Big\{
\begin{array}{llrll}\\[-20pt]
\scriptstyle0& &\scriptstyle {\rm NS}\;{\rm sector}\\   [-6pt]
{\textstyle \scriptstyle{1\over 16}}& &\scriptstyle{\rm R}\;{\rm sector}
\end{array}
\\
L_m &=&{\textstyle{1\over 4}}\sum\limits_{k}(2k-m)\,\psi_{m-k}\psi_k\, ,
\end{eqnarray*}
where $k \in \mathbb{Z}+{1\over 2}$ in  the NS sector
and $k \in \mathbb{Z}$ in the R sector.
Each Fock space can be decomposed into Virasoro Verma modules which leads to the
decomposition of the total Hilbert space of the chiral fermion into Virasoro Verma modules with conformal weights
$0, {1\over 2}, {1\over 16}, {1\over 16}$:
\begin{equation}
\label{deco}
{\cal H}_{\rm F}={\cal F}_{\rm NS}\oplus {\cal F}_{\rm R}=
{\cal U}_{\,0} \oplus {\cal U}_{1\over 2}\oplus {\cal U}^+_{1\over 16} \oplus {\cal U}^-_{1\over 16}.
\end{equation}
In this decomposition ${\cal U}_{\,0} \oplus {\cal U}^+_{1\over 16}$ and
${\cal U}_{1\over 2} \oplus {\cal U}^-_{1\over 16}$ are the positive and the negative eigenspaces of
the partity operator $(-1)^F$, respectively.

\noindent We introduce new chiral fields  $\sigma^\pm(z)$ satisfying:
\begin{eqnarray}
\label{def_sigma}
\psi(z)\sigma^{\pm}(w)
&=&
\frac{{\rm e}^{\mp
 {i\over 4}\pi}\sigma^{\mp}(w)}{\sqrt{2(z-w)}} +\dots\ ,
\\[4pt]
\nonumber
 &&\hspace{-70pt}
  \langle \Omega_{\rm F}\, |\sigma^\pm(z)\sigma^\pm(w)|\Omega_{\rm F}\,\rangle
  \;=\;
   (z-w)^{-{1\over 8}}\;,
\\[4pt]
   \nonumber
\psi (z)\sigma^\pm (w)
&=&
\mp\epsilon \sigma^\pm(w)\psi(z)
,\;\;\;\;\;\;\;\;\epsilon =
 \Big\{
\begin{array}[c]{lllllllll}
\scriptstyle  +1&\scriptstyle {\rm for } &  \scriptstyle {\rm Arg}(z-w) >0   \\[-4pt]
\scriptstyle  -1&\scriptstyle  {\rm for } &\scriptstyle   {\rm Arg}(z-w) <0
\end{array}
 \; .
\end{eqnarray}
The operators $\sigma^\pm$ are uniquely determined by the relations  above. One could in principle
calculate them in terms of modes. It is however more convenient to use decomposition (\ref{deco})
and to express all the fields in terms of the (Virasoro) chiral vertex operators:
\begin{eqnarray}
\nonumber
     \sigma^{+}(z ) &=&
      \begin{pmatrix}
      0&0           & 	\V{1}{\sigma}{\sigma}{z} &0  \\
      0&0  &0  & \frac{{\rm e}^{-{i\over 4}\pi}}{\sqrt{2}}\V{\varepsilon}{\sigma}{\sigma}{z} \\
      \V{\sigma}{\sigma}{1}{z} &0  & 0 &0  \\
      0& \frac{{\rm e}^{{i\over 4}\pi}}{\sqrt{2}}\V{\sigma}{\sigma}{\varepsilon}{z} &  0& 0
      \end{pmatrix}
\\[8pt]
 \label{def:chralnych operatorow}
     \sigma^{-}(z ) &=&
     \begin{pmatrix}
      0&0  &  0&	\V{1}{\sigma}{\sigma}{z}  \\
      0&0  & \frac{{\rm e}^{{i\over 4}\pi}}{\sqrt{2}}\V{\varepsilon}{\sigma}{\sigma}{z} &0  \\
      0& \frac{{\rm e}^{-{i\over 4}\pi}}{\sqrt{2}}\V{\sigma}{\sigma}{\varepsilon}{z}   & 0 &0  \\
	 \V{\sigma}{\sigma}{1}{z} & 0&  0& 0
      \end{pmatrix}
\\[8pt]
\nonumber
     \psi(z) &=&
    \begin{pmatrix}
     0 & \V{1}{\varepsilon}{\varepsilon}{z}  &  0&0  \\
    \V{\varepsilon}{\varepsilon}{1}{z}   &0  &  0&  0\\
     0 &0  &0  &\frac{{\rm e}^{{i\over 4}\pi}}{\sqrt{2}}\V{\sigma}{\varepsilon}{\sigma}{z} \\
      0&0  & 	\frac{{\rm e}^{-{i\over 4}\pi}}{\sqrt{2}}\V{\sigma}{\varepsilon}{\sigma}{z} 	  &0
    \end{pmatrix}
\end{eqnarray}
where we have applied the standard notation $1, \varepsilon, \sigma$ for
the  Ising  chiral fields with the weights $0,{1\over 2}, {1\over 16}$,
respectively.
The representation above can be easily verified using well known properties of the Ising
chiral vertices summarized in Appendix A.

\noindent In a similar way one can calculate the braiding relation we shall need in the following:
\begin{eqnarray}
\nonumber
\sigma^+(z)\sigma^+(w)
& = &
\frac{ {\rm e}^{{i\pi\epsilon\over 8}}}{\sqrt{2}}\,\sigma^+(w)\sigma^+(z)
+
\frac{{\rm e}^{-{3i\pi\epsilon\over 8}}}{\sqrt{2}}\,\sigma^-(w)\sigma^-(z),
\\[2pt]
\nonumber
\sigma^-(z)\sigma^-(w)
& = &
\frac{{\rm e}^{{i\pi\epsilon\over 8}}}{\sqrt{2}}\,\sigma^-(w)\sigma^-(z)
+
\frac{ {\rm e}^{-{3i\pi\epsilon\over 8}}}{\sqrt{2}}\,\sigma^+(w)\sigma^+(z),
\\[-6pt]
\label{braiding:sigma}
\\[-6pt]
\nonumber
\sigma^+(z)\sigma^-(w)
& = &
\frac{ {\rm e}^{{i\pi\epsilon\over 8}}}{\sqrt{2}}\,\sigma^+(w)\sigma^-(z)
+
i\frac{{\rm e}^{-{3i\pi\epsilon\over 8}}}{\sqrt{2}}\,\sigma^-(w)\sigma^+(z),
\\[2pt]
\nonumber
\sigma^-(z)\sigma^+(w)
& = &
\frac{{\rm e}^{{i\pi\epsilon\over 8}}}{\sqrt{2}}\,\sigma^-(w)\sigma^+(z)
-
i\frac{ {\rm e}^{-{3i\pi\epsilon\over 8}}}{\sqrt{2}}\,\sigma^+(w)\sigma^-(z).
\end{eqnarray}
It follows from (\ref{def:chralnych operatorow}) that
\begin{equation}
\label{sigma:matrix:elements:even}
\bra{w^+}\sigma^+(z)\ket{\,\xi\,} = \bra{w^-}\sigma^-(z)\ket{\,\xi\,},\;\;\;
\bra{\,\xi\,}\sigma^+(z)\ket{w^+} = \bra{\,\xi\,}\sigma^-(z)\ket{w^-},
\end{equation}
for even $\xi\in {\cal V}$ and
\begin{equation}
\label{sigma:matrix:elements:odd}
i\bra{w^+}\sigma^-(z)\ket{\,\xi\,} =\bra{w^-}\sigma^+(z)\ket{\,\xi\,},\;\;\;
\bra{\,\xi\,}\sigma^-(z)\ket{w^+} = i\bra{\,\xi\,}\sigma^+(z)\ket{w^-},
\end{equation}
for odd $\xi\in {\cal V}$.
One also has
\begin{equation}
\label{psi:matrix:elements}
\bra{w^+}\psi(z)\ket{w^-} = i\bra{w^-}\psi(z)\ket{w^+}.
\end{equation}

It is probably worth to mention that in the chiral fermion theory described above
the algebra of chiral fields $1,\psi, \sigma^+,\sigma^-$ does not close with respect to OPE.
For instance  the OPE of $\sigma^+$ with itself  contains a new local operator which is
neither a conformal primary nor a descendent of a primary field.
In this respect the chiral fermion with a partity operator in both sectors is not a complete chiral CFT.
In consequence the corresponding superscalar model we shall describe in the next subsection is not a complete
chiral $N=1$ superconformal field theory.
Nevertheless,  it provides an appropriate representation of the chiral vertex operators in the $N=1$ superconformal theory.

\subsection{Chiral  fields}

The chiral boson field with periodic boundary can be defined in terms of the decomposition
\begin{eqnarray*}
\label{chiral:scalar}
\varphi(z)
& = &
{\sf q}
-i
 \ln(z)\;{\sf p}
+
\varphi_<(z)
+
\varphi_>(z),\\
\varphi_<(z)
& = &
i \sum\limits_{n = -\infty}^{-1} \frac{{\sf a}_n}{n}\, z^{-n},
\hspace{10pt}
\varphi_>(z)
\; = \;
i\sum\limits_{n = 1}^{\infty} \frac{{\sf a}_n}{n}\, z^{-n},
\end{eqnarray*}
where the modes satisfy
\begin{equation}
\label{modes:conjugation:bosonic}
[{\sf q}, {\sf p}] = i,
\hspace{10pt}
[{\sf a}_m,{\sf a}_n] = m\delta_{m+n},
\hspace{10pt}
{\sf p}^{\dagger} = {\sf p},
\hspace{10pt}
{\sf q}^{\dagger} = {\sf q},
\hspace{10pt}
{\sf a}^{\dagger}_{n} = {\sf a}_{-n}.
\end{equation}
They are realized on the Hilbert space ${\cal H}_{\rm B} = L^2({\mathbb R})\otimes {\cal F}_{\rm B},$
where ${\cal F}_{\rm B}$ is the Fock space with the
vacuum state $|\Omega_{\rm B} \rangle$
defined by   the conditions $
{\sf a}_n|\Omega_{\rm B}\rangle = 0,\; n> 0,\; \langle\Omega_{\rm B}| \Omega_{\rm B} \rangle=1.
$
The superscalar Hilbert space is defined as a  tensor product
$$
{\cal H}= {\cal H}_{\rm B}\otimes {\cal H}_{\rm F}=
\left(L^2({\mathbb R})\otimes {\cal F}_{\rm B}\otimes {\cal F}_{\rm NS}\right)\,
\oplus\,
\left(
L^2({\mathbb R})\otimes {\cal F}_{\rm B}\otimes {\cal F}_{\rm R}
\right)\
$$
and carries the representation of the $N=1$ superconformal algebra
with the central charge $c = \frac32 + 3Q^2$ :
\begin{eqnarray}
\nonumber
L_0 & = & \frac18Q^2 + \frac12 {\sf p}^2 +
\sum\limits_{m \geqslant 1}{\sf a}_{-m}{\sf a}_m +
\sum\limits_{k>0}\left(k+\textstyle{1\over 2}\right)\psi_{-k}\psi_k + \Big\{
\begin{array}{llrll}\\[-20pt]
\scriptstyle0& &\scriptstyle {\rm NS}\;{\rm sector}\\   [-6pt]
{\textstyle \scriptstyle{1\over 16}}& &\scriptstyle{\rm R}\;{\rm sector}
\end{array}
\ ,
\\
\label{N=1:alebra:rep}
L_n & = & \left({\sf p}+\frac{inQ}{2}\right){\sf a}_n +
\frac12\sum\limits_{m\neq 0,n}\!{\sf a}_{n-m}{\sf a}_m +
{\textstyle{1\over 4}}\sum\limits_{k}(2k-m)\,\psi_{m-k}\psi_k\, ,
\\
\nonumber
S_k & = & ({\sf p}+ iQk)\psi_k +\sum\limits_{m\neq 0}{\sf a}_{m}\psi_{k-m},
\end{eqnarray}
where $k,l \in \mathbb{Z}+{1\over 2}$ in  the NS sector
and $k,l \in \mathbb{Z}$ in the R sector.

 The superscalar Hilbert space ${\cal H}$ can be seen as a direct integral over the specturm of the operator $\sf  p$:
$$
{\cal H} = \int\limits_{\mathbb{R}}^\oplus ({\cal H}^p_{\rm NS} \oplus {\cal H}^p_{\rm R})\, dp
$$
where
$
{\cal H}^p_{\rm NS} = |p\rangle \otimes {\cal F}_{\rm B} \otimes {\cal F}_{\rm NS}
$
,
$
{\cal H}^p_{\rm R}=|p\rangle \otimes {\cal F}_{\rm B} \otimes {\cal F}_{\rm NS},
$
and
${\sf p} |p\rangle = p |p\rangle$.
The representation (\ref{N=1:alebra:rep}) defines on ${\cal H}^p_{\rm NS}$ the structure of the NS Verma module
${\cal V}_\Delta$ with the conformal weight $\Delta= \frac18Q^2 + \frac12p^2$,
and the structure of the R Verma module ${\cal W}_\Delta$
with the conformal weight $\Delta=\frac18Q^2+ \frac{1}{16}+ \frac12p^2$ on
${\cal H}^p_{\rm R}$.

For our purposes it is convenient to work with the boson and the fermion fields on the unit circle, transformed
back to the zero time slice of the infinite cylinder:
\begin{eqnarray*}
\varphi_<(\sigma)
& = &
i \!\!\sum\limits_{n = -\infty}^{-1}\!\! \frac{{\sf a}_n}{n}\, {\rm e}^{-in\sigma},
\hspace{8pt}
\varphi_>(\sigma)
\; = \;
i\sum\limits_{n = 1}^{\infty} \frac{{\sf a}_n}{n}\, {\rm e}^{-in\sigma},
\hspace{8pt}
\psi(\sigma)
\;=\;
\!\!\sum\limits_{k } \psi_k\, {\rm e}^{-ik\sigma},
\hspace{8pt}\sigma^\pm(\sigma).
\end{eqnarray*}
In terms of these fields we construct the ordered exponential
\begin{eqnarray*}
\label{exponent}
{\sf E}^\alpha(\sigma)
&=&
{\rm e}^{\frac12\alpha{\sf q}}\,
{\rm e}^{\alpha\varphi_{<}(\sigma)}\,
{\rm e}^{\alpha\sigma{\sf p}}\,
{\rm e}^{\alpha\varphi_{>}(\sigma)}\,
{\rm e}^{\frac12\alpha{\sf q}},
\end{eqnarray*}
and the screening charge operators in both sectors:
\begin{eqnarray*}
\label{screening:charge}
{\sf Q}(\sigma)
& = &
\int\limits_\sigma^{\sigma+2\pi}\!\!\!dx\ \psi(x){\sf E}^b(x),\;\;\;Q=b+{1\over b},
\end{eqnarray*}
satisfying
\begin{equation}
\label{comm:Ln:Epsi}
\begin{array}{rcll}
\left[L_n,{\sf E}^\alpha(\sigma)\right]
& = &
\displaystyle
{\rm e}^{in\sigma}\left(
-i\frac{d}{d\sigma}
+
n\Delta_\alpha
\right){\sf E}^\alpha(\sigma),
\hspace{10pt}
\Delta_\alpha=\frac12\alpha(Q-\alpha),
\\     [6pt]
\left[S_k,{\sf E}^\alpha(\sigma)\right]
&=&
-i\alpha\, {\rm e}^{ik\sigma}\,\psi(\sigma)\,{\sf E}^\alpha(\sigma)
 ,
 \\[6pt]
                  \left[L_n,{\sf Q}(\sigma)\right]
&=&\left\{S_k,{\sf Q}(\sigma)\right\}\;=\;0,
\end{array}
\end{equation}
where $k\in \mathbb{Z}+{1\over 2}$ in the NS sector and $k\in \mathbb{Z}$ in the R sector.
For real $b$ the screening charge ${\sf Q}(\sigma)$ is  hermitian, its square is  positive and
$\left[{\sf Q}(\sigma)^2\right]^t$ may be uniquely defined for complex $t.$
Following \cite{Chorazkiewicz:2008es} we define the ``even'' and the ``odd'' complex powers of the screening charge
\begin{equation}
\label{even:and:odd:complex:powers}
{\sf Q}^{s}_{\rm e}(\sigma)
=
\left({\sf Q}^2(\sigma)\right)^{\frac{s}{2}} \;,\;\;\;
{\sf Q}^{s}_{\rm o}(\sigma)
=
{\sf Q}(\sigma)\left({\sf Q}^2(\sigma)\right)^{\frac{s-1}{2}},
\end{equation}
and the compositions
\begin{equation}
\label{g:definition}
{\sf g}_{{\scriptscriptstyle \rm\bf f}\,s}^{\hskip 4pt\alpha}\!(\sigma)
= {\sf E}^{\alpha}(\sigma){\sf Q}^{s}_{\scriptscriptstyle \rm \bf f}(\sigma).
\end{equation}
The chiral NS fields are diagonal with respect to the sector decomposition
\begin{equation}
\label{chiral:NSfields}
\begin{array}{rcll}
{\sf V}_{{\scriptscriptstyle \rm \bf f}\;s}^{\;\;\alpha}(\sigma)
& = &
\left[\begin{array}{cc}
{\sf g}_{{\scriptscriptstyle \rm\bf f}\,s}^{\hskip 4pt\alpha}\!(\sigma)
& 0\\[2pt]
0 & 
{\sf g}_{{\scriptscriptstyle \rm\bf f}\,s}^{\hskip 4pt\alpha}\!(\sigma)
\end{array}\right],
\\[14pt]
{\sf V}_{{\scriptscriptstyle \rm \bf f}\;s}^{*\,\alpha}(\sigma)
& = &
\left[\begin{array}{cc}
[S_{-{1\over 2}},{\sf E}^{\alpha}(\sigma)]
{\sf Q}^{s}_{\scriptscriptstyle \rm \bar{\bf f}}(\sigma)& 0\\[2pt]
0 & [S_0,{\sf E}^{\alpha}(\sigma)]
{\sf Q}^{s}_{\scriptscriptstyle \rm \bar{\bf f}}(\sigma)
\end{array}\right]\; =
\\[14pt]
& = &
\left[\begin{array}{cc}
-i\alpha\,{\rm e}^{-i\frac{\sigma}{2}}\psi(\sigma)
{\sf g}_{\bar{\scriptscriptstyle \rm  \bf f}\,s}^{\hskip 4pt\alpha}\!(\sigma)
& 0\\[2pt]
0 & -i\alpha\psi(\sigma)
{\sf g}_{\bar{\scriptscriptstyle \rm \bf f}\,s}^{\hskip 4pt\alpha}\!(\sigma)
\end{array}\right],
\end{array}
\end{equation}
where ${\scriptstyle\rm\bf f}= {\rm e,\,o}$ and for ${\scriptstyle\rm\bf f}= {\rm e}$ (resp.\ ${\scriptstyle\rm\bf f}= {\rm o})$
we have $\bar{\scriptstyle \rm\bf f}= {\rm o}$ (resp.\ $\bar{\scriptstyle\rm\bf f}= {\rm e}$).
The chiral R fields are off-diagonal:
\begin{equation}
\label{chiral:Rfields}
\begin{array}{rcll}
{\sf W}_{\;{\scriptscriptstyle \rm \bf f}\;s}^{+\,\beta}(\sigma)
&=&
\sigma^+(\sigma)
{\sf g}_{{\scriptscriptstyle \rm \bf f}\,s}^{{Q\over 2}-\sqrt{2}\beta}\!(\sigma),
\\
{\sf W}_{\;{\scriptscriptstyle \rm \bf f}\;s}^{-\,\beta}(\sigma)
&=&
\sigma^-(\sigma)
{\sf g}_{\bar{\scriptscriptstyle \rm \bf f}\,s}^{{Q\over 2}-\sqrt{2}\beta}\!(\sigma).
\end{array}
\end{equation}
The fields  can be extended to Euclidean fields on the whole cylinder
 by the analytic con\-ti\-nu\-a\-tion to the imaginary time
\begin{equation}
\label{Euclidean:fields}
\begin{array}{rcll}
{\sf V}_{{\scriptscriptstyle \rm \bf f}\;s}^{\;\;\alpha}(w)
&=&
{\rm e}^{\tau L_0}\,
{\sf V}_{{\scriptscriptstyle \rm \bf f}\;s}^{\;\;\alpha}(\sigma)
\,{\rm e}^{-\tau L_0}    ,
\\
{\sf V}_{{\scriptscriptstyle \rm \bf f}\;s}^{*\,\alpha}(w)
&=&
{\rm e}^{\tau L_0}\,
{\sf V}_{{\scriptscriptstyle \rm \bf f}\;s}^{*\,\alpha}(\sigma)
\,{\rm e}^{-\tau L_0}           ,
\\
{\sf W}_{\;{\scriptscriptstyle \rm\bf f}\;s}^{\pm\,\beta}(w)
&=&
{\rm e}^{\tau L_0}\,
{\sf W}_{\;{\scriptscriptstyle \rm\bf f}\;s}^{\pm\,\beta}(\sigma)\,{\rm e}^{-\tau L_0},
\hspace{10pt}
w = \tau+i\sigma.
\end{array}
\end{equation}
They
are related to the fields  on the complex plane $z={\rm e}^w$ via
\begin{equation}
\label{complex:plane:fields}
\begin{array}{rcllllllllll}
{\sf V}_{{\scriptscriptstyle \rm \bf f}\;s}^{\;\;\alpha}(w)
&=&
 {z}^{\Delta_\alpha}\,
{\sf V}_{{\scriptscriptstyle \rm \bf f}\;s}^{\;\;\alpha}(z),
&&
\\
{\sf V}_{{\scriptscriptstyle \rm \bf f}\;s}^{*\,\alpha}(w)
&=&
 {z}^{\Delta_\alpha+{1\over 2}}\,
{\sf V}_{{\scriptscriptstyle \rm \bf f}\;s}^{*\,\alpha}(z),
&&
\Delta_\alpha =   {\textstyle {1\over 2} }\alpha (Q-\alpha),
\\
{\sf W}_{\;{\scriptscriptstyle \rm\bf f}\;s}^{\pm\,\beta}(w)
&=&
{z}^{\Delta_\beta}\,{\sf W}_{\;{\scriptscriptstyle \rm\bf f}\;s}^{\pm\,\beta}(z),
&&
\Delta_\beta = \frac{c}{24}  -\beta^2.
\end{array}
\end{equation}
The chiral fields
satisfy a simple
braiding relation with functions of $\sf p$:
\begin{equation*}
\label{commutation:with:p}
\begin{array}{rcllllll}
{\sf V}_{{\scriptscriptstyle \rm \bf f}\;s}^{\;\;\alpha}(w)
f({\sf p})
&=&
f\left({\sf p}-i\left(\alpha+bs\right)\right)
{\sf V}_{{\scriptscriptstyle \rm \bf f}\;s}^{\;\;\alpha}(w)
 ,
 \\
{\sf V}_{{\scriptscriptstyle \rm \bf f}\;s}^{*\,\alpha}(w)
f({\sf p})
&=&
f\left({\sf p}-i\left(\alpha+bs\right)\right)
{\sf V}_{{\scriptscriptstyle \rm \bf f}\;s}^{*\,\alpha}(w)
,
\\
{\sf W}_{\;{\scriptscriptstyle \rm\bf f}\;s}^{\pm\,\beta}(w)f({\sf p})
&=&
f\left({\sf p}-i\left({\textstyle{Q\over 2}}-\sqrt{2} \beta+bs\right)\right){\sf W}_{\;{\scriptscriptstyle \rm\bf f}\;s}^{\pm\,\beta}(w).
\end{array}
\end{equation*}
An important feature of the fields  ${\sf V}_{{\scriptscriptstyle \rm \bf f}\;s}^{\;\;\alpha}(z),
{\sf V}_{{\scriptscriptstyle \rm \bf f}\;s}^{*\,\alpha}(z)$
 is that for the conformal weights
\begin{eqnarray*}
\Delta_1&=& {Q^2\over 8}+ {1\over 2}p^2,\;\;\;
\Delta_2 \;=\;  {1\over 2} \alpha(Q-\alpha),\;\;\;
\Delta_3\;=\; {Q^2\over 8}+{1\over 2}\left(p-i\left[\alpha +bs\right]\right)^2,
\end{eqnarray*}
 there exists a unique form
$
 \varrho_{\scriptscriptstyle \rm NN}(\xi_3, \xi_2, \xi_1|z):
\mathcal{V}_{\Delta_3} \times \mathcal{V}_{\Delta_2} \times \mathcal{V}_{\Delta_1} \rightarrow \mathbb{C},
$
satisfying the Ward identities
(\ref{ward:NN})
 such that
\begin{eqnarray*}
\begin{array}{rclllll}
  \varrho_{\scriptscriptstyle \rm NN}(\xi_3, \nu_2, \xi_1|z)
  &=&
  \bra{\xi_3}    {\sf V}_{{\rm e}\;s}^{\;\;\alpha}(z) \ket{\xi_1}
  \\
  \varrho_{\scriptscriptstyle \rm NN}(\xi_3, *\nu_2, \xi_1|z)
  &=&
  \bra{\xi_3}    {\sf V}_{{\rm e}\;s}^{*\,\alpha}(z) \ket{\xi_1}
\end{array}
&&{\rm for}\hspace{5pt} |\xi_3|+|\xi_1| \;\;{\rm even},
\\
\begin{array}{rcl}
  \varrho_{\scriptscriptstyle \rm NN}(\xi_3, \nu_2, \xi_1|z)
  &=&
  \bra{\xi_3}    {\sf V}_{{\rm o}\;s}^{\;\;\alpha}(z) \ket{\xi_1}
  \\
  \varrho_{\scriptscriptstyle \rm NN}(\xi_3, *\nu_2, \xi_1|z)
  &=&
  \bra{\xi_3}    {\sf V}_{{\rm o}\;s}^{*\,\alpha}(z) \ket{\xi_1}
\end{array}
&&{\rm for}\hspace{5pt} |\eta_3|+|\eta_1| \;\;{\rm odd}.
\end{eqnarray*}
It thus follows from  definition (\ref{chiral:vertex:NN}) that one can use
  ${\sf V}_{{\scriptscriptstyle \rm \bf f}\;s}^{\;\;\alpha}(z)$ and
${\sf V}_{{\scriptscriptstyle \rm \bf f}\;s}^{*\,\alpha}(z)$
to represent the chiral vertex operators in the NS sector:
\begin{equation}
\label{chiral:matrix:elements:NN}
\begin{array}{rcl}
\langle \xi_3|V_{\rm e}\chiral{\Delta_2}{\Delta_3}{\Delta_1}
(z)|\xi_1\rangle
&=&        \displaystyle
{\bra{\xi_3}    {\sf V}_{{\rm e}\;s}^{\;\;\alpha}(z) \ket{\xi_1}
\over
\bra{\nu_3}    {\sf V}_{{\rm e}\;s}^{\;\;\alpha}(1) \ket{\nu_1}} ,
\\   [12pt]
\langle \xi_3|
V_{\rm o}\chiral{\Delta_2}{\Delta_3}{\Delta_1}
(z)|\xi_1\rangle
&=&       \displaystyle
{\bra{\xi_3}    {\sf V}_{{\rm o}\;s}^{\;\;\alpha}(z) \ket{\xi_1}
\over
\bra{\nu_3}    {\sf V}_{{\rm e}\;s}^{*\;\alpha}(1) \ket{\nu_1}} ,
\\        [12pt]
\langle \xi_3|
V_{\rm e}\chiral{*\Delta_2}{\Delta_3}{\Delta_1}
(z)|\xi_1\rangle
&=&        \displaystyle
{\bra{\xi_3}    {\sf V}_{{\rm e}\;s}^{*\;\alpha}(z) \ket{\xi_1}
\over
\bra{\nu_3}    {\sf V}_{{\rm e}\;s}^{*\;\alpha}(1) \ket{\nu_1}}    ,
\\   [12pt]
\langle \xi_3|
V_{\rm o}\chiral{*\Delta_2}{\Delta_3}{\Delta_1}
(z)|\xi_1\rangle
&=&              \displaystyle
{\bra{\xi_3}    {\sf V}_{{\rm o}\;s}^{*\;\alpha}(z) \ket{\xi_1}
\over
\bra{\nu_3}    {\sf V}_{{\rm e}\;s}^{\;\;\alpha}(1) \ket{\nu_1}}    \ .
\end{array}
\end{equation}
A similar
property holds in the R sector.
For the conformal weights
\begin{eqnarray*}
\Delta_1&=& {c\over 24}+ {1\over 2}p^2,\;\;\;
\Delta_2 \;=\;  {1\over 2} \alpha(Q-\alpha),\;\;\;
\Delta_3\;=\; {c\over 24}+{1\over 2}\big(p-i(\alpha +bs)\big)^2,
\end{eqnarray*}
there exists a unique form
$
 \varrho_{\scriptscriptstyle \rm RR}(\eta_3, \xi_2, \eta_1|z):
\mathcal{W}_{\Delta_3} \times \mathcal{V}_{\Delta_2} \times \mathcal{W}_{\Delta_1} \rightarrow \mathbb{C},
$
satisfying the Ward identities
(\ref{ward:RR})
 such that
\begin{eqnarray*}
\begin{array}{rclllll}
  \varrho_{\scriptscriptstyle \rm RR}(\eta_3, \nu_2, \eta_1|z)
  &=&
  \bra{\eta_3}    {\sf V}_{{\rm e}\;s}^{\;\;\alpha}(z) \ket{\eta_1}
  \\
  \varrho_{\scriptscriptstyle \rm RR}(\eta_3, *\nu_2, \eta_1|z)
  &=&
  \bra{\eta_3}    {\sf V}_{{\rm e}\;s}^{*\,\alpha}(z) \ket{\eta_1}
\end{array}
&&{\rm for}\hspace{5pt} |\eta_3|+|\eta_1| \;\;{\rm even},
\\
\begin{array}{rcl}
  \varrho_{\scriptscriptstyle \rm RR}(\eta_3, \nu_2, \eta_1|z)
  &=&
  \bra{\eta_3}    {\sf V}_{{\rm o}\;s}^{\;\;\alpha}(z) \ket{\eta_1}
  \\
  \varrho_{\scriptscriptstyle \rm RR}(\eta_3, *\nu_2, \eta_1|z)
  &=&
  \bra{\eta_3}    {\sf V}_{{\rm o}\;s}^{*\,\alpha}(z) \ket{\eta_1}
\end{array}
&&{\rm for}\hspace{5pt} |\eta_3|+|\eta_1| \;\;{\rm odd}.
\end{eqnarray*}
From definitions (\ref{chiral:NSfields}), (\ref{Euclidean:fields}), (\ref{complex:plane:fields})  and relation
(\ref{psi:matrix:elements}) one gets
\begin{equation}
\label{xx}
\begin{array}{rcrrrlllll}
\bra{\,w^+}{\sf V}_{{\rm e}\;s}^{\;\;\alpha}(z)\ket{\,w^+}
&=&
\bra{\,w^-}{\sf V}_{{\rm e}\;s}^{\;\;\alpha}(z)\ket{\,w^-}
,
\\
\bra{\,w^+}{\sf V}_{{\rm o}\;s}^{\;\;\alpha}(z)\ket{\,w^-}
&=&
i\bra{\,w^-}{\sf V}_{{\rm o}\;s}^{\;\;\alpha}(z)\ket{\,w^+}.
\end{array}
\end{equation}
Chiral vertices  (\ref{chiral:vertex:RNR}) can then be represented as follows
\begin{equation}
\label{chiral:matrix:elements:RR}
\begin{array}{rcl}
\langle \eta_3|V^{+}_{{\rm e}}\chiral{\Delta_2}{\beta_3}{\beta_1}(z)|\eta_1\rangle
&=&
\displaystyle
{ \bra{\eta_3}{\sf V}_{{\rm e}\;s}^{\;\;\alpha}(z)\ket{\eta_1}
\over
\bra{w^+_3}{\sf V}_{{\rm e}\;s}^{\;\;\alpha}(1)\ket{w^+_1}},
\\
[12pt]
\langle S_Iw^+_3|
V^{-}_{{\rm e}}\chiral{\Delta_2}{\beta_3}{\beta_1}
(z)
|S_Jw^+_1\rangle
&=&
(-1)^{\#I}
\displaystyle
{ \bra{S_Iw^-_3}{\sf V}_{{\rm o}\;s}^{\;\;\alpha}(z)\ket{S_Jw^+_1}
\over
\bra{w^-_3}{\sf V}_{{\rm o}\;s}^{\;\;\alpha}(1)\ket{w^+_1}}
\\
&=&
(-1)^{\#J}
\displaystyle
{ \bra{S_Iw^+_3}{\sf V}_{{\rm o}\;s}^{\;\;\alpha}(z)\ket{S_Jw^-_1}
\over
\bra{w^+_3}{\sf V}_{{\rm o}\;s}^{\;\;\alpha}(1)\ket{w^-_1}},
\\
[12pt]
\langle S_Iw^+_3|
V^{+}_{{\rm o}}\chiral{\Delta_2}{\beta_3}{\beta_1}
(z)
|S_Jw^+_1\rangle
&=&
(-1)^{\#I}
\displaystyle
{ \bra{S_Iw^-_3}{\sf V}_{{\rm e}\;s}^{\;\;\alpha}(z)\ket{S_Jw^+_1}
\over
\bra{w^-_3}{\sf V}_{{\rm e}\;s}^{\;\;\alpha}(1)\ket{w^-_1}}
\\
&=&
i(-1)^{\#J}
\displaystyle
{ \bra{S_Iw^+_3}{\sf V}_{{\rm e}\;s}^{\;\;\alpha}(z)\ket{S_Jw^-_1}
\over
\bra{w^+_3}{\sf V}_{{\rm e}\;s}^{\;\;\alpha}(1)\ket{w^+_1}},
\\  [12pt]
\langle \eta_3|
V^{-}_{{\rm o}}\chiral{\Delta_2}{\beta_3}{\beta_1}
(z)
|\eta_1\rangle
&=&
\displaystyle
{ \bra{\eta_3}{\sf V}_{{\rm o}\;s}^{\;\;\alpha}(z)\ket{\eta_1}
\over
\bra{w^+_3}{\sf V}_{{\rm o}\;s}^{\;\;\alpha}(1)\ket{w^-_1}},
\end{array}
\end{equation}
and
\begin{equation}
\label{chiral:matrix:elements:RR*}
\begin{array}{rcl}
\langle \eta_3|
V^{+}_{{\rm o}}\chiral{*\Delta_2}{\beta_3}{\beta_1}
(z)
|\eta_1\rangle
&=&
\displaystyle
{ \bra{\eta_3}{\sf V}_{{\rm o}\;s}^{*\,\alpha}(z)\ket{\eta_1}
\over
\bra{w^+_3}{\sf V}_{{\rm e}\;s}^{\;\;\alpha}(1)\ket{w^+_1}},
 \\
[12pt]
\langle S_Iw^+_3|
V^{-}_{{\rm e}}\chiral{*\Delta_2}{\beta_3}{\beta_1}
(z)
|S_Jw^+_1\rangle
&=&
(-1)^{\#I}
\displaystyle
{ \bra{S_Iw^-_3}{\sf V}_{{\rm o}\;s}^{*\,\alpha}(z)\ket{S_Jw^+_1}
\over
\bra{w^-_3}{\sf V}_{{\rm o}\;s}^{\;\;\alpha}(1)\ket{w^+_1}}
\\
&=&
(-1)^{\#J}
\displaystyle
{ \bra{S_Iw^+_3}{\sf V}_{{\rm o}\;s}^{*\,\alpha}(z)\ket{S_Jw^-_1}
\over
\bra{w^+_3}{\sf V}_{{\rm o}\;s}^{\;\;\alpha}(1)\ket{w^-_1}},
\\
[12pt]
\langle S_Iw^+_3|
V^{+}_{{\rm o}}\chiral{*\Delta_2}{\beta_3}{\beta_1}
(z)
|S_Jw^+_1\rangle
&=&
(-1)^{\#I}
\displaystyle
{ \bra{S_Iw^-_3}{\sf V}_{{\rm e}\;s}^{*\,\alpha}(z)\ket{S_Jw^+_1}
\over
\bra{w^-_3}{\sf V}_{{\rm e}\;s}^{\;\;\alpha}(1)\ket{w^-_1}}
\\
&= &
i(-1)^{\#J}
\displaystyle
{ \bra{S_Iw^+_3}{\sf V}_{{\rm e}\;s}^{*\,\alpha}(z)\ket{S_Jw^-_1}
\over
\bra{w^+_3}{\sf V}_{{\rm e}\;s}^{\;\;\alpha}(1)\ket{w^+_1}},
\\  [12pt]
\langle \eta_3|
V^{-}_{{\rm e}}\chiral{*\Delta_2}{\beta_3}{\beta_1}
(z)
|\eta_1\rangle
&=&
\displaystyle
{ \bra{\eta_3}{\sf V}_{{\rm e}\;s}^{*\,\alpha}(z)\ket{\eta_1}
\over
\bra{w^+_3}{\sf V}_{{\rm o}\;s}^{\;\;\alpha}(1)\ket{w^-_1}}.
\end{array}
\end{equation}
One can repeat the above considerations for the R chiral fields. For the conformal weights
\begin{eqnarray*}
\Delta_1&=& {c\over 24}+ {1\over 2}p^2,\;\;\;
\Delta_2 \;=\; {c\over 24} - \beta^2,\;\;\;
\Delta_3\;=\;{Q^2\over 8} +{1\over 2}\left(p-i\left[{Q\over 2}-\sqrt{2}\beta +bs\right]\right)^2,
\end{eqnarray*}
there exists a unique form
$
 \varrho_{\scriptscriptstyle \rm NR}(\xi_3, \eta_2, \eta_1|z):
\mathcal{V}_{\Delta_3} \times \mathcal{W}_{\Delta_2} \times \mathcal{W}_{\Delta_1} \rightarrow \mathbb{C},
$
satisfying the Ward identities
(\ref{Ward_rhoNR})
 such that
\begin{eqnarray*}
\begin{array}{rclllll}
  \varrho_{\scriptscriptstyle \rm NR}(\xi_3, w^+_2, \eta_1|z)
  &=&
  \bra{\xi_3}     {\sf W}_{\;{\rm e}\;s}^{+\,\beta}(z) \ket{\eta_1}
  \\
  \varrho_{\scriptscriptstyle \rm NR}(\xi_3, w^-_2, \eta_1|z)
  &=&
  \bra{\xi_3}    {\sf W}_{\;{\rm e}\;s}^{-\,\beta}(z) \ket{\eta_1}
\end{array}
&&{\rm for}\hspace{5pt} |\xi_3|+|\eta_1| \;\;{\rm even},
\\
\begin{array}{rcl}
  \varrho_{\scriptscriptstyle \rm NR}(\xi_3, w^+_2, \eta_1|z)
  &=&
  \bra{\xi_3}     {\sf W}_{\;{\rm o}\;s}^{+\,\beta}(z) \ket{\eta_1}
  \\
  \varrho_{\scriptscriptstyle \rm NR}(\xi_3, w^-_2, \eta_1|z)
  &=&
  \bra{\xi_3}    {\sf W}_{\;{\rm o}\;s}^{-\,\beta}(z) \ket{\eta_1}
\end{array}
&&{\rm for}\hspace{5pt} |\xi_3|+|\eta_1| \;\;{\rm odd}.
\end{eqnarray*}
In a similar way, for the conformal weights
\begin{eqnarray*}
\Delta_1&=&{Q^2\over 8} + {1\over 2}p^2,\;\;\;
\Delta_2 \;=\; {c\over 24} - \beta^2,\;\;\;
\Delta_3\;=\; {c\over 24} +{1\over 2}\left(p-i\left[{Q\over 2}-\sqrt{2}\beta +bs\right]\right)^2,
\end{eqnarray*}
the relations
\begin{eqnarray*}
\begin{array}{rclllll}
  \varrho_{\scriptscriptstyle \rm RN}(\eta_3, w^+_2, \xi_1|z)
  &=&
  \bra{\eta_3}     {\sf W}_{\;{\rm e}\;s}^{+\,\beta}(z) \ket{\xi_1}
  \\
  \varrho_{\scriptscriptstyle \rm RN}(\eta_3, w^-_2, \xi_1|z)
  &=&
  \bra{\eta_3}    {\sf W}_{\;{\rm e}\;s}^{-\,\beta}(z) \ket{\xi_1}
\end{array}
&&{\rm for}\hspace{5pt} |\eta_3|+|\xi_1| \;\;{\rm even},
\\
\begin{array}{rcl}
  \varrho_{\scriptscriptstyle \rm RN}(\eta_3, w^+_2, \xi_1|z)
  &=&
  \bra{\eta_3}     {\sf W}_{\;{\rm o}\;s}^{+\,\beta}(z) \ket{\xi_1}
  \\
  \varrho_{\scriptscriptstyle \rm RN}(\eta_3, w^-_2, \xi_1|z)
  &=&
  \bra{\eta_3}    {\sf W}_{\;{\rm o}\;s}^{-\,\beta}(z) \ket{\xi_1}
\end{array}
&&{\rm for}\hspace{5pt} |\eta_3|+|\xi_1| \;\;{\rm odd},
\end{eqnarray*}
uniquely define the form
$
 \varrho_{\scriptscriptstyle \rm RN}(\eta_3, \eta_2, \nu_1|z):
\mathcal{W}_{\Delta_3} \times \mathcal{W}_{\Delta_2}  \times \mathcal{V}_{\Delta_1}\rightarrow \mathbb{C}
$
satisfying the Ward identities (\ref{Ward_rhoRN}).
From definitions (\ref{chiral:Rfields}), (\ref{Euclidean:fields}), (\ref{complex:plane:fields})  and relations
(\ref{sigma:matrix:elements:even}), (\ref{sigma:matrix:elements:odd}) one gets
\begin{equation}
\label{xxx}
\begin{array}{rcllllllll}
\bra{\,\nu\,}{\sf W}_{\;{\rm e}\;s}^{+\,\beta}(z)\ket{\,w^+}
&=&
\bra{\,\nu\,}{\sf W}_{\;{\rm o}\;s}^{-\,\beta}(z)\ket{\,w^-}
,
\\
i\bra{\,\nu\,}{\sf W}_{\;{\rm o}\;s}^{+\,\beta}(z)\ket{\,w^-}
&=&
\bra{\,\nu\,}{\sf W}_{\;{\rm e}\;s}^{-\,\beta}(z)\ket{\,w^+},
\\
\bra{w^+}{\sf W}_{\;{\rm e}\;s}^{+\,\beta}(z)\ket{\,\nu\,}
&=&
\bra{w^-}{\sf W}_{\;{\rm o}\;s}^{-\,\beta}(z)\ket{\,\nu\,}
,
\\
i\bra{w^-}{\sf W}_{\;{\rm o}\;s}^{+\,\beta}(z)\ket{\,\nu\,}
&=&
\bra{w^+}{\sf W}_{\;{\rm e}\;s}^{-\,\beta}(z)\ket{\,\nu\,}.
\end{array}
\end{equation}
Using identities (\ref{tNR1}), (\ref{tRN1}) and (\ref{xx}) one can express universal forms
$\rho_{\scriptscriptstyle }^{\pm\pm},\rho_{\scriptscriptstyle }^{\pm\mp}$
and then the matrix elements of chiral vertex operators
(\ref{chiral:vertex:definition}) in terms of  matrix elements
of  operators ${\sf W}_{\;{\scriptscriptstyle \rm\bf f}\;s}^{\pm\,\beta}(z)$.
The result is:
\begin{equation}
\label{chiral:matrix:elements}
\begin{array}{rcl}
\langle \xi_3|
V^{\pm}_{{\rm e}}\chiral{\beta_2}{\Delta_3}{\beta_1}
(z)|\eta_1\rangle
&=&
\displaystyle
{ \bra{\xi_3}{\sf W}_{\;{\rm e}\;s}^{\pm\,\beta_2}(z)\ket{\eta_1}
\over
\bra{\nu_3}{\sf W}_{\;{\rm e}\;s}^{\pm\,\beta_2}(z)\ket{w^+_1}},
\\  [12pt]
\langle \xi_3|
V^{\pm}_{{\rm o}}\chiral{\beta_2}{\Delta_3}{\beta_1}
(z)|\eta_1\rangle
&=&
\displaystyle
{ \bra{\xi_3}{\sf W}_{\;{\rm o}\;s}^{\mp\,\beta_2}(z)\ket{\eta_1}
\over
\bra{\nu_3}{\sf W}_{\;{\rm o}\;s}^{\mp\,\beta_2}(z)\ket{w^-_1}},
\\      [12pt]
\langle \eta_3|
V^{\pm}_{{\rm e}}\chiral{\beta_2}{\beta_3}{\Delta_1}
(z)|\xi_1\rangle
&=&
\displaystyle
{ \bra{\eta_3}{\sf W}_{\;{\rm e}\;s}^{\pm\,\beta_2}(z)\ket{\xi_1}
\over
\bra{w^+_3}{\sf W}_{\;{\rm e}\;s}^{\pm\,\beta_2}(z)\ket{\nu_1}},
\\            [12pt]
\langle \eta_3|
V^{\pm}_{{\rm o}}\chiral{\beta_2}{\beta_3}{\Delta_1}
(z)|\xi_1\rangle
&=&
\displaystyle
{ \bra{\eta_3}{\sf W}_{\;{\rm o}\;s}^{\mp\,\beta_2}(z)\ket{\xi_1}
\over
\bra{w^-_3}{\sf W}_{\;{\rm o}\;s}^{\mp\,\beta_2}(z)\ket{\nu_1}}.
\end{array}
\end{equation}

\subsection{Matrix elements}
In this subsection we shall calculate the matrix elements
\[
\langle \nu_3|{\sf W}_{\;{\scriptscriptstyle \rm \bf f}\;s}^{\pm\,\beta}(1)|w_1\rangle,
\hskip .7cm
\langle w_3|{\sf W}_{\;{\scriptscriptstyle \rm \bf f}\;s}^{\pm\,\beta}(1)|\nu_1\rangle
\hskip .5cm
{\rm and}
\hskip 7mm
\langle w_3|{\sf V}_{\:{\scriptscriptstyle \rm \bf f}\;s}^{\underline{\hspace*{4pt}}\alpha}(1)|w_1\rangle,
\]
using a suitable modification of the procedure proposed in \cite{Teschner:2003en} and adapted to the NS sector of the $N=1$ superconformal theory in \cite{Chorazkiewicz:2008es}.
The idea is to find an explicit form of an appropriate four-point, chiral  correlator containing a degenerate Ramond field and then, by studying its different limits,
to express the matrix elements we are after through the matrix elements of the chiral NS field computed in \cite{Chorazkiewicz:2008es}.

For $\beta = \beta_+ \equiv \frac{1}{2\sqrt2}\left(b^{-1} + 2b\right) $
the Ramond supermodule ${\cal W}_{\beta_+}$
is degenerate (with respect to  scalar product (\ref{scalar:products})). The vector
\[
\chi_{+} \; = \; \left(\kappa_+ L_{-1} - S_{-1}S_0\right)w_{\beta_+},
\hskip 1cm
\kappa_+ = \sqrt{2}b^{-1}\beta_+
\]
is orthogonal  to all vectors in ${\cal W}_{\beta_+}.$
Using (\ref{action:of:S0}) we can rewrite the condition that $\chi_+$ is null  in the form of a pair of operator equations
\begin{equation}
\label{null:vector:equations}
\frac{\sqrt 2}{b}\frac{\partial}{\partial z} {\sf W}_{\;{ \rm e}\;s}^{+\,\beta_+}(z)
 = i\,{\rm e}^{-\frac{i\pi}{4}}{S}_{-1}{\sf W}_{\;{ \rm o}\;s}^{-\,\beta_+}(z),
\hskip 1cm
\frac{\sqrt 2}{b}\frac{\partial}{\partial z} {\sf W}_{\;{ \rm o}\;s}^{-\,\beta_+}(z)
= i\,{\rm e}^{\frac{i\pi}{4}} {S}_{-1}  {\sf W}_{\;{ \rm e}\;s}^{+\,\beta_+}(z),
\end{equation}
which should hold in arbitrary correlation function. Introducing
\begin{eqnarray}
\nonumber
f_{+}(z) & = &
\bra{\nu_4}
  {\sf W}_{\; {\rm e}\;s_3}^{+\,\beta_3}(1)
  {\sf W}_{\; { \rm e}\; }^{+\,\beta_+}(z)
\ket{\nu_1},
\\[4pt]
\nonumber
f_{-}(z) & = &
\bra{\nu_4}
  {\sf W}_{\;{ \rm o}\;s_3}^{-\,\beta_3}(1)
  {\sf W}_{\;{ \rm o}\; }^{-\,\beta_+}(z)
\ket{\nu_1},
\\[-8pt]
\\[-4pt]
\nonumber
g_{+}(z) & = &
{\rm e}^{\frac{i\pi}{4}}
\bra{\nu_4}
  {\sf W}_{\;{ \rm o}\;s_3}^{-\,\beta_3}(1)
  {\sf W}_{\;{ \rm e}\; }^{+\,\beta_+}(z)
|S_{-\frac12}\nu_1\rangle,
\\[4pt]
\nonumber
g_{-}(z) & = &
{\rm e}^{-\frac{i\pi}{4}}\langle\nu_4|
  {\sf W}_{\,{\rm e}\;s_3}^{\,+\,\beta_3}(1)
  {\sf W}_{\,{ \rm o}\; }^{\,-\,\beta_+}(z)
|S_{-\frac12}\nu_1\rangle,
\end{eqnarray}
we get in particular
\begin{eqnarray}
\nonumber
\left(\frac{\sqrt{2}}{b}\frac{\partial}{\partial z} - \frac{\frac12\beta_+}{z-1}\right)\!f_+(z) + \frac{\beta_3}{z-1}f_-(z)
& = &
-\frac{g_{-}(z)}{\sqrt{z(1-z)}},
\\[4pt]
\label{diff:eq:1}
\left(\frac{\sqrt{2}}{b}\frac{\partial}{\partial z} - \frac{\frac12\beta_+}{z-1}\right)\!f_-(z) + \frac{\beta_3}{z-1}f_+(z)
& = &
\frac{g_{+}(z)}{\sqrt{z(1-z)}},
\\[4pt]
\nonumber
\left(\frac{\sqrt{2}}{b}\frac{\partial}{\partial z} - \frac{\frac12\beta_+}{z-1}\right)\!g_+(z) + \frac{\beta_3}{z-1}g_-(z)
& = &
-\frac{1}{\sqrt{z(1-z)}}\left[\frac{\Delta_1}{z} + \Delta_{2+3-4} + (z-1)\frac{\partial}{\partial z}\right]\!f_{-}(z),
\\[4pt]
\nonumber
\left(\frac{\sqrt{2}}{b}\frac{\partial}{\partial z} - \frac{\frac12\beta_+}{z-1}\right)\!g_-(z) + \frac{\beta_3}{z-1}g_+(z)
& = &
\frac{1}{\sqrt{z(1-z)}}\left[\frac{\Delta_1}{z} + \Delta_{2+3-4} + (z-1)\frac{\partial}{\partial z}\right]\!f_{+}(z).
\end{eqnarray}
For the new function $h(z)$:
\[
f_+(z) + f_-(z) = (1-z)^{A}\,h(z)\;,\;\;\;A = \frac{b\alpha_3}{2}-\frac18\;,
\]
we obtain from (\ref{diff:eq:1})
\begin{equation}
\label{df:1}
	\frac{\partial^2 h(z)}{\partial z^2}
	+
	\frac12\left(\frac{1-b^2}{z} + \frac{2\mathfrak{A}}{z-1}\right)\frac{\partial h(z) }{\partial z}
	=
 	\frac{b^2}{2z(z-1)}\left[\frac{\Delta_1}{z} + \Delta_{2+3-4} + A\right]h(z),
\end{equation}
where $\mathfrak{A} = b\alpha_3-\frac12b^2.$

The solution of (\ref{df:1})  corresponding to the sum of correlators $f_+(z) + f_-(z)$
can be singled  out  by its
leading behavior at $z \to 0.$ It follows from the momentum conservation
that all the states obtained by the action
of ${\sf W}_{\;{ \scriptscriptstyle \rm\bf f}\; }^{\pm\,\beta_+}(z)$ on the vector $\nu_1$ have momenta equal to
\begin{equation}
\label{q:momentum}
q = p_1 + \frac{ib}{2}.
\end{equation}
The small $z$ behavior of $f_\pm(z)$ can thus be calculated  by inserting  a projection on the  highest weight states of
${\cal W}_{q}$ with an appropriately chosen parity. This gives
\begin{eqnarray}
\nonumber
\bra{\nu_4}
  {\sf W}_{\;{ \rm e}\;s_3}^{+\,\beta_3}(1)
  {\sf W}_{\;{ \rm e}\; }^{+\,\beta_+}(z)
\ket{\nu_1}
& = &
\bra{\nu_4}
  {\sf W}_{\;{ \rm e}\;s_3}^{+\,\beta_3}(1)
|w_q^+\rangle\, z^{\frac{b\alpha_1}{2}}\Big(1+ {\cal O}(z)\Big),
\\[-6pt]
\label{asym:1a}
\\[-6pt]
\nonumber
\bra{\nu_4}
  {\sf W}_{\;{ \rm o}\;s_3}^{-\,\beta_3}(1)
  {\sf W}_{\;{ \rm o}\; }^{-\,\beta_+}(z)
\ket{\nu_1}
& = &
\bra{\nu_4}
  {\sf W}_{\;{ \rm o}\;s_3}^{-\,\beta_3}(1)
|w_q^-\rangle\, z^{\frac{b\alpha_1}{2}}\Big(1+ {\cal O}(z)\Big),
\end{eqnarray}
where we have used the identities
\begin{eqnarray*}
\langle w_q^+|  {\sf W}_{\;{ \rm e}\; }^{+\,\beta_+}(1)|\nu_1\rangle
&=&
\langle q|{\sf E}^{-\frac{b}{2}}(1)|p_1\rangle\,
\langle \sigma^+\big|\sigma^{+}(1)|0\rangle
=1
,
\\
\langle w_q^-|  {\sf W}_{\;{ \rm o}\; }^{-\,\beta_+}(1)|\nu_1\rangle
&=&
\langle q|{\sf E}^{-\frac{b}{2}}(1)|p_1\rangle\,
\langle \sigma^-\big|\sigma^{-}(1)|0\rangle
=1.
\end{eqnarray*}
It follows from (\ref{xx}) that the asymptotics in (\ref{asym:1a}) are identical, hence
\begin{eqnarray}
\label{expl:sol:1}
f_+(z) + f_-(z)
& = &
\bra{\nu_4}
  {\sf W}_{\;{ \rm e}\;s_3}^{+\,\beta_3}(1)
|w_q^+\rangle\,
z^{\frac{b\alpha_1}{2}}\,(1-z)^{\frac{b\alpha_3}{2}-\frac18}\,
\\[6pt]
\nonumber
&& \hspace*{-7mm}
\times  \, {}_2F_1\left({\textstyle\frac{b}{4}}(2a_1+2\alpha_3-2\alpha_4-b),{\textstyle\frac{b}{4}}(2a_1+2\alpha_3-2\bar \alpha_4-b);{\textstyle\frac{1}{2}}(1-b^2+2b\alpha_1);z\right)\!,
\end{eqnarray}
where ${}_2F_1$ denotes the hypergeometric function.

In a similar way one can deduce from (\ref{diff:eq:1}) the differential equation satisfied by the difference $f_+(z)-f_-(z).$
Extracting  a suitable power of $(1-z)$ from $f_+(z)-f_-(z)$
we  obtain (\ref{df:1}) with different values of the parameters $A, \mathfrak{A}$.
 Equations (\ref{asym:1a}) and  (\ref{xx})
then give
\[
f_+(z)-f_-(z)= 0
\]
and consequently
\begin{eqnarray}
\label{expl:sol:2}
\nonumber
&&
\hskip -1.4cm
\bra{\nu_4}
  {\sf W}_{\;{ \rm e}\;s_3}^{+\,\beta_3}(1)
  {\sf W}_{\;{ \rm e}\; }^{+\,\beta_+}(z)
\ket{\nu_1}
=
\bra{\nu_4}
  {\sf W}_{\;{ \rm o}\;s_3}^{-\,\beta_3}(1)
  {\sf W}_{\;{ \rm o}\; }^{-\,\beta_+}(z)
\ket{\nu_1}
\\[6pt]
&= &
\bra{\nu_4}\,
  {\sf W}_{\;{ \rm e}\;s_3}^{+\,\beta_3}(1)
|w_q^+\rangle\,
z^{\frac{b\alpha_1}{2}}\,(1-z)^{\frac{b\alpha_3}{2}-\frac18}\,
\\[6pt]
\nonumber
&&
 \times \, {}_2F_1\left({\textstyle\frac{b}{4}}(2a_1+2\alpha_3-2\alpha_4-b),{\textstyle\frac{b}{4}}(2a_1+2\alpha_3-2\bar \alpha_4-b);{\textstyle\frac{1}{2}}(1-b^2+2b\alpha_1);z\right).
\end{eqnarray}
We shall now determine $\bra{\nu_4}\,{\sf W}_{\;{ \rm e}\;s_3}^{+\,\beta_3}(1)|w_q^+\rangle$
by comparing the leading behavior of the left and the right hand side
of (\ref{expl:sol:2}) for $z \to 1.$
We start by calculating the leading term of the OPE of
${\sf W}_{\;{ \rm e}\;s_3}^{\,+\,\beta_3}(z_3) = \sigma^+(z_3){\sf E}^{\alpha_3}(z_3){\sf Q}^{s_3}_{\rm e}(z_3)$
and
${\sf W}_{\;{ \rm e}\; }^{\,+\,\beta_+}(z_2) = \sigma^+(z_2){\sf E}^{-\frac{b}{2}}(z_2).$ Since
\[
{\sf Q}^{s_3}_{\rm e}(z_3)\sigma^+(z_2){\sf E}^{-\frac{b}{2}}(z_2)
=
{\sf Q}^{s_3}_{\rm e}(z_2)\sigma^+(z_2){\sf E}^{-\frac{b}{2}}(z_2)
+
\left({\sf Q}^{s_3}_{\rm e}(z_3)- {\sf Q}^{s_3}_{\rm e}(z_2)\right)\sigma^+(z_2){\sf E}^{-\frac{b}{2}}(z_2)
\]
we have for $z_3 \to z_2:$
\[
{\sf Q}^{s_3}_{\rm e}(z_3)\sigma^+(z_2){\sf E}^{-\frac{b}{2}}(z_2)
\sim
{\sf Q}^{s_3}_{\rm e}(z_2)\sigma^+(z_2){\sf E}^{-\frac{b}{2}}(z_2)
=
\sigma^+(z_2){\sf Q}^{s_3}_{\rm e}(z_2){\sf E}^{-\frac{b}{2}}(z_2)
\]
where the equality follows form the definition of the screening charge $\sf Q,$ its even power (\ref{even:and:odd:complex:powers})
and  braiding property (\ref{def_sigma}). Further, from the definition of the screening charge and  braiding properties of the
normal ordered exponentials (see the next section for details) one has,
\[
{\sf Q}^{s_3}_{\rm e}(z_2){\sf E}^{-\frac{b}{2}}(z_2)
=
{\rm e}^{\frac{i\pi}{2}b^2s_3}\,{\sf E}^{-\frac{b}{2}}(z_2){\sf Q}^{s_3}_{\rm e}(z_2),
\]
hence
\[
{\sf W}_{\;{ \rm e}\;s_3}^{\,+\,\beta_3}(z_3)
{\sf W}_{\;{ \rm e}\; }^{\,+\,\beta_+}(z_2)
\; \sim \;
{\rm e}^{\frac{i\pi}{2}b^2s_3}\, \sigma^+(z_3)\sigma^+(z_2){\sf E}^{\alpha_3}(z_3){\sf E}^{-\frac{b}{2}}(z_2){\sf Q}^{s_3}_{\rm e}(z_2).
\]
Finally, from (\ref{def:chralnych operatorow}) and (\ref{exponent}):
\[
 \sigma^+(z_3)\sigma^+(z_2) \; \sim \; (z_3-z_2)^{-\frac18},
 \hskip 1cm
 {\sf E}^{\alpha_3}(z_3){\sf E}^{-\frac{b}{2}}(z_2) \; \sim \;(z_3-z_2)^{\frac{b\alpha_3}{2}}{\sf E}^{\alpha_3-\frac{b}{2}}(z_2)
\]
and we get
\begin{eqnarray*}
{\sf W}_{\;{ \rm e}\;s_3}^{\,+\,\beta_3}(z_3)
{\sf W}_{\;{ \rm e}\; }^{\,+\,\beta_+}(z_2)
& \sim &
{\rm e}^{\frac{i\pi}{2}b^2s_3}\,
(z_3-z_2)^{\frac{b\alpha_3}{2}-\frac18}\,{\sf V}_{{\rm e}\;s_3}^{\;\;\alpha_3-\frac{b}{2}}(z_2),
\end{eqnarray*}
so that
\begin{eqnarray}
{\sf W}_{\;{ \rm e}\;s_3}^{\,+\,\beta_3}(1)
{\sf W}_{\;{ \rm e}\; }^{\,+\,\beta_+}(z)
& \sim  &
{\rm e}^{\frac{i\pi}{2}b^2s_3}\,
(1-z)^{\frac{b\alpha_3}{2}-\frac18}\,{\sf V}_{{\rm e}\;s_3}^{\;\;\alpha_3-\frac{b}{2}}(z)
\\[4pt]
\nonumber
& \sim &
{\rm e}^{\frac{i\pi}{2}b^2s_3}\,
(1-z)^{\frac{b\alpha_3}{2}-\frac18}\,{\sf V}_{{\rm e}\;s_3}^{\;\;\alpha_3-\frac{b}{2}}(1).
\end{eqnarray}
In conclusion, for $z \to 1:$
\begin{equation}
\label{sing:at:1:a}
\bra{\nu_4}
  {\sf W}_{\;{ \rm e}\;s_3}^{+\,\beta_3}(1)
  {\sf W}_{\;{ \rm e}\; }^{+\,\beta_+}(z)
\ket{\nu_1}
\; \sim \;
{\rm e}^{\frac{i\pi}{2}b^2s_3}\,(1-z)^{\frac{b\alpha_3}{2}-\frac18}\,
\bra{\nu_4}
{\sf V}_{{\rm e}\;s_3}^{\;\;\alpha_3-\frac{b}{2}}(1)
\ket{\nu_1}
\end{equation}
where the matrix element on the r.h.s.\ was calculated in \cite{Chorazkiewicz:2008es} and reads:
\begin{eqnarray}
\label{3pt:NS}
\bra{\nu_3}{\sf V}_{{\rm e}\;s}^{\;\;\alpha_2}(1)\ket{\nu_1}
& =&{\cal M}_{\alpha_3,\alpha_2,\alpha_1}^{\rm N \,N\, N},
\\[4pt]
\nonumber
&&\hspace{-100pt}   {\cal M}_{\alpha_3,\alpha_2,\alpha_1}^{\rm N \,N\, N}\equiv
{\cal N}_{\alpha_3\alpha_2\alpha_1}
\frac{\Gamma_{\rm NS}(Q+\alpha_{1-2-3})\Gamma_{\rm NS}(\alpha_{1+3-2})\Gamma_{\rm NS}(Q+\alpha_{3-1-2})\Gamma_{\rm NS}(2Q-\alpha_{1+2+3})}
  {\Gamma_{\rm NS}(Q)\Gamma_{\rm NS}(2\alpha_1)\Gamma_{\rm NS}(2\alpha_2)\Gamma_{\rm NS}(2Q-2\alpha_3)},
\end{eqnarray}
with $\alpha_3 = \alpha_1 + \alpha_2 + bs,$ $\alpha_{1-2-3} \equiv \alpha_1 - \alpha_2 - \alpha_3$ etc.\ and
\begin{equation}
{\cal N}_{\alpha_3\alpha_2\alpha_1} =
\left[ \frac12 \Gamma\left( \frac{bQ}{2} \right) b^{-\frac{bQ}{2}}\right]^{\frac{\alpha_3-\alpha_2-\alpha_1}{b}}
  {\rm e}^{\frac{i\pi}{2}\left( \alpha_3-\alpha_2-\alpha_1 \right)(Q-\alpha_3+\alpha_2-\alpha_1)}.
\end{equation}

On the other hand, we can analyze the $z\to 1$ behavior of the r.h.s.\ of (\ref{expl:sol:2}). Using
the analytic continuation formula for the hypergeometric function,
\begin{eqnarray}
\label{hyper:anal:cont}
{}_2F_1(a,b;c;z)
& = &
\frac{\Gamma(c)\Gamma(c-a-b)}{\Gamma(c-a)\Gamma(c-b)}\,
{}_2F_1(a,b;1+a+b-c;1-z)
\\[4pt]
\nonumber
& + &
\frac{\Gamma(c)\Gamma(a+b-c)}{\Gamma(a)\Gamma(b)}\,
(1-z)^{c-a-b}\,
{}_2F_1(c-a,c-b;1+c-a-b;1-z).
\end{eqnarray}
we obtain a coefficient in front of
$(1-z)^{\frac{b\alpha_3}{2}-\frac18}.$ It has the form
\begin{equation}
\label{coeff:rhs}
\frac{\Gamma \left(-\frac{b^2}{2}+b\alpha_1+\frac{1}{2}\right) \Gamma \left(\frac{b^2}{2}-b\alpha_3 +1\right)}
{\Gamma \left(\frac12 +\frac{b}{4} \left(2 \alpha_{1+4-3}-b\right)\right) \Gamma \left(1 +\frac{b}{4} \left(2 \alpha_{1-4-3}+b\right)\right)}
\bra{\nu_4}\,{\sf W}_{\;{ \rm e}\;s_3}^{+\,\beta_3}(1)|w_q^+\rangle,
\end{equation}
with  $q$ given by (\ref{q:momentum}). Comparing (\ref{coeff:rhs}) with the corresponding coefficient  in (\ref{sing:at:1:a}) we arrive at the identity
\begin{equation}
\begin{aligned}
\label{norm1:almost:final}
\bra{\nu_4}\,{\sf W}_{\;{ \rm e}\;s_3}^{+\,\beta_3}(1)|w_q^+\rangle
&=\\
&\hspace{-2.2cm}{\rm e}^{\frac{i\pi}{2}bs_3}\,
\frac{\Gamma \left(\frac12 +\frac{b}{4} \left(2 \alpha_{1+4-3}-b\right)\right) \Gamma \left(1 +\frac{b}{4} \left(2 \alpha_{1-4-3}+b\right)\right)}
{\Gamma \left(-\frac{b^2}{2}+b\alpha_1+\frac{1}{2}\right) \Gamma \left(\frac{b^2}{2}-b\alpha_3 +1\right)}
\bra{\nu_4}
{\sf V}_{{\rm e}\;s_3}^{\;\;\alpha_3-\frac{b}{2}}(1)
\ket{\nu_1}
\end{aligned}
\end{equation}
Using (\ref{3pt:NS}) and the ``shift identities'' for the Barnes functions (see Appendix \ref{Appendix:Barnes}) we finally get
\begin{eqnarray}
\label{norm1:final}
\bra{\nu_3}\,{\sf W}_{\;{ \rm e}\;s}^{+\,\beta_2}(1)|w_1^+\rangle&=&
\bra{\nu_3}\,{\sf W}_{\;{ \rm o}\;s}^{-\,\beta_2}(1)|w_1^-\rangle
\;=\;
{\cal M}_{\alpha_3,\alpha_2,\alpha_1}^{\rm R \,R\, N},
\\[4pt]
\nonumber
&&\hspace{-110pt} {\cal M}_{\alpha_3,\alpha_2,\alpha_1}^{\rm R \,R\, N}\equiv
{\cal N}_{\alpha_3\alpha_2\alpha_1}
\frac{\Gamma_{\rm R}(Q+\alpha_{1-2-3})\Gamma_{\rm R}(\alpha_{1+3-2})\Gamma_{\rm NS}(Q+\alpha_{3-1-2})\Gamma_{\rm NS}(2Q-\alpha_{1+2+3})}
{\Gamma_{\rm NS}(Q)\Gamma_{\rm R}(2\alpha_1)\Gamma_{\rm R}(Q-2\alpha_2)\Gamma_{\rm NS}(2Q-2\alpha_3)}
\end{eqnarray}
where the first relation of (\ref{xxx}) has been added for completeness.

The same procedure can be applied to other matrix elements. For instance, in order to compute
\(
\bra{\nu_3}\,{\sf W}_{\;{ \rm o}\;s}^{+\,\beta_2}(1)|w_1^-\rangle
\)
we start from  the function
$
\bra{\nu_4}\,{\sf W}_{\;{ \rm o}\;s_3}^{+\,\beta_3}(1)
{\sf W}_{\;{ \rm o}}^{-\,\beta_+}(z)|\nu_1\rangle
$.
The correlators
\[
\bra{\nu_4}\,{\sf W}_{\;{ \rm e}}^{+\,\beta_+}(z){\sf W}_{\;{ \rm e}\;s_2}^{+\,\beta_2}(1)|\nu_1\rangle
\hskip 1cm
{\rm and}
\hskip 1cm
\bra{\nu_4}\,{\sf W}_{\;{ \rm o}}^{-\,\beta_+}(z){\sf W}_{\;{ \rm o}\;s_2}^{+\,\beta_2}(1)|\nu_1\rangle,
\]
give the formulae for
\(
\langle w_3^+|\,{\sf W}_{\;{ \rm e}\;s}^{+\,\beta_2}(1)\ket{\nu_1}
\)
and
\(
\langle w_3^-|\,{\sf W}_{\;{ \rm o}\;s}^{+\,\beta_2}(1)\ket{\nu_1},
\)
respectively, and the matrix elements
\(
\langle w_3^+|\,{\sf V}_{{ \rm e}\;s}^{\;\;\alpha_2}(1)|w_1^+\rangle
\)
and
\(
\langle w_3^-|\,{\sf V}_{{ \rm o}\;s}^{\;\;\alpha_2}(1)|w_1^+\rangle
\)
can be obtained from the correlators
\[
\langle w_4^+|\,{\sf V}_{{ \rm e}\;s_3}^{\;\;\alpha_3}(1){\sf W}_{\;{ \rm e}}^{+\,\beta_+}(z)|\nu_1\rangle
\hskip 1cm
{\rm and}
\hskip 1cm
\langle w_4^-|\,{\sf V}_{{ \rm o}\;s_3}^{\;\;\alpha_3}(1){\sf W}_{\;{ \rm e}}^{+\,\beta_+}(z)|\nu_1\rangle.
\]
The result reads
\begin{equation}
\begin{array}{rlrll}
\hskip -5mm
\bra{\nu_3}\,{\sf W}_{\,{ \rm o}\;s}^{+\,\beta_2}(1)|w_1^-\rangle
& = &
-i\bra{\nu_3}\,{\sf W}_{\,{ \rm e}\;s}^{-\,\beta_2}(1)|w_1^+\rangle
& = &
{\cal M}_{\alpha_3,\alpha_2,\alpha_1}^{\rm N \,N\, R}\, ,
\\[6pt]
\langle w_3^+|\,{\sf W}_{\,{ \rm e}\;s}^{+\,\beta_2}(1)\ket{\nu_1}
& = &
\langle w_3^-|\,{\sf W}_{\,{ \rm o}\;s}^{-\,\beta_2}(1)\ket{\nu_1}
& = &
{\cal M}_{\alpha_3,\alpha_2,\alpha_1}^{\rm R \,N\, N}\, ,
\\[8pt]
\langle w_3^-|\,{\sf W}_{\,{ \rm o}\;s}^{+\,\beta_2}(1)\ket{\nu_1}
& = &
-i \langle w_3^+|\,{\sf W}_{\,{ \rm e}\;s}^{-\,\beta_2}(1)\ket{\nu_1}
& = &
{\cal M}_{\alpha_3,\alpha_2,\alpha_1}^{\rm N \,R\, R}\, ,
\\[8pt]
\langle w_3^+|\,{\sf V}_{{ \rm e}\;s}^{\;\,\alpha_2}(1)|w_1^+\rangle
& = &
\langle w_3^-|\,{\sf V}_{{ \rm e}\;s}^{\;\,\alpha_2}(1)|w_1^-\rangle
& = &
{\cal M}_{\alpha_3,\alpha_2,\alpha_1}^{\rm N \,R\, N}\, ,
\\[8pt]
\langle w_3^-|\,{\sf V}_{{ \rm o}\;s}^{\;\;\alpha_2}(1)|w_1^+\rangle
& = &
-i \langle w_3^+|\,{\sf V}_{{ \rm o}\;s}^{\;\;\alpha_2}(1)|w_1^-\rangle
& = &
{\cal M}_{\alpha_3,\alpha_2,\alpha_1}^{\rm R \,N\, R}\, ,
\end{array}
\end{equation}
where
\begin{eqnarray}
\nonumber
{\cal M}_{\alpha_3,\alpha_2,\alpha_1}^{\rm N \,N\, R}
 \!& \equiv  &\!
\frac{{\rm e}^{-\frac{3i\pi}{4}}{\cal N}_{\alpha_3\alpha_2\alpha_1}}{\sqrt{2}}
\frac{\Gamma_{\rm NS}(Q+\alpha_{1-2-3})\Gamma_{\rm NS}(\alpha_{1+3-2})\Gamma_{\rm R}(Q+\alpha_{3-1-2})\Gamma_{\rm R}(2Q-\alpha_{1+2+3})}
{\Gamma_{\rm NS}(Q)\Gamma_{\rm R}(2\alpha_1)\Gamma_{\rm R}(Q-2\alpha_2)\Gamma_{\rm NS}(2Q-2\alpha_3)},
\\[8pt]
\nonumber
{\cal M}_{\alpha_3,\alpha_2,\alpha_1}^{\rm R \,N\, N}
 \!& \equiv  &\!
{\cal N}_{\alpha_3\alpha_2\alpha_1}
\frac{\Gamma_{\rm R}(Q+\alpha_{1-2-3})\Gamma_{\rm NS}(\alpha_{1+3-2})\Gamma_{\rm NS}(Q+\alpha_{3-1-2})\Gamma_{\rm R}(2Q-\alpha_{1+2+3})}
{\Gamma_{\rm NS}(Q)\Gamma_{\rm NS}(2\alpha_1)\Gamma_{\rm R}(Q-2\alpha_2)\Gamma_{\rm R}(2Q-2\alpha_3)},
\\[8pt]
\nonumber
{\cal M}_{\alpha_3,\alpha_2,\alpha_1}^{\rm N \,R\, R}
 \!& \equiv  &\!
\frac{{\rm e}^{-\frac{i\pi}{4}}{\cal N}_{\alpha_3\alpha_2\alpha_1}}{\sqrt{2}}
\frac{\Gamma_{\rm NS}(Q+\alpha_{1-2-3})\Gamma_{\rm R}(\alpha_{1+3-2})\Gamma_{\rm R}(Q+\alpha_{3-1-2})\Gamma_{\rm NS}(2Q-\alpha_{1+2+3})}
{\Gamma_{\rm NS}(Q)\Gamma_{\rm NS}(2\alpha_1)\Gamma_{\rm R}(Q-2\alpha_2)\Gamma_{\rm R}(2Q-2\alpha_3)},
\\
\label{normalizations2}
\\[-6pt]
\nonumber
{\cal M}_{\alpha_3,\alpha_2,\alpha_1}^{\rm N \,R\, N}
 \!& \equiv  &\!
{\cal N}_{\alpha_3\alpha_2\alpha_1}
\frac{\Gamma_{\rm NS}(Q+\alpha_{1-2-3})\Gamma_{\rm R}(\alpha_{1+3-2})\Gamma_{\rm NS}(Q+\alpha_{3-1-2})\Gamma_{\rm R}(2Q-\alpha_{1+2+3})}
{\Gamma_{\rm NS}(Q)\Gamma_{\rm R}(2\alpha_1)\Gamma_{\rm NS}(Q-2\alpha_2)\Gamma_{\rm R}(2Q-2\alpha_3)},
\\[8pt]
\nonumber
{\cal M}_{\alpha_3,\alpha_2,\alpha_1}^{\rm R \,N\, R}
 \!& \equiv  &\!
\frac{{\rm e}^{-\frac{i\pi}{4}}{\cal N}_{\alpha_3\alpha_2\alpha_1}}{\sqrt{2}}
\frac{\Gamma_{\rm R}(Q+\alpha_{1-2-3})\Gamma_{\rm NS}(\alpha_{1+3-2})\Gamma_{\rm R}(Q+\alpha_{3-1-2})\Gamma_{\rm NS}(2Q-\alpha_{1+2+3})}
{\Gamma_{\rm NS}(Q)\Gamma_{\rm R}(2\alpha_1)\Gamma_{\rm NS}(Q-2\alpha_2)\Gamma_{\rm R}(2Q-2\alpha_3)}.
\end{eqnarray}
Matrix elements (\ref{3pt:NS}), (\ref{norm1:final}), (\ref{normalizations2}), supplemented by the matrix element
\begin{eqnarray}
\label{3pt:NS*}
\bra{\nu_3}{\sf V}_{{\rm e}\hskip 4pt s}^{*\,\alpha_2}(1)\ket{\nu_1}
& =&{\cal M}_{\alpha_3,\alpha_2,\alpha_1}^{\rm R \,R\, R}
\\[4pt]
\nonumber
&  &\hspace{-70pt} \equiv\;
2{\cal N}_{\alpha_3\alpha_2\alpha_1}\,
\frac{\Gamma_{\rm R}(Q+\alpha_{1-2-3})\Gamma_{\rm R}(\alpha_{1+3-2})\Gamma_{\rm R}(Q+\alpha_{3-1-2})\Gamma_{\rm R}(2Q-\alpha_{1+2+3})}
{\Gamma_{\rm NS}(Q)\Gamma_{\rm R}(2\alpha_1)\Gamma_{\rm NS}(Q-2\alpha_2)\Gamma_{\rm R}(2Q-2\alpha_3)}
\end{eqnarray}
calculated in \cite{Chorazkiewicz:2008es} form a complete set of eight independent normalizations
required for  representation of all chiral vertices in both sectors.

\section{Braiding relations}
\label{section:braid:rel}

\subsection{Braiding of normal ordered exponentials and screening charges}

In this subsection we shall calculate
the braiding matrix for  operators (\ref{g:definition}).
We follow the procedure proposed in \cite{Teschner:2003en}
and extended to the NS sector of the N=1 superconformal theory in \cite{Chorazkiewicz:2008es}.
Let us assume the existence of a braiding relation of the form
\begin{equation}
\label{unn:braiding:def}
{\sf g}_{{\scriptscriptstyle \rm\bf f}_3\,s_3}^{\hskip 7pt\alpha_3}(\sigma_3)
{\sf g}_{{\scriptscriptstyle \rm\bf f}_2\,s_2}^{\hskip 7pt\alpha_2}(\sigma_2)
=
\sum_{\vec{\scriptscriptstyle \rm\bf g}}
\int\!\!d\mu(\vec t)\,
B^\epsilon_\natural(\vec\alpha;\vec s,\vec{\scriptstyle \rm\bf f};\vec t,\vec{\scriptstyle \rm\bf h})
{\sf g}_{{\scriptscriptstyle \rm\bf g}_2\,t_2}^{\hskip 7pt\alpha_2}(\sigma_2)
{\sf g}_{{\scriptscriptstyle \rm\bf g}_3\,t_3}^{\hskip 7pt\alpha_3}(\sigma_3),
\end{equation}
where $\epsilon = {\rm sgn}(\sigma_3-\sigma_2),$
$\vec{\scriptstyle \rm\bf f} = ({\scriptstyle \rm\bf f}_2,{\scriptstyle \rm\bf f}_3),\
\vec s = (s_2,s_3)$ etc. The sum over the indices $\vec {\scriptstyle \rm\bf g}$ is restricted
by the parity conservation which holds in the chiral superscalar model
and the integration measure $d\mu(\vec t)$ is proportional (by the momentum conservation)
to the Dirac delta $\delta(t_2+t_3-s_2-s_3).$
The additional subscript $\natural = {\rm R}, {\rm NS}$
denotes the sector in which  relation (\ref{unn:braiding:def}) is considered.

It follows from  commutation relations (\ref{modes:conjugation:bosonic}) that the ordered exponentials satisfy the braiding relation:
\begin{equation}
\label{braiding:elementary}
{\sf E}^\alpha(x){\sf E}^\beta(y) = {\rm e}^{-i\pi\alpha\beta\,{\rm sgn}(x-y)}\,{\sf E}^\beta(y){\sf E}^\alpha(x).
\end{equation}
Let, for $\sigma_3 > \sigma_2,$  $I = [\sigma_2,\sigma_3],\ I_c = [\sigma_3,\sigma_2 +2\pi],\ I' = [\sigma_2+2\pi,\sigma_3+2\pi]$
and define:
\begin{equation}
\label{Q:s}
\nonumber
{\mathsf Q}_{\rm\scriptscriptstyle I}
=
\int_I\!dx\ {\mathsf E}^b(x)\psi(x),
\hskip 10mm
{\mathsf Q}_{\rm\scriptscriptstyle I}^c
=
\int_{I^c}\!dx\ {\mathsf E}^b(x)\psi(x),
\hskip 10mm
\nonumber
{\mathsf Q}'_{\rm\scriptscriptstyle I}
=
\int_{I'}\!dx\ {\mathsf E}^b(x)\psi(x),
\end{equation}
so that
\[
{\sf Q}(\sigma_2) = {\mathsf Q}_{\rm\scriptscriptstyle I}^c + {\mathsf Q}_{\rm\scriptscriptstyle I},
\hskip 1cm
{\sf Q}(\sigma_3) = {\mathsf Q}_{\rm\scriptscriptstyle I}^c + {\mathsf Q}'_{\rm\scriptscriptstyle I}.
\]
Using (\ref{braiding:elementary}) we  thus get
\begin{eqnarray*}
{\sf Q}(\sigma_3){\sf E}^{\alpha_2}(\sigma_2)
& = &
{\rm e}^{-i\pi b\alpha_2}{\sf E}^{\alpha_2}(\sigma_2)
\left({\mathsf Q}_{\rm\scriptscriptstyle I}^c + {\rm e}^{-2i\pi b\alpha_2}{\mathsf Q}'_{\rm\scriptscriptstyle I}\right),
\\[6pt]
{\sf Q}(\sigma_2){\sf E}^{\alpha_3}(\sigma_3)
& = &
{\rm e}^{-i\pi b\alpha_3}{\sf E}^{\alpha_3}(\sigma_3)
\left({\mathsf Q}_{\rm\scriptscriptstyle I}^c + {\rm e}^{2i\pi b\alpha_3}{\mathsf Q}_{\rm\scriptscriptstyle I}\right),
\end{eqnarray*}
and consequently
\begin{eqnarray}
\nonumber
{\sf g}_{{\scriptscriptstyle \rm\bf f}_3\,s_3}^{\hskip 7pt\alpha_3}(\sigma_3)
{\sf g}_{{\scriptscriptstyle \rm\bf f}_2\,s_2}^{\hskip 7pt\alpha_2}(\sigma_2)
& = &
{\sf E}^{\alpha_2}(\sigma_2){\sf E}^{\alpha_3}(\sigma_3)\,
{\rm e}^{-i\pi\alpha_2\alpha_3 - i\pi\alpha_2 b s_3}
\left({\mathsf Q}_{\rm\scriptscriptstyle I}^c + {\rm e}^{-2i\pi b\alpha_2}{\mathsf Q}'_{\rm\scriptscriptstyle I}\right)^{s_3}_{{\scriptscriptstyle \rm\bf f}_3}
\left({\mathsf Q}_{\rm\scriptscriptstyle I}^c + {\mathsf Q}_{\rm\scriptscriptstyle I}\right)^{s_2}_{{\scriptscriptstyle \rm\bf f}_2},
\\[-6pt]
\label{normal:ordering:step:1}
\\[-6pt]
\nonumber
{\sf g}_{{\scriptscriptstyle \rm\bf g}_2\,t_2}^{\hskip 7pt\alpha_2}(\sigma_2)
{\sf g}_{{\scriptscriptstyle \rm\bf g}_3\,t_3}^{\hskip 7pt\alpha_3}(\sigma_3)
& = &
{\sf E}^{\alpha_2}(\sigma_2){\sf E}^{\alpha_3}(\sigma_3)\,
{\rm e}^{- i\pi\alpha_3 b t_2}
\left({\mathsf Q}_{\rm\scriptscriptstyle I}^c + {\rm e}^{2i\pi b\alpha_3}{\mathsf Q}_{\rm\scriptscriptstyle I}\right)^{t_2}_{{\scriptscriptstyle \rm\bf g}_2}
\left({\mathsf Q}_{\rm\scriptscriptstyle I}^c + {\mathsf Q}'_{\rm\scriptscriptstyle I}\right)^{t_3}_{{\scriptscriptstyle \rm\bf g}_3}.
\end{eqnarray}
Braiding relation (\ref{unn:braiding:def}) is then equivalent to the equation
\begin{equation}
\label{braiding:Qs}
\begin{aligned}
\left({\mathsf Q}_{\rm\scriptscriptstyle I}^c + {\rm e}^{-2i\pi b\alpha_2}{\mathsf Q}'_{\rm\scriptscriptstyle I}\right)^{s_3}_{{\scriptscriptstyle \rm\bf f}_3}
\left({\mathsf Q}_{\rm\scriptscriptstyle I}^c + {\mathsf Q}_{\rm\scriptscriptstyle I}\right)^{s_2}_{{\scriptscriptstyle \rm\bf f}_2}
&=\\
&\hspace{-2cm}\sum_{\vec{\scriptscriptstyle \rm\bf g}}
\int\!\!d\mu(\vec t)\,
\tilde B^+_\natural(\vec\alpha;\vec s,\vec{\scriptstyle \rm\bf f};\vec t,\vec{\scriptstyle \rm\bf g})
\left({\mathsf Q}_{\rm\scriptscriptstyle I}^c + {\rm e}^{2i\pi b\alpha_3}{\mathsf Q}_{\rm\scriptscriptstyle I}\right)^{t_2}_{{\scriptscriptstyle \rm\bf g}_2}
\left({\mathsf Q}_{\rm\scriptscriptstyle I}^c + {\mathsf Q}'_{\rm\scriptscriptstyle I}\right)^{t_3}_{{\scriptscriptstyle \rm\bf g}_3},\end{aligned}
\end{equation}
where
\[
\tilde B^+_\natural(\vec\alpha;\vec s,\vec{\scriptstyle \rm\bf f};\vec t,\vec{\scriptstyle \rm\bf g})
=
{\rm e}^{i\pi\alpha_2\alpha_3 + i\pi\alpha_2 b s_3- i\pi\alpha_3 b t_2}
B^+_\natural(\vec\alpha;\vec s,\vec{\scriptstyle \rm\bf f};\vec t,\vec{\scriptstyle \rm\bf g}).
\]
If the product of fields in (\ref{normal:ordering:step:1}) act on the NS state,
the fermion field $\psi(x)$ appearing in the definition of
the screening charges  is anti-periodic, $\psi(x+2\pi) = -\psi(x)$, and
\[
{\sf Q}_{\rm\scriptscriptstyle I}'
=
-{\rm e}^{i\pi b^2}\,{\rm e}^{2\pi b{\mathsf p}}\,{\sf Q}_{\rm\scriptscriptstyle I}.
\]
On the other hand, while acting on the Ramond state the
fermion field is periodic, $\psi(x+2\pi) = \psi(x)$. In this case
\[
{\sf Q}_{\rm\scriptscriptstyle I}'
=
{\rm e}^{i\pi b^2}\,{\rm e}^{2\pi b{\mathsf p}}\,{\sf Q}_{\rm\scriptscriptstyle I}.
\]

It follows from the definition of the screening charges and from braiding relation
(\ref{braiding:elementary})  that the operators ${\mathsf Q}_{\rm\scriptscriptstyle I},$  ${\mathsf Q}_{\rm\scriptscriptstyle I}^c$ and ${\rm e}^{\pi b{\sf p}}$
satisfy the Weyl-type algebra
\begin{equation}
\label{braiding:2}
{\mathsf Q}_{\rm\scriptscriptstyle I} \,
{\mathsf Q}_{\rm\scriptscriptstyle I}^c
\; = \;
-\,{\rm e}^{i\pi b^2}\,
{\mathsf Q}_{\rm\scriptscriptstyle I}^c\,
{\mathsf Q}_{\rm\scriptscriptstyle I},
\hskip 1cm
{\mathsf Q}_{\rm\scriptscriptstyle I}^c \,
{\rm e}^{\pi b{\sf p}}
\; = \;
{\rm e}^{i\pi b^2}\,
{\rm e}^{\pi b{\sf p}}\,
{\mathsf Q}_{\rm\scriptscriptstyle I}^c,
\hskip 1cm
{\mathsf Q}_{\rm\scriptscriptstyle I}\,
{\rm e}^{\pi b{\sf p}}
\; = \;
{\rm e}^{i\pi b^2}\,
{\rm e}^{\pi b{\sf p}}\,
{\mathsf Q}_{\rm\scriptscriptstyle I}.
\end{equation}

Formula (\ref{braiding:Qs}) can be seen as a relation in the algebra generated by the elements ${\mathsf Q}_{\rm\scriptscriptstyle I},
{\mathsf Q}_{\rm\scriptscriptstyle I}^c$ and ${\sf p},$ satisfying (\ref{braiding:2}). To compute the braiding matrix $\tilde B^{+}_\natural$
we shall choose a convenient representation of this algebra.
Let us introduce an auxiliary Hilbert space ${\cal H}_{\rm\scriptscriptstyle aux} = {\mathbb C}^2 \otimes L^2\big({\mathbb R}^2\big)$
and consider operators
${\sf p, x, t} \in {\rm End}({\cal H}_{\rm\scriptscriptstyle aux})$
satisfying commutation relations
\begin{equation}
\label{commutations:2}
[{\sf p},{\sf x}] \; =\; -i,
\hskip 1cm
[{\sf p},{\mathsf t}] \; = \; [{\sf x},{\mathsf t}] \; = \; 0,
\end{equation}
together with  conjugation properties ${\mathsf p}^\dagger ={\mathsf p}^\dagger,\ {\mathsf x}^\dagger ={\mathsf x}^\dagger$ and ${\mathsf t}^\dagger = - {\mathsf t}.$
One easily  checks that the operators
\(
\tilde{\mathsf Q}_{\rm\scriptscriptstyle I}^c, \tilde{\mathsf Q}_{\rm\scriptscriptstyle I}\in {\rm End}({\cal H}_{\rm\scriptscriptstyle aux}),
\)
defined by
\begin{eqnarray}
\label{Weyl:representation}
\begin{array}{rclllllllllll}
\tilde{\mathsf Q}_{\rm\scriptscriptstyle I}^c & = & \tau_1\, {\rm e}^{b {\mathsf x}}\, {\rm e}^{-\frac12 i\pi b {\mathsf t}},
&&
\tau_1 &=& \left(^{0\;1}_{1\;0}\right),
\\     [5pt]
\tilde{\mathsf Q}_{\rm\scriptscriptstyle I}
& = &
\tau_2\,
{\rm e}^{\frac12 b {\mathsf x}}\,
{\rm e}^{-\pi b {\mathsf p}}\,
{\rm e}^{\frac12 b {\mathsf x}}\,
{\rm e}^{\frac12 i\pi b{\mathsf t}},
&&
\tau_2 &=& \left(^{\hskip 4pt 0\hskip 5pt i}_{-i\:\, 0}\right),
\end{array}
\end{eqnarray}
 form a representation
of (\ref{braiding:2}). In this representation
\[
\tilde{\mathsf Q}'_{\rm\scriptscriptstyle I}
=
\eta_\natural\,
\tau_2\,
{\rm e}^{\frac12 b {\mathsf x}}\,
{\rm e}^{\pi b {\mathsf p}}\,
{\rm e}^{\frac12 b {\mathsf x}}\,
{\rm e}^{\frac12 i\pi b{\mathsf t}},\;\;\;\eta_{\rm NS} = -1, \;\;\;\eta_{\rm R} = +1.
\]

Representation (\ref{Weyl:representation}) allows to transform the powers of sums of $\tilde{\sf Q}-$s
into the normal ordered form, with
 $\sf x$ operator to the left of $\sf p$ and $\sf t$ operators. Using the
shift property of the Barnes functions:
\[
G_{\rm NS} (y+b) = \left(1+{\rm e}^{i\pi b y}\right)G_{\rm R}(y),
\hskip 1cm
G_{\rm R} (y+b) = \left(1-{\rm e}^{i\pi b y}\right)G_{\rm NS}(y),
\]
and
\[
{\rm e}^{\alpha\sf x}\, f({\sf p}) = f({\sf p}+i\alpha)\,{\rm e}^{\alpha\sf x},
\]
one has
\begin{eqnarray*}
{\rm e}^{b\sf x}\,{\rm e}^{-\frac12i\pi b {\sf t}} + i\, {\rm e}^{\frac12b\sf x}\,{\rm e}^{-\pi b {\sf p}}\, {\rm e}^{\frac12b\sf x}\,{\rm e}^{\frac12i\pi b {\sf t}}
& = &
{\rm e}^{-\frac12i\pi b {\sf t}}\,
{\rm e}^{\frac12b\sf x}
\left(1+{\rm e}^{i\pi b\left(ip + t + \frac12 b^{-1}\right)}\right)
{\rm e}^{\frac12b\sf x}
\\[4pt]
& = &
{\rm e}^{-\frac12i\pi b {\sf t}}\,
{\rm e}^{\frac12b\sf x}\,
\frac{G_{\rm NS}\left(ip + t + \frac12 b^{-1}+b\right)}{G_{\rm R}\left(ip + t + \frac12 b^{-1}\right)}\,
{\rm e}^{\frac12b\sf x}
\\[4pt]
& = &
{\rm e}^{-\frac12i\pi b {\sf t}}\,
G_{\rm NS}\left(ip + t + {\textstyle\frac12} Q\right)\,
{\rm e}^{b\sf x}\,
\frac{1}{G_{\rm R}\left(ip + t + \frac12 Q\right)}
\end{eqnarray*}
with $Q = b + b^{-1}$ and similarly
\begin{eqnarray*}
{\rm e}^{b\sf x}\,{\rm e}^{-\frac12i\pi b {\sf t}} - i\, {\rm e}^{\frac12b\sf x}\,{\rm e}^{-\pi b {\sf p}}\, {\rm e}^{\frac12b\sf x}\,{\rm e}^{\frac12i\pi b {\sf t}}
& = &
{\rm e}^{-\frac12i\pi b {\sf t}}\,
{\rm e}^{\frac12b\sf x}
\left(1-{\rm e}^{i\pi b\left(ip + t + \frac12 b^{-1}\right)}\right)
{\rm e}^{\frac12b\sf x}
\\[4pt]
& = &
{\rm e}^{-\frac12i\pi b {\sf t}}\,
{\rm e}^{\frac12b\sf x}\,
\frac{G_{\rm R}\left(ip + t + \frac12 b^{-1}+b\right)}{G_{\rm NS}\left(ip + t + \frac12 b^{-1}\right)}\,
{\rm e}^{\frac12b\sf x}
\\[4pt]
& = &
{\rm e}^{-\frac12i\pi b {\sf t}}\,
G_{\rm R}\left(ip + t + {\textstyle\frac12} Q\right)\,
{\rm e}^{b\sf x}\,
\frac{1}{G_{\rm NS}\left(ip + t + \frac12 Q\right)}\ .
\end{eqnarray*}
Introducing a matrix notation
\[
{\mathbb G}_{\rm NS}\left(z\right)
\; = \;
\left(
\hskip -3pt
\begin{array}{cc}
G_{\rm NS}(z) & 0
\\
0 & G_{\rm R}(z)
\end{array}
\hskip -3pt
\right),
\hskip .5cm
{\mathbb G}_{\rm R}\left(z\right)
\; = \;
\left(
\hskip -3pt
\begin{array}{cc}
G_{\rm R}(z) & 0
\\
0 & G_{\rm NS}(z)
\end{array}
\hskip -3pt
\right)
\; = \; \tau_1\cdot {\mathbb G}_{\rm NS}\left(z\right) \cdot\tau_1,
\]
one can present the result of the calculations above in the compact form
\begin{eqnarray*}
\tilde{\sf Q}_{\rm\scriptscriptstyle I}^c +\tilde{\sf Q}_{\rm\scriptscriptstyle I}
& = &
{\rm e}^{-\frac12{i\pi b {\sf t}}}\,
{\mathbb G}_{\rm NS}\!\left(i{\sf p} + {\sf t} +{\textstyle\frac12} Q \right)\,
\tau_1\, {\rm e}^{b{\sf x}}\,
{\mathbb G}^{-1}_{\rm NS}\!\left(i{\sf p} + {\sf t} +{\textstyle\frac12} Q \right).
\end{eqnarray*}
In a similar way one obtains
\begin{eqnarray*}
\tilde{\sf Q}_{\rm\scriptscriptstyle I}^c +{\rm e}^{2\pi i b\alpha_3}\tilde{\sf Q}_{\rm\scriptscriptstyle I}
& = &
{\rm e}^{-\frac12{i\pi b {\sf t}}}\,
{\mathbb G}_{\rm NS}\!\left(i{\sf p} + {\sf t} + 2\alpha_3 + {\textstyle\frac12} Q \right)\,
\tau_1\, {\rm e}^{b{\sf x}}\,
{\mathbb G}^{-1}_{\rm NS}\!\left(i{\sf p} + {\sf t} + 2\alpha_3 + {\textstyle\frac12} Q \right),
\\[6pt]
\tilde{\sf Q}_{\rm\scriptscriptstyle I}^c +\tilde{\sf Q}_{\rm\scriptscriptstyle I}'
& = &
{\rm e}^{-\frac12{i\pi b {\sf t}}}\,
{\mathbb G}^{-1}_{\natural}\!\left(-i{\sf p} + {\sf t} +{\textstyle\frac12} Q \right)\,
\tau_1\, {\rm e}^{b{\sf x}}\,
{\mathbb G}_{\natural}\!\left(-i{\sf p} + {\sf t} +{\textstyle\frac12} Q \right),
\\[6pt]
\tilde{\sf Q}_{\rm\scriptscriptstyle I}^c +{\rm e}^{-2\pi i b\alpha_2}\tilde{\sf Q}_{\rm\scriptscriptstyle I}'
& = &
{\rm e}^{-\frac12{i\pi b {\sf t}}}\,
{\mathbb G}^{-1}_{\natural}\!\left(-i{\sf p} + {\sf t} -2\alpha_2 + {\textstyle\frac12} Q \right)\,
\tau_1\, {\rm e}^{b{\sf x}}\,
{\mathbb G}_{\natural}\!\left(-i{\sf p} + {\sf t} -2\alpha_2 + {\textstyle\frac12} Q \right).
\end{eqnarray*}
Defining an even and an odd complex power of $\tau_1$
\[
\big(\tau_1\big)^s_\rho \; \equiv  \;  {\mathbf 1}_\rho \; = \;
\left\{
\begin{array}{ccl}
{\mathbf 1}, & ~& \rho = {\rm e},
\\[2pt]
\tau_1, && \rho = {\rm o},
\end{array}
\right.
\]
we get:
\begin{eqnarray}
\label{QQ:products}
\nonumber
&&
\hskip -1.5cm
(\tilde{\mathsf Q}_{\rm\scriptscriptstyle I}^c + {\rm e}^{-2i\pi b \alpha_2}\tilde{\mathsf Q}_{\rm\scriptscriptstyle I}')_{{\scriptscriptstyle \rm\bf f}_3}^{s_3}
(\tilde{\mathsf Q}_{\rm\scriptscriptstyle I}^c +\tilde{\mathsf Q}_{\rm\scriptscriptstyle I})_{{\scriptscriptstyle \rm\bf f}_2}^{s_2}
\; = \;
{\rm e}^{b(s_2+s_3){\mathsf x}}\,
{\rm e}^{-\frac12 i\pi b(s_2+s_3) {\mathsf t}}
\\[6pt]
\nonumber
& \times &
{\mathbb G}_{\natural}^{-1}\!\left(-i{\mathsf p}+{\mathsf t} -2\alpha_2 -bs_2-bs_3 + {\textstyle\frac12} Q \right)\,
 {\mathbf 1}_{{\scriptscriptstyle \rm\bf f}_3}\
{\mathbb G}_{\natural}\!\left(-i{\mathsf p}+{\mathsf t} -2\alpha_2 -bs_2 + {\textstyle\frac12} Q \right)
\\[6pt]
\nonumber
& \times &
{\mathbb G}_{\rm NS}\!\left(i{\mathsf p}+{\mathsf t}  +bs_2 + {\textstyle\frac12} Q \right)\,
{\mathbf 1}_{{\scriptscriptstyle \rm\bf f}_2}\
{\mathbb G}_{\rm NS}^{-1}\!\left(i{\mathsf p}+{\mathsf t}  + {\textstyle\frac12} Q \right),
\\[-2pt]
\\[-2pt]
\nonumber
&&
\hskip -1.5cm
(\tilde{\mathsf Q}_{\rm\scriptscriptstyle I}^c +{\rm e}^{2i\pi b \alpha_3}\tilde{\mathsf Q}_{\rm\scriptscriptstyle I})^{t_2}_{{\scriptscriptstyle \rm\bf g}_2}
(\tilde{\mathsf Q}_{\rm\scriptscriptstyle I}^c +\tilde{\mathsf Q}'_{\rm\scriptscriptstyle I})^{t_3}_{{\scriptscriptstyle \rm\bf g}_3}
\; =\;
{\rm e}^{b(t_2+t_3){\mathsf x}}\,
{\rm e}^{-\frac12 i\pi b(t_2+t_3) {\mathsf t}}
\\[6pt]
\nonumber
& \times &
{\mathbb G}_{\rm NS}\!\left(i{\mathsf p}+{\mathsf t} + 2\alpha_3 + bt_2 + bt_3+ {\textstyle\frac12} Q \right)\,
{\mathbf 1}_{{\scriptscriptstyle \rm\bf g}_2}\
{\mathbb G}_{\rm NS}^{-1}\!\left(i{\mathsf p}+{\mathsf t} +2\alpha_3 + bt_3 + {\textstyle\frac12} Q \right)
\\[6pt]
\nonumber
& \times &
{\mathbb G}_{\natural}^{-1}\!\left(-i{\mathsf p}+{\mathsf t} -bt_3 + {\textstyle\frac12} Q \right)\,
{\mathbf 1}_{{\scriptscriptstyle \rm\bf g}_3}\
{\mathbb G}_{\natural}\,\left(-i{\mathsf p}+{\mathsf t} + {\textstyle\frac12} Q \right).
\end{eqnarray}
Evaluated on a common eigenstate of  operators $\sf p$ and $\sf t,$
\[
{\sf p}\ket{p,\tau} = p \ket{p,\tau},
\hskip 1cm
{\sf t}\ket{p,\tau} = \tau \ket{p,\tau},\;\; \tau \in i{\mathbb R},
\]
 the right hand sides of  formulae
(\ref{QQ:products}) take the form of analytic functions of $p$ and $\tau,$ multiplied by the operator
\[
{\rm e}^{(s_1+s_2){\sf x}} = {\rm e}^{(t_1+t_2){\sf x}}.
\]
Stripping off this factor from both sides of  formula (\ref{braiding:Qs}) one gets a relation between
analytic functions of $p$ and $\tau.$
In terms of the Barnes S functions\footnote{See Appendix \ref{Appendix:Barnes} for the relation between the Barnes G and S functions.}
\[
{\mathbb S}_{\rm NS}\left(z\right)
 =
\left(
\begin{array}{cc}
\hskip -4pt S_{\rm NS}(z) & \hskip -6pt 0
\\
\hskip -4pt 0 & \hskip -6pt S_{\rm R}(z) \hskip -4pt
\end{array}
\right),
\hskip 1cm
{\mathbb S}_{\rm R}\left(z\right)
=
\left(
\begin{array}{cc}
\hskip -4pt S_{\rm R}(z) & \hskip -6pt 0
\\
\hskip -4pt 0 & \hskip -6pt S_{\rm NS}(z) \hskip -4pt
\end{array}
\right)
= \tau_1\cdot {\mathbb S}_{\rm NS}\left(z\right) \cdot\tau_1,
\]
this relation
takes the form
{\small
\begin{eqnarray}
\nonumber
&&
\hskip -.7cm
{\rm e}^{i\chi}\!
\begin{array}{r}
    {\mathbb S}_{\natural}\!\left(\frac{Q}{2}+iA_2\!-\!\tau\right)
    \end{array}\!\!\!
    \left(F_{{\scriptscriptstyle \rm\bf f}_3}^{\natural}\right)^{T}\!\!
    \begin{array}{r}
    {\mathbb S}^{-1}_{\natural}\!\left(Q-iC_2+ip_s\!-\!\tau\right)
    \end{array}\!\!
    \begin{array}{r}
    {\mathbb S}^{-1}_{\rm NS}\!\left(Q-iC_2-ip_s\!-\!\tau\right)
    \end{array}\!\!
    F_{{\scriptscriptstyle \rm\bf f}_2}^{\rm NS}
    \begin{array}{r}
    {\mathbb S}_{\rm NS}\!\left(\frac{Q}{2}+iB_2\!-\!\tau\right)
    \end{array}
\\[6pt]
\label{braiding:temp:1}
& = &
\sum\limits_{\vec{\scriptscriptstyle \rm\bf g}}\,
\int\! d\mu(t_3)\ {\rm e}^{-\frac{i\pi }{2}(bt_3)^2 +\pi  p_1 bt_3}\
B_{\natural}(\vec\alpha;\vec s,\vec{\scriptstyle \rm\bf f};s-t_3,t_3,\vec{\scriptstyle \rm\bf g})
\\[2pt]
\nonumber
    && \hskip -1mm \times
    \begin{array}{r}
    {\mathbb S}_{\rm NS}\!\left(\frac{Q}{2}+iA_3\!+\!\tau\right)
    \end{array}\!\!
    F_{{\scriptscriptstyle \rm\bf g}_2}^{\rm NS}\!
    \begin{array}{r}
    {\mathbb S}^{-1}_{\rm NS}\!\left(Q-iC_3+ip_u\!+\!\tau\right)
    \end{array}\!\!\!
    \begin{array}{r}
    {\mathbb S}^{-1}_{\natural}\!\!\left(Q-iC_3-ip_u\!+\!\tau\right)
    \end{array}\!\!\!
    \left(F_{{\scriptscriptstyle \rm\bf g}_3}^{\natural}\right)^{T}\!\!
    \begin{array}{r}
    {\mathbb S}_{\natural}\!\left(\frac{Q}{2}+iB_3\!+\!\tau\right)\!,
    \end{array}
\end{eqnarray}
}
where
$
i\chi = \frac{i\pi b^2}{2}(s_2^2 - s^2) +\pi b p_1 s_{2} -i\pi\alpha_2(\alpha_3 +2bs_3),$
$s = s_2 + s_3,$
    \[
    F_{\rm e}^{\rm NS}\; = \;
    \left(\begin{array}{cc} 1 & 0 \\ 0 & 1 \end{array}\right),
    \hskip 5mm
    F_{\rm o}^{\rm NS}\; = \;
    \left(\begin{array}{cc} 0 & {\rm e}^{\frac{i\pi}{4}} \\ {\rm e}^{-\frac{i\pi}{4}} & 0 \end{array}\right),
    \hskip 10mm
    F_{\scriptscriptstyle \rm\bf f}^{\rm R}\; = \;\left(F_{\scriptscriptstyle \rm\bf f}^{\rm NS}\right)^T,
    \]
and
\begin{eqnarray}
\nonumber
\label{parameters}
A_2 & = & p_1 - 2i\alpha_2 -ibs,
\hskip 2mm
B_2 \; = \; -p_1,
\hskip 2mm
C_2 \; = \; i(\alpha_2 - Q/2),
\hskip 2mm
p_s \; = \; p_1-i(\alpha_2 + bs_2),
\\[-6pt]
\\[-6pt]
\nonumber
A_3 & = & p_1 - 2i\alpha_3 -ibs,
\hskip 2mm
B_3 \; = \; -p_1,
\hskip 2mm
C_3 \; = \; i(\alpha_3 - Q/2),
\hskip 2mm
p_u \; = \; p_1-i(\alpha_3 + bt_3).
\end{eqnarray}
The Barnes functions in the integrand of (\ref{braiding:temp:1}) depend on the integration variable $t_3$
only via the parameter $p_u$.
Multiplying  both sides of  equation (\ref{braiding:temp:1})
by
{\small
\(
{\mathbb S}_{\rm NS}\!\left(\frac{Q}{2}+iA_3\!+\!\tau\right)^{-1} \hskip -5pt = {\mathbb S}_{\rm NS}\!\left(\frac{Q}{2}-iA_3\!-\!\tau\right)
\)
}
(from the left)
and by
{\small
\(
{\mathbb S}_{\natural}\!\left(\frac{Q}{2}+iB_3\!+\!\tau\right)^{-1} \hskip -5pt = {\mathbb S}_{\natural}\!\left(\frac{Q}{2}-iB_3\!-\!\tau\right)
\)
}
(from the right) and choosing the integration measure to be $d\mu(t_3) = \theta(p_u)\,dp_u$  one gets
\begin{eqnarray}
\label{braiding:temp:2}
&&
\hskip -5mm
{\rm e}^{i\chi}\,
\displaystyle
\frac{
{\mathbb S}_{\natural}\!\left(\frac{Q}{2}+iA_2-\tau\right)
}{
{\mathbb S}_{\rm NS}\!\left(\frac{Q}{2}+iA_3+\tau\right)
}
\left(F_{{\scriptscriptstyle \rm\bf f}_3}^{\natural}\right)^{T}
\frac{
{\mathbb S}_{\rm NS}\!\left(ip_s+\tau+iC_2\right)
}{
{\mathbb S}_{\natural}\!\left(Q+ip_s-\tau-iC_2\right)
}
F_{{\scriptscriptstyle \rm\bf f}_2}^{\rm NS}\,
\frac{
{\mathbb S}_{\rm NS}\!\left(\frac{Q}{2}+iB_2-\tau\right)
}{
{\mathbb S}_{\natural}\!\left(\frac{Q}{2}+iB_3+\tau\right)
}
\\
\nonumber
& = &
\sum\limits_{\vec{\scriptscriptstyle \rm\bf g}}\,
\int\limits_0^\infty\! dp_u\ {\rm e}^{-\frac{i\pi }{2}(bt_3)^2 +\pi  p_1 bt_3}\
B^+_{\natural}(\vec\alpha;\vec s,\vec{\scriptstyle \rm\bf f};s-t_3,t_3,\vec{\scriptstyle \rm\bf g})\
F_{{\scriptscriptstyle \rm\bf g}_2}^{\rm NS}\!
\frac{
{\mathbb S}_{\natural}\!\left(ip_u-\tau+iC_3\right)
}{
{\mathbb S}_{\rm NS}\!\left(Q+ip_u+\tau-iC_3\right)
}
\left(F_{{\scriptscriptstyle \rm\bf g}_3}^{\natural}\right)^{T}
\end{eqnarray}
where
\(
\frac{
{\mathbb S}_{\natural}\!\left(\frac{Q}{2}+iA_2-\tau\right)
}{
{\mathbb S}_{\rm NS}\!\left(\frac{Q}{2}+iA_3+\tau\right)
}
\)
stands for
\(
{\mathbb S}_{\natural}\!\left(\frac{Q}{2}+iA_2-\tau\right)
{\mathbb S}_{\rm NS}^{-1}\!\left(\frac{Q}{2}+iA_3+\tau\right)
\)
etc.

\parskip 10pt
In order to calculate the braiding kernel
from (\ref{braiding:temp:2})
one can then use the orthogonality relation\footnote{
A derivation of (\ref{orthogonality:matrix:form}) is presented in  Appendix \ref{Appendix:Orthogonality}.}:
\begin{eqnarray}
\nonumber
&&
\hskip -0.5cm
\int\limits_{i\mathbb R}\!\frac{d\tau}{i}\
\!\!{\rm Tr}
\left\{
\Big[\!
 F_{{\scriptscriptstyle \rm\bf h}_2}^{\rm NS}
 \frac{{\mathbb S}_{\natural}\!\left(ip_u\!-\!\tau+iC_3\right)}
      {{\mathbb S}_{\rm NS}\!\left(Q+ip_u\!+\!\tau-iC_3\right)}
 \left(F_{{\scriptscriptstyle \rm\bf h}_3}^{\natural}\right)^{T}
\Big]
\!\!
\Big[
 F_{{\scriptscriptstyle \rm\bf g}_2}^{\rm NS}
 \frac{{\mathbb S}_{\natural}\!\left(ip_u'\!-\!\tau+iC_3\right)}
      {{\mathbb S}_{\rm NS}\!\left(Q+ip_u'\!+\!\tau-iC_3\right)}
 \left(F_{{\scriptscriptstyle \rm\bf g}_3}^{\natural}\right)^{T}
\Big]^\dag
\right\}
\\[6pt]
\label{orthogonality:matrix:form}
&= &
{\cal N}_{\natural}^{-1}(p_u)\delta_{\vec{\scriptscriptstyle \rm\bf h},\vec{\scriptscriptstyle \rm\bf g}}\,\delta(p_u-p'_u)
,\;\;\;\;\;\;\;\;\;\;\;p_u,p_u'\in {\mathbb R}_+,
\end{eqnarray}
where
\begin{equation}
\label{normalizations}
{\cal N}_{\rm NS}(p_u) = \sinh\pi b p_u\sinh \frac{\pi p_u}{b},
\hskip 1cm
{\cal N}_{\rm R}(p_u) = \cosh\pi b p_u\cosh \frac{\pi p_u}{b}.
\end{equation}
This yields
\begin{eqnarray}
\nonumber
&&
\hskip -1.3cm
B^+_{\natural}(\vec\alpha;\vec s,\vec{\scriptstyle \rm\bf f};s-t_3,t_3,\vec{\scriptstyle \rm\bf g})
=
{\cal N}_{\natural}(p_u){\rm e}^{i\chi + \frac{i\pi }{2}(bt_3)^2 -\pi  p_1 bt_3}
\\[6pt]
\label{braiding:unnorm}
&\times &
\int\limits_{i\mathbb R}\!\frac{d\tau}{i}\
{\rm Tr}
\left\{
\frac{
{\mathbb S}_{\natural}\!\left(Q/2+iA_2-\tau\right)
}{
{\mathbb S}_{\rm NS}\!\left(Q/2+iA_3+\tau\right)
}
\left(F_{{\scriptscriptstyle \rm\bf f}_3}^{\natural}\right)^{T}
\frac{
{\mathbb S}_{\rm NS}\!\left(ip_s+\tau+iC_2\right)
}{
{\mathbb S}_{\natural}\!\left(Q+ip_s-\tau-iC_2\right)
}
F_{{\scriptscriptstyle \rm\bf f}_2}^{\rm NS}
\right.
\\[4pt]
\nonumber
&&
\hskip 17mm
\left.
\frac{
{\mathbb S}_{\natural}\!\left(Q/2-iB_3-\tau\right)
}{
{\mathbb S}_{\rm NS}\!\left(Q/2-iB_2+\tau\right)
}
\Big[
 F_{{\scriptscriptstyle{\rm\bf g}_2}}^{\rm NS}
 \frac{{\mathbb S}_{\natural}\!\left(ip_u\!-\!\tau+iC_3\right)}
      {{\mathbb S}_{\rm NS}\!\left(Q+ip_u\!+\!\tau-iC_3\right)}
 \left(F_{{\scriptscriptstyle{\rm\bf g}_3}}^{\natural}\right)^{T}
\Big]^\dag
\right\}.
\end{eqnarray}
For further applications it is convenient to regard the braiding kernel as a function
of new variables:
\begin{equation}
\alpha_\imath = \frac{Q}{2} + ip_\imath,\;\; \imath = 1,s,u,
\hskip 1cm
\alpha_4 = \alpha_3+\alpha_2+\alpha_1 + bs.
\end{equation}
In order to make it more readable we introduce the notation
\begin{equation}
\label{un:braiding:new:notation}
\mathbb{B}^{\epsilon,\,\natural}_{\alpha_s\alpha_u}
\!\!\left[^{\alpha_3\:\alpha_2}_{\alpha_4\:\alpha_1}\right]^{\scriptscriptstyle\rm\bf g_2\,g_3}_{\scriptscriptstyle\rm\bf f_3\,f_2}
\equiv
B^\epsilon_{\natural}(\vec\alpha;\vec s,\vec{\scriptstyle \rm\bf f};s-t_3,t_3,\vec{\scriptstyle \rm\bf g})
\end{equation}
so that
\begin{eqnarray}
\nonumber
\hskip-0.7cm\mathbb{B}^{+,\, \natural}_{\alpha_s\alpha_u}
\!\!\left[^{\alpha_3\:\alpha_2}_{\alpha_4\:\alpha_1}\right]^{\scriptscriptstyle\rm\bf g_2\,g_3}_{\scriptscriptstyle\rm\bf f_3\,f_2}
& = &
{\textstyle \frac14}
{\rm e}^{\delta}\,
S_{\natural}(2\alpha_u)S_{\natural}(2Q-2\alpha_u)
\\[4pt]
& \times &\!\!\!
\int\limits_{i\mathbb R}\!\frac{d\tau}{i}\
{\rm Tr}
\left\{
\frac{
{\mathbb S}_{\natural}\!\left(\alpha_4-\alpha_3+\alpha_2-\tau\right)
}{
{\mathbb S}_{\rm NS}\!\left(\alpha_4+\alpha_3-\alpha_2+\tau\right)
}
\left(F_{{\scriptscriptstyle \rm\bf f}_3}^{\natural}\right)^{T}
\frac{
{\mathbb S}_{\rm NS}\!\left(\alpha_s-\alpha_2+\tau\right)
}{
{\mathbb S}_{\natural}\!\left(\alpha_s+\alpha_2-\tau\right)
}
F_{{\scriptscriptstyle \rm\bf f}_2}^{\rm NS}
\right.
\\[4pt]
\nonumber
&&
\hskip 25mm
\left.
\frac{
{\mathbb S}_{\natural}\!\left(\alpha_1-\tau\right)
}{
{\mathbb S}_{\rm NS}\!\left(\alpha_1+\tau\right)
}
\Big[
 F_{{\scriptscriptstyle{\rm\bf g}_2}}^{\rm NS}
 \frac{{\mathbb S}_{\natural}\!\left(\alpha_u-\alpha_3 -\tau\right)}
      {{\mathbb S}_{\rm NS}\!\left(\alpha_u+\alpha_3 +\tau\right)}
 \left(F_{{\scriptscriptstyle{\rm\bf g}_3}}^{\natural}\right)^{T}
\Big]^\dag
\right\}.
\end{eqnarray}
where
\begin{eqnarray*}
\delta
&=&i\chi+\frac{i\pi }{2}(bt_3)^2 -\pi  p_1 bt_3
\\
&=&
{i\pi\over 2}\Big[
\alpha_4(Q-\alpha_4) + \alpha_1(Q-\alpha_1)-\alpha_u(Q-\alpha_u)-\alpha_s(Q-\alpha_s)
\\
&&\hskip 4.3cm +\;2\alpha_3(\alpha_4-\alpha_u) - 2\alpha_2(\alpha_4-\alpha_s)
\Big].
\end{eqnarray*}
Repeating the steps above for $\epsilon = {\rm sgn}(\sigma_3-\sigma_2) < 0$ one gets
\begin{eqnarray}
\mathbb{B}^{-,\,\natural}_{\alpha_s\alpha_u}
\!\!\left[^{\alpha_3\:\alpha_2}_{\alpha_4\:\alpha_1}\right]^{\scriptscriptstyle\rm\bf g_2\,g_3}_{\scriptscriptstyle\rm\bf f_3\,f_2}
& = &
{\textstyle \frac14}
{\rm e}^{\delta^{(-)}}\,
S_{\natural}(2\alpha_u)S_{\natural}(2Q-2\alpha_u)\,
\\[4pt]\nonumber
&\times &
\int\limits_{i\mathbb R}\!\frac{d\tau}{i}\
{\rm Tr}
\left\{
\frac{
{\mathbb S}_{\rm NS}\!\left(\alpha_4-\alpha_3+\alpha_2-\tau\right)
}{
{\mathbb S}_{\natural}\!\left(\alpha_4+\alpha_3-\alpha_2+\tau\right)
}
F_{{\scriptscriptstyle \rm\bf f}_3}^{\rm NS}
\frac{
{\mathbb S}_{\natural}\!\left(\alpha_s-\alpha_2+\tau\right)
}{
{\mathbb S}_{\rm NS}\!\left(\alpha_s+\alpha_2-\tau\right)
}
\left(F_{{\scriptscriptstyle \rm\bf f}_2}^{\natural}\right)^{T}
\right.
\\[4pt]
\nonumber
&&
\hskip 17mm
\left.
\frac{
{\mathbb S}_{\rm NS}\!\left(\alpha_1-\tau\right)
}{
{\mathbb S}_{\natural}\!\left(\alpha_1+\tau\right)
}
\Big[
 \left(F_{{\scriptscriptstyle{\rm\bf g}_2}}^{\natural}\right)^{T}
 \frac{{\mathbb S}_{\rm NS}\!\left(\alpha_u-\alpha_3-\tau\right)}
      {{\mathbb S}_{\natural}\!\left(\alpha_u+\alpha_3+\tau\right)}
F_{{\scriptscriptstyle{\rm\bf g}_3}}^{\rm NS}
\Big]^\dag
\right\}.
\end{eqnarray}
where
\[
\delta^{(-)}
=
\delta-i\pi\big(\alpha_4(Q-\alpha_4) + \alpha_1(Q-\alpha_1)-\alpha_u(Q-\alpha_u)-\alpha_s(Q-\alpha_s)\big).
\]

The explicit form of the braiding matrix $\mathbb B$
can be read off directly from  formula (\ref{braiding:unnorm}).
For instance
\begin{eqnarray}
\mathbb{B}^{+,\,\rm NS}_{\alpha_s\alpha_u}\!\!\left[^{\alpha_3\:\alpha_2}_{\alpha_4\:\alpha_1}\right]^{\rm e\, e}_{\rm e\,e}
& = &
{\textstyle \frac14}
{\rm e}^{\delta}\,
S_{\rm NS}(2\alpha_u)S_{\rm NS}(2Q-2\alpha_u)
\\[4pt]
\nonumber
&  &\hspace*{-1.5cm}
\times\int\limits_{i\mathbb R}\!\frac{d\tau}{i}
\left[
\frac{
S_{\rm NS}(\tau+\alpha_1)S_{\rm NS}(\tau+\bar\alpha_1)S_{\rm NS}(\tau+\alpha_4-\alpha_3+\alpha_2)S_{\rm NS}(\tau+\bar\alpha_4-\alpha_3+\alpha_2)
}{
S_{\rm NS}(\tau +\alpha_s+\alpha_2)S_{\rm NS}(\tau +\bar\alpha_s+\alpha_2)S_{\rm NS}(\tau +\alpha_u+\bar\alpha_3)S_{\rm NS}(\tau +\bar\alpha_u+\bar\alpha_3)
}
\right.
\\[4pt]
\nonumber
&&
\left.
\hspace*{-0.5cm}
+\,
\frac{
S_{\rm R}(\tau+\alpha_1)S_{\rm R}(\tau+\bar\alpha_1)S_{\rm R}(\tau+\alpha_4-\alpha_3+\alpha_2)S_{\rm R}(\tau+\bar\alpha_4-\alpha_3+\alpha_2)
}{
S_{\rm R}(\tau +\alpha_s+\alpha_2)S_{\rm R}(\tau +\bar\alpha_s+\alpha_2)S_{\rm R}(\tau +\alpha_u+\bar\alpha_3)S_{\rm R}(\tau +\bar\alpha_u+\bar\alpha_3)
}
\right],
\\[6pt]
 \label{braidingBB:explicit:1}
\mathbb{B}^{+,\,\rm NS}_{\alpha_s\alpha_u}\!\!\left[^{\alpha_3\:\alpha_2}_{\alpha_4\:\alpha_1}\right]^{\rm o\, o}_{\rm e\,e}
&=&
{\textstyle \frac14}
{\rm e}^{-\frac{i\pi}{2}}\,
{\rm e}^{\delta}\,
S_{\rm NS}(2\alpha_u)S_{\rm NS}(2Q-2\alpha_u)
\\[4pt]
\nonumber
&  &\hspace*{-1.5cm}
\times\int\limits_{i\mathbb R}\!\frac{d\tau}{i}
\left[
\frac{
S_{\rm NS}(\tau+\alpha_1)S_{\rm NS}(\tau+\bar\alpha_1)S_{\rm NS}(\tau+\alpha_4-\alpha_3+\alpha_2)S_{\rm NS}(\tau+\bar\alpha_4-\alpha_3+\alpha_2)
}{
S_{\rm NS}(\tau +\alpha_s+\alpha_2)S_{\rm NS}(\tau +\bar\alpha_s+\alpha_2)S_{\rm R}(\tau +\alpha_u+\bar\alpha_3)S_{\rm R}(\tau +\bar\alpha_u+\bar\alpha_3)
}
\right.
\\[4pt]
\nonumber
&&
\left.
\hspace*{-0.5cm}
-\,
\frac{
S_{\rm R}(\tau+\alpha_1)S_{\rm R}(\tau+\bar\alpha_1)S_{\rm R}(\tau+\alpha_4-\alpha_3+\alpha_2)S_{\rm R}(\tau+\bar\alpha_4-\alpha_3+\alpha_2)
}{
S_{\rm R}(\tau +\alpha_s+\alpha_2)S_{\rm R}(\tau +\bar\alpha_s+\alpha_2)S_{\rm NS}(\tau +\alpha_u+\bar\alpha_3)S_{\rm NS}(\tau +\bar\alpha_u+\bar\alpha_3)
}
\right],
\end{eqnarray}
or
\begin{eqnarray}
\label{braidingBB:explicit:2}
\mathbb{B}^{+,\,\rm R}_{\alpha_s\alpha_u}\!\!\left[^{\alpha_3\:\alpha_2}_{\alpha_4\:\alpha_1}\right]^{\rm o\, e}_{\rm e\,o}
& = &
{\textstyle \frac14}
{\rm e}^{\frac{i\pi}{2}}\,
{\rm e}^{\delta}\,
S_{\rm R}(2\alpha_u)S_{\rm R}(2Q-2\alpha_u)
\\[4pt]
\nonumber
&  &\hspace*{-1.5cm}
\times\int\limits_{i\mathbb R}\!\frac{d\tau}{i}
\left[
\frac{
S_{\rm NS}(\tau+\alpha_1)S_{\rm R}(\tau+\bar\alpha_1)S_{\rm R}(\tau+\alpha_4-\alpha_3+\alpha_2)S_{\rm NS}(\tau+\bar\alpha_4-\alpha_3+\alpha_2)
}{
S_{\rm R}(\tau +\alpha_s+\alpha_2)S_{\rm NS}(\tau +\bar\alpha_s+\alpha_2)S_{\rm NS}(\tau +\alpha_u+\bar\alpha_3)S_{\rm R}(\tau +\bar\alpha_u+\bar\alpha_3)
}
\right.
\\[4pt]
\nonumber
&&
\left.
\hspace*{-0.5cm}
-\,
\frac{
S_{\rm R}(\tau+\alpha_1)S_{\rm NS}(\tau+\bar\alpha_1)S_{\rm NS}(\tau+\alpha_4-\alpha_3+\alpha_2)S_{\rm R}(\tau+\bar\alpha_4-\alpha_3+\alpha_2)
}{
S_{\rm NS}(\tau +\alpha_s+\alpha_2)S_{\rm R}(\tau +\bar\alpha_s+\alpha_2)S_{\rm R}(\tau +\alpha_u+\bar\alpha_3)S_{\rm NS}(\tau +\bar\alpha_u+\bar\alpha_3)
}
\right].
\end{eqnarray}
The other cases differ from the expressions above only by constant phases and  NS/R indices of the Barnes functions.

\subsection{Braiding of chiral vertex operators}
In this subsection we shall derive the braiding properties of the chiral
vertex operators. Rather than presenting the general formula (which would be quite clumsy due to a plethora of
indices) we will discuss several examples choosing vertex operators from different sectors.
All other cases  can be easily obtained in a similar way.

\subsubsection{NS-NS braiding}
The braiding properties of the Neveu-Schwarz vertex operators in the NS sector
were already discussed in \cite{Chorazkiewicz:2008es}. In this subsection  we shall calculate
two examples of braiding matrices for the NS operators in the Ramond sector.
Let us first consider the composition
\[
V_{\rm e}^+\chiral{\Delta_3}{\beta_4}{\beta_s}(z_3)
V_{\rm e}^+\chiral{\Delta_2}{\beta_s}{\beta_1}(z_2):
\hspace{5pt}
{\cal W}_{\beta_1}\to {\cal W}_{\beta_4}\,,
\]
which in representation (\ref{chiral:matrix:elements:RR}) takes the form
\begin{eqnarray*}
V^+_{\rm e}\chiral{\Delta_3}{\beta_4}{\beta_s}(z_3)
V^+_{\rm e}\chiral{\Delta_2}{\beta_s}{\beta_1}(z_2)
& = &
\frac{{\sf V}_{{\rm e}\;s_3}^{\;\;\alpha_3}(z_3)
{\sf V}_{{\rm e}\;s_2}^{\;\;\alpha_2}(z_2)\big|_{\rm R}}{\bra{w^+_4}{\sf V}_{{\rm e}\;s_3}^{\;\;\alpha_3}(1)\ket{w^+_s}\bra{w^+_s}{\sf V}_{{\rm e}\;s_2}^{\;\;\alpha_2}(1)\ket{w^+_1}}
.
\end{eqnarray*}
The notation $\big|_{\rm R}$  indicates that the product of chiral fields
\(
{\sf V}_{{\rm e}\;s_3}^{\;\;\alpha_3}(z_3)
{\sf V}_{{\rm e}\;s_2}^{\;\;\alpha_2}(z_2)
\)
acts on the states from the Ramond sector.
It follows from (\ref{chiral:NSfields}) that for this product one can
apply  braiding relation (\ref{unn:braiding:def}) derived in the previous subsection.
In notation (\ref{un:braiding:new:notation}) we  get
\begin{eqnarray*}
{\sf V}_{{\rm e}\;s_3}^{\;\;\alpha_3}(z_3)
{\sf V}_{{\rm e}\;s_2}^{\;\;\alpha_2}(z_2)\big|_{\rm R}
& = &
\int\limits_{\mathbb S}\!\frac{d\alpha_u}{2i} \sum\limits_{\vec{\scriptscriptstyle \rm\bf g}}
\mathbb{B}^{\epsilon,\,\rm R}_{\alpha_s\alpha_u}\!\!\left[^{\alpha_3\:\alpha_2}_{\alpha_4\:\alpha_1}\right]^{\scriptscriptstyle\rm\bf g_2g_3}_{\scriptstyle\rm\, e\hskip 3pt e}\
{\sf V}_{{\scriptscriptstyle \rm\bf g_2}\;t_2}^{\hskip 8pt\alpha_2}(z_2)
{\sf V}_{{\scriptscriptstyle \rm\bf g_3}\;t_3}^{\hskip 8pt\alpha_3}(z_3)\big|_{\rm R}.
\end{eqnarray*}
Using (\ref{chiral:matrix:elements:RR}) one can express the r.h.s.\   in terms of chiral vertex operators
\begin{eqnarray*}
{\sf V}_{{\rm e}\;t_2}^{\hskip 4pt\alpha_2}(z_2)
{\sf V}_{{\rm e}\;t_3}^{\hskip 4pt\alpha_3}(z_3)\big|_{\rm R}
& = &
\bra{w_4^+}{\sf V}_{{\rm e}\;t_2}^{\hskip 4pt\alpha_2}(1)\ket{w_u^+}
\bra{w^+_u}{\sf V}_{{\rm e}\;t_3}^{\hskip 4pt\alpha_3}(1)\ket{w^+_1}
V_{\rm e}^+\chiral{\Delta_2}{\beta_4}{\beta_u}(z_2)
V_{\rm e}^+\chiral{\Delta_3}{\beta_u}{\beta_1}(z_3),
\\[8pt]
{\sf V}_{{\rm o}\;t_2}^{\hskip 4pt\alpha_2}(z_2)
{\sf V}_{{\rm o}\;t_3}^{\hskip 4pt\alpha_3}(z_3)\big|_{\rm R}
& = &
\bra{w_4^+}{\sf V}_{{\rm o}\;t_2}^{\hskip 4pt\alpha_2}(1)\ket{w_u^-}
\bra{w^+_u}{\sf V}_{{\rm o}\;t_3}^{\hskip 4pt\alpha_3}(1)\ket{w^-_1}
V_{\rm o}^-\chiral{\Delta_2}{\beta_4}{\beta_u}(z_2)
V_{\rm o}^-\chiral{\Delta_3}{\beta_u}{\beta_1}(z_3).
\end{eqnarray*}
Thus, if we define the braiding matrix by the relation
\begin{eqnarray*}
V_{\rm e}^+\chiral{\Delta_3}{\beta_4}{\beta_s}(z_3)
V_{\rm e}^+\chiral{\Delta_2}{\beta_s}{\beta_1}(z_2)
& = &
\int\limits_{\mathbb S}\!\frac{d\alpha_u}{2i}
\left(
{\sf B}^\epsilon_{\alpha_s\alpha_u}\!\left[^{\hskip 4pt \Delta_3\hskip 5pt\Delta_2}_{+\beta_4 \:+\beta_1}\right]^{\rm e\,e}_{\rm e\,e}
V_{\rm e}^+\chiral{\Delta_2}{\beta_4}{\beta_u}(z_2)
V_{\rm e}^+\chiral{\Delta_3}{\beta_u}{\beta_1}(z_3)
\right.
\\[4pt]
&&
\hskip .9cm
\left.
+\;
{\sf B}^\epsilon_{\alpha_s\alpha_u}\!\left[^{\hskip 4pt \Delta_3\hskip 5pt\Delta_2}_{+\beta_4 \:+\beta_1}\right]^{\rm o\,o}_{\rm e\,e}
V_{\rm o}^-\chiral{\Delta_2}{\beta_4}{\beta_u}(z_2)
V_{\rm o}^-\chiral{\Delta_3}{\beta_u}{\beta_1}(z_3)
\right),
\end{eqnarray*}
then
\begin{equation}
\label{braidingI}
\begin{array}{rrrrr}
{\sf B}^\epsilon_{\alpha_s\alpha_u}\!\left[^{\hskip 4pt \Delta_3\hskip 5pt\Delta_2}_{+\beta_4 \:+\beta_1}\right]^{\rm e\,e}_{\rm e\,e}
& = &                 \displaystyle
\frac{
{\cal M}^{\rm N\,R\,N}_{\alpha_4,\alpha_2,\alpha_u}\,
{\cal M}^{\rm N\,R\,N}_{\alpha_u,\alpha_3,\alpha_1}
}{
{\cal M}^{\rm N\,R\,N}_{\alpha_4,\alpha_3,\alpha_s}\,
{\cal M}^{\rm N\,R\,N}_{\alpha_s,\alpha_2,\alpha_1}
}\
\mathbb{B}^{\epsilon,\,\rm R}_{\alpha_s\alpha_u}\!\!\left[^{\alpha_3\:\alpha_2}_{\alpha_4\:\alpha_1}\right]^{\rm e\,e}_{\rm e\,e},
\\[18pt]
{\sf B}^\epsilon_{\alpha_s\alpha_u}\!\left[^{\hskip 4pt \Delta_3\hskip 5pt\Delta_2}_{+\beta_4 \:+\beta_1}\right]^{\rm o\,o}_{\rm e\,e}
& = &   \displaystyle
-\frac{
{\cal M}^{\rm R\,N\,R}_{\alpha_4,\alpha_2,\alpha_u}\,
{\cal M}^{\rm R\,N\,R}_{\alpha_u,\alpha_3,\alpha_1}
}{
{\cal M}^{\rm N\,R\,N}_{\alpha_4,\alpha_3,\alpha_s}\,
{\cal M}^{\rm N\,R\,N}_{\alpha_s,\alpha_2,\alpha_1}
}\
\mathbb{B}^{\epsilon,\,\rm R}_{\alpha_s\alpha_u}\!\!\left[^{\alpha_3\:\alpha_2}_{\alpha_4\:\alpha_1}\right]^{\rm o\,o}_{\rm e\,e}.
\end{array}
\end{equation}
The braiding of all other pairs of the NS chiral vertex operators can be calculated in an essentially the same way.
For instance, for the composition
\begin{equation}
\label{RRvertex:chiral:fields:1}
V_{\rm o}^+\chiral{\Delta_3}{\beta_4}{\beta_s}(z_3)
V_{\rm e}^-\chiral{\Delta_2}{\beta_s}{\beta_1}(z_2)
=
\frac{i\,{\sf V}_{{\rm e}\;s_3}^{\;\;\alpha_3}(z_3)
{\sf V}_{{\rm o}\;s_2}^{\;\;\alpha_2}(z_2)\big|_{\rm R}}{\bra{w_4^+}{\sf V}_{{\rm e}\;s_3}^{\;\;\alpha_3}(1)\ket{w_s^+}
\bra{w_s^-}{\sf V}_{{\rm o}\;s_2}^{\;\;\alpha_2}(1)\ket{w_1^+}}\,
\end{equation}
the relevant braiding relation (\ref{unn:braiding:def})  reads
\begin{eqnarray}
\label{br1}
{\sf V}_{{\rm e}\;s_3}^{\;\;\alpha_3}(z_3)
{\sf V}_{{\rm o}\;s_2}^{\;\;\alpha_2}(z_2)\big|_{\rm R}
& = &
\int\limits_{\mathbb S}\!\frac{d\alpha_u}{2i} \sum\limits_{\vec{\scriptscriptstyle \rm\bf g}}
\mathbb{B}^{\epsilon,\,\rm R}_{\alpha_s\alpha_u}\!\!\left[^{\alpha_3\:\alpha_2}_{\alpha_4\:\alpha_1}\right]^{\scriptscriptstyle\rm\bf g_2g_3}_{\scriptstyle\rm\, e\hskip 3pt o}\
{\sf V}_{{\scriptscriptstyle \rm\bf g_2}\;t_2}^{\hskip 8pt\alpha_2}(z_2)
{\sf V}_{{\scriptscriptstyle \rm\bf g_3}\;t_3}^{\hskip 8pt\alpha_3}(z_3)\big|_{\rm R}.
\end{eqnarray}
As before, one can express  the r.h.s.\ of (\ref{RRvertex:chiral:fields:1})  in terms of chiral vertex operators:
\begin{equation}
\label{re1}
\begin{array}{rcllll}
{\sf V}_{{\rm e}\;t_2}^{\hskip 4pt\alpha_2}(z_2)
{\sf V}_{{\rm o}\;t_3}^{\hskip 4pt\alpha_3}(z_3)\big|_{\rm R}
& = &
-i\bra{w_4^+}{\sf V}_{{\rm e}\;t_2}^{\hskip 4pt\alpha_2}(1)\ket{w_u^+}\bra{w^-_u}{\sf V}_{{\rm o}\;t_3}^{\hskip 4pt\alpha_3}(1)\ket{w^+_1}
V^+_{\rm o}\chiral{\Delta_2}{\beta_4}{\beta_u}(z_2)V^-_{\rm e}\chiral{\Delta_3}{\beta_u}{\beta_1}(z_3),
\\[6pt]
{\sf V}_{{\rm o}\;t_2}^{\hskip 4pt\alpha_2}(z_2)
{\sf V}_{{\rm e}\;t_3}^{\hskip 4pt\alpha_3}(z_3)\big|_{\rm R}
& = &
i\bra{w_4^-}{\sf V}_{{\rm o}\;t_2}^{\hskip 4pt\alpha_2}(1)\ket{w_u^+}\bra{w^-_u}{\sf V}_{{\rm e}\;t_3}^{\hskip 4pt\alpha_3}(1)\ket{w^-_1}
V_{\rm e}^-\chiral{\Delta_2}{\beta_4}{\beta_u}(z_2)V_{\rm o}^+\chiral{\Delta_3}{\beta_u}{\beta_1}(z_3).
\end{array}
\end{equation}
Equations (\ref{RRvertex:chiral:fields:1}), (\ref{br1}) and (\ref{re1}) suggest the following definition
of the braiding matrix
\begin{eqnarray*}
V_{\rm o}^+\chiral{\Delta_3}{\beta_4}{\beta_s}(z_3)
V_{\rm e}^-\chiral{\Delta_2}{\beta_s}{\beta_1}(z_2)
& = &
\int\limits_{\mathbb S}\!\frac{d\alpha_u}{2i}
\left(
{\sf B}^\epsilon_{\alpha_s\alpha_u}\!\left[^{\hskip 4pt \Delta_3\hskip 5pt\Delta_2}_{+\beta_4 \:-\beta_1}\right]^{\rm o\,e}_{\rm o\,e}
V_{\rm o}^+\chiral{\Delta_2}{\beta_4}{\beta_u}(z_2)
V_{\rm e}^-\chiral{\Delta_3}{\beta_u}{\beta_1}(z_3)
\right.
\\[4pt]
&&
\hskip .9cm
\left.
+\;
{\sf B}^\epsilon_{\alpha_s\alpha_u}\!\left[^{\hskip 4pt \Delta_3\hskip 5pt\Delta_2}_{+\beta_4 \:-\beta_1}\right]^{\rm e\,o}_{\rm o\,e}
V_{\rm e}^-\chiral{\Delta_2}{\beta_4}{\beta_u}(z_2)
V_{\rm o}^+\chiral{\Delta_3}{\beta_u}{\beta_1}(z_3)
\right).
\end{eqnarray*}
Then the results above yield
\begin{equation}
\label{braidingII}
\begin{array}{rrrr}
{\sf B}^\epsilon_{\alpha_s\alpha_u}\!
\left[^{\hskip 4pt \Delta_3\hskip 5pt\Delta_2}_{+\beta_4 \:-\beta_1}\right]^{\rm o\,e}_{\rm o\,e}
& = &    \displaystyle
 \,\frac{
{\cal M}^{\rm N\,R\,N}_{\alpha_4,\alpha_2,\alpha_u}\,
{\cal M}^{\rm R\,N\,R}_{\alpha_u,\alpha_3,\alpha_1}
}{
{\cal M}^{\rm N\,R\,N}_{\alpha_4,\alpha_3,\alpha_s}\,
{\cal M}^{\rm R\,N\,R}_{\alpha_s,\alpha_2,\alpha_1}
}\
\mathbb{B}^{\epsilon,\,\rm R}_{\alpha_s\alpha_u}\!\!
\left[^{\alpha_3\:\alpha_2}_{\alpha_4\:\alpha_1}\right]^{\rm e\,o}_{\rm e\,o},
\\   [18pt]
{\sf B}^\epsilon_{\alpha_s\alpha_u}\!\left[^{\hskip 4pt \Delta_3\hskip 5pt\Delta_2}_{+\beta_4 \:-\beta_1}\right]^{\rm e\,o}_{\rm o\,e}
& = &      \displaystyle
-\,\frac{
{\cal M}^{\rm R\,N\,R}_{\alpha_4,\alpha_2,\alpha_u}\,
{\cal M}^{\rm N\,R\,N}_{\alpha_u,\alpha_3,\alpha_1}
}{
{\cal M}^{\rm N\,R\,N}_{\alpha_4,\alpha_3,\alpha_s}\,
{\cal M}^{\rm R\,N\,R}_{\alpha_s,\alpha_2,\alpha_1}
}\
\mathbb{B}^{\epsilon,\,\rm R}_{\alpha_s\alpha_u}\!\!\left[^{\alpha_3\:\alpha_2}_{\alpha_4\:\alpha_1}\right]^{\rm o\,e}_{\rm e\,o}.
\end{array}
\end{equation}
\subsubsection{NS-R braiding}
As an example of the braiding between the NS and the R fields consider the composition
\[
V_{\rm o}^-\chiral{\Delta_3}{\beta_4}{\beta_s}(z_3)
V_{\rm e}^-\chiral{\beta_2}{\beta_s}{\Delta_1}(z_2):
\hspace{5pt}
{\cal V}_{\alpha_1} \to {\cal W}_{\beta_4}\,,
\]
which can be represented,  (\ref{chiral:matrix:elements:RR}), (\ref{chiral:matrix:elements}), as:
\begin{eqnarray*}
V_{\rm o}^-\chiral{\Delta_3}{\beta_4}{\beta_s}(z_3)
V_{\rm e}^-\chiral{\beta_2}{\beta_s}{\Delta_1}(z_2)
& = &
\frac{{\sf V}_{{\rm o}\;s_3}^{\;\;\alpha_3}(z_3){\sf W}_{{\rm e}\hskip 4pt s_2}^{-\, \beta_2}(z_2)\big|_{\rm NS}}
{\bra{w_4^+}{\sf V}_{{\rm o}\;s_3}^{\;\;\alpha_3}(1)\ket{w_s^-}\bra{w_s^+}{\sf W}_{{\rm e}\hskip 4pt s_2}^{-\, \beta_2}(1)\ket{\nu_1}}\,.
\end{eqnarray*}
By definitions (\ref{chiral:NSfields}), (\ref{chiral:Rfields})
\begin{eqnarray*}
{\sf V}_{{\rm o}\;s_3}^{\;\;\alpha_3}(\sigma_3){\sf W}_{{\rm e}\hskip 4pt s_2}^{-\, \beta_2}(\sigma_2)\big|_{\rm NS}
& = &
{\sf g}_{\,{\rm o}\,s_3}^{\hskip 5pt\alpha_3}\!(\sigma_3)\,\sigma^-(\sigma_2){\sf g}_{\,{\rm o}\,s_2}^{\;\;\alpha_2}\!(\sigma_2)\big|_{\rm NS},
\end{eqnarray*}
where $\alpha_2 = {Q\over 2}-\sqrt{2}\beta_2$.
From (\ref{def_sigma}) one gets the braiding of  ${\sf g}_{\,{\rm o}\,s_3}^{\hskip 5pt\alpha_3}\!(\sigma_3)$ and $\sigma^-(\sigma_2)$
\begin{eqnarray*}
{\sf g}_{\,{\rm o}\,s_3}^{\hskip 5pt\alpha_3}\!(\sigma_3)\sigma^-(\sigma_2)
& = &
{\rm e}^{\frac{i\pi}{2}(\epsilon-1)}\,\sigma^-(\sigma_2){\sf g}_{\,{\rm o}\,s_3}^{\hskip 5pt\alpha_3}\!(\sigma_3).
\end{eqnarray*}
Then using braiding relation (\ref{unn:braiding:def}) and notation (\ref{un:braiding:new:notation})
we obtain
\begin{eqnarray*}
&&
\hskip -2cm
{\sf g}_{\,{\rm o}\,s_3}^{\hskip 5pt\alpha_3}\!(\sigma_3)\,\sigma^-(\sigma_2){\sf g}_{\,{\rm o}\,s_2}^{\hskip 5pt\alpha_2}\!(\sigma_2)\big|_{\rm NS}
\; = \;
{\rm e}^{\frac{i\pi}{2}(\epsilon-1)}\,\sigma^-(\sigma_2){\sf g}_{\,{\rm o}\,s_3}^{\hskip 5pt\alpha_3}\!(\sigma_3)\,{\sf g}_{\,{\rm o}\,s_2}^{\hskip 5pt\alpha_2}\!(\sigma_2)\big|_{\rm NS}
\\[4pt]
& = &
\int\limits_{\mathbb S}\!\frac{d\alpha_u}{2i} \sum\limits_{\vec{\scriptscriptstyle \rm\bf g}}
{\rm e}^{\frac{i\pi}{2}(\epsilon-1)}\,\mathbb{B}^{\epsilon,\,\rm NS}_{\alpha_s\alpha_u}\!\!\left[^{\alpha_3\:\alpha_2}_{\alpha_4\:\alpha_1}\right]^{\scriptscriptstyle\rm\bf g_2g_3}_{\scriptstyle\rm\, o\hskip 3pt o}\,
\sigma^-(\sigma_2){\sf g}_{\,{\scriptscriptstyle\rm\bf g_2}\,t_2}^{\hskip 8.5pt\alpha_2}\!(\sigma_2)
\,{\sf g}_{\,{\scriptscriptstyle\rm\bf g_3}\,t_3}^{\hskip 8.5pt\alpha_3}\!(\sigma_3)\big|_{\rm NS}
\end{eqnarray*}
and
\begin{eqnarray*}
{\sf V}_{{\rm o}\;s_3}^{\;\;\alpha_3}(z_3){\sf W}_{{\rm e}\hskip 4pt s_2}^{-\, \beta_2}(z_2)\big|_{\rm NS}
& = &
\int\limits_{\mathbb S}\!\frac{d\alpha_u}{2i} \sum\limits_{\vec{\scriptscriptstyle \rm\bf g}}
{\rm e}^{\frac{i\pi}{2}(\epsilon-1)}\,\mathbb{B}^{\epsilon,\,\rm NS}_{\alpha_s\alpha_u}\!\!\left[^{\alpha_3\:\alpha_2}_{\alpha_4\:\alpha_1}\right]^{\scriptscriptstyle\rm\bf g_2g_3}_{\scriptstyle\rm\, o\hskip 3pt o}\,
{\sf W}_{\overline{\scriptscriptstyle\rm\bf g}_2\, t_2}^{-\, \beta_2}(z_2)
{\sf V}_{{\scriptscriptstyle\rm\bf g_3}\,t_3}^{\hskip 7.5pt\alpha_3}(z_3)\big|_{\rm NS}.
\end{eqnarray*}
By formulae (\ref{chiral:matrix:elements:NN}) and  (\ref{chiral:matrix:elements})
the r.h.s. can be expressed in terms of chiral vertex operators
\begin{eqnarray*}
{\sf W}_{{\rm o}\hskip 4pt t_2}^{-\, \beta_2}(z_2)
{\sf V}_{{\rm e}\;t_3}^{\;\;\alpha_3}(z_3)\big|_{\rm NS}
& = &
\bra{w_4^-}{\sf W}_{{\rm o}\hskip 4pt t_2}^{-\, \beta_2}(1)\ket{\nu_u}
\bra{\nu_u}{\sf V}_{{\rm e}\;t_3}^{\;\;\alpha_3}(1)\ket{\nu_1}\,
V_{\rm o}^+\chiral{\beta_2}{\beta_4}{\Delta_u}(z_2)
V_{\rm e}\chiral{ \Delta_3}{\Delta_u}{\Delta_1}(z_3),
\\[6pt]
{\sf W}_{{\rm e}\hskip 4pt t_2}^{-\, \beta_2}(z_2)
{\sf V}_{{\rm o}\;t_3}^{\;\;\alpha_3}(z_3)\big|_{\rm NS}
& = &
\bra{w_4^+}{\sf W}_{\;{\rm e}\;t_2}^{-\,\beta_2}(z_2)\ket{\nu_u}
\bra{\nu_u}{\sf V}_{{\rm e}\;t_3}^{*\hskip 1pt \alpha_3}(1)\ket{\nu_1}\,
V_{\rm e}^-\chiral{\beta_2}{\beta_4}{\Delta_u}(z_2)
V_{\rm o}\chiral{ \Delta_3}{\Delta_u}{\Delta_1}(z_3).
\end{eqnarray*}
This leads to the braiding relation
\begin{eqnarray*}
V_{\rm o}^-\chiral{\Delta_3}{\beta_4}{\beta_s}(z_3)
V_{\rm e}^-\chiral{\beta_2}{\beta_s}{\Delta_1}(z_2)
& = &
\int\limits_{\mathbb S}\!\frac{d\alpha_u}{2i}
\left(
{\sf B}^\epsilon_{\alpha_s\alpha_u}\!\left[^{\hskip 3pt\Delta_3\:-\beta_2}_{ -\beta_4\hskip 4pt \Delta_1}\right]^{\rm o\,e}_{\rm o\,e}
V_{\rm o}^+\chiral{\beta_2}{\beta_4}{\Delta_u}(z_2)
V_{\rm e}\chiral{\Delta_3}{\Delta_u}{\Delta_1}(z_3)
\right.
\\[4pt]
&&
\hskip .9cm
\left.
+\;
{\sf B}^\epsilon_{\alpha_s\alpha_u}\!\left[^{\hskip 3pt\Delta_3\:-\beta_2}_{ -\beta_4\hskip 4pt \Delta_1}\right]^{\rm e\,o}_{\rm o\,e}
V_{\rm e}^-\chiral{\beta_2}{\beta_4}{\Delta_u}(z_2)
V_{\rm o}\chiral{ \Delta_3}{\Delta_u}{\Delta_1}(z_3)
\right),
\end{eqnarray*}
where
\begin{equation}
\begin{array}{rrrr}
{\sf B}^\epsilon_{\alpha_s\alpha_u}\!\left[^{\hskip 3pt\Delta_3\:-\beta_2}_{ -\beta_4\hskip 4pt \Delta_1}\right]^{\rm o\,e}_{\rm o\,e}
& = & \displaystyle
i{\rm e}^{\frac{i\pi\epsilon}{2}}\,
\frac{
{\cal M}^{\rm R\,N\,N}_{\alpha_4,\alpha_2,\alpha_u}\,                     
{\cal M}^{\rm N\,N\,N}_{\alpha_u,\alpha_3,\alpha_1}                       
}{
{\cal M}^{\rm R\,N\,R}_{\alpha_4,\alpha_3,\alpha_s}\,                     
{\cal M}^{\rm N\,R\,R}_{\alpha_s,\alpha_2,\alpha_1}                       
}\
\mathbb{B}^{\epsilon,\,\rm NS}_{\alpha_s\alpha_u}\!\!
\left[^{\alpha_3\:\alpha_2}_{\alpha_4\:\alpha_1}\right]^{\rm e\,e}_{\rm o\,o} & ,
\\[18pt]
{\sf B}^\epsilon_{\alpha_s\alpha_u}\!\left[^{\hskip 3pt\Delta_3\:-\beta_2}_{ -\beta_4\hskip 4pt \Delta_1}\right]^{\rm e\,o}_{\rm o\,e}
& = & \displaystyle
{\rm e}^{-\frac{i\pi\epsilon}{2}}\,
\frac{
{\cal M}^{\rm N\,R\,R}_{\alpha_4,\alpha_2,\alpha_u}\,                     
{\cal M}^{\rm R\,R\,R}_{\alpha_u,\alpha_3,\alpha_1}                       
}{
{\cal M}^{\rm R\,N\,R}_{\alpha_4,\alpha_3,\alpha_s}\,                     
{\cal M}^{\rm N\,R\,R}_{\alpha_s,\alpha_2,\alpha_1}                       
}\
\mathbb{B}^{\epsilon,\,\rm NS}_{\alpha_s\alpha_u}\!\!
\left[^{\alpha_3\:\alpha_2}_{\alpha_4\:\alpha_1}\right]^{\rm o\,o}_{\rm o\,o} & .
\end{array}
\end{equation}
\subsubsection{R-R braiding}
We shall start from the composition of Ramond vertex operators in the NS sector
\[
V_{\rm e}^+\chiral{\beta_3}{\Delta_4}{\beta_s}(z_3)
V_{\rm o}^+\chiral{\beta_2}{\beta_s}{\Delta_1}(z_2):
\hskip 5mm
{\cal V}_{\alpha_1} \to {\cal V}_{\alpha_4}.
\]
In representation (\ref{chiral:matrix:elements:RR}):
\begin{eqnarray*}
V_{\rm e}^+\chiral{\beta_3}{\Delta_4}{\beta_s}(z_3)
V_{\rm o}^+\chiral{\beta_2}{\beta_s}{\Delta_1}(z_2)
& = &
\frac{{\sf W}_{{\rm e}\hskip 4.5pt s_3}^{+\, \beta_3}(z_3){\sf W}_{{\rm o}\hskip 4.5pt s_2}^{-\, \beta_2}(z_2)\big|_{\rm NS}}
{\bra{\nu_4}{\sf W}_{{\rm e}\hskip 4.5pt s_3}^{+\,\beta_3}(1)\ket{w_s^+}
\bra{w_s^-}{\sf W}_{{\rm o}\hskip 4.5pt s_2}^{-\, \beta_2}(1)\ket{\nu_1}}.
\end{eqnarray*}
From definition  (\ref{chiral:Rfields}) and  braiding properties
(\ref{def_sigma}), (\ref{braiding:sigma}) one gets
\begin{eqnarray*}
{\sf W}_{{\rm e}\hskip 4.5pt s_3}^{+\, \beta_3}(\sigma_3){\sf W}_{{\rm o}\hskip 4.5pt s_2}^{-\, \beta_2}(\sigma_2)\big|_{\rm NS}
& = &
\sigma^+(\sigma_3){\sf g}_{\,{\rm e}\,s_3}^{\hskip 5pt\alpha_3}\!(\sigma_3)
\sigma^-(\sigma_2){\sf g}_{\,{\rm e}\,s_2}^{\hskip 5pt\alpha_2}\!(\sigma_2)\big|_{\rm NS}
\\[4pt]
& = &
\sigma^+(\sigma_3)\sigma^-(\sigma_2)
{\sf g}_{\,{\rm e}\,s_3}^{\hskip 5pt\alpha_3}\!(\sigma_3){\sf g}_{\,{\rm e}\,s_2}^{\hskip 5pt\alpha_2}\!(\sigma_2)\big|_{\rm NS}
\\[4pt]
& = &
\frac{1}{\sqrt2}{\rm e}^{\frac{i\pi}{8}}
\left(\sigma^-(\sigma_2)\sigma^+(\sigma_3) + \sigma^+(\sigma_2)\sigma^-(\sigma_3)\right)
{\sf g}_{\,{\rm e}\,s_3}^{\hskip 5pt\alpha_3}\!(\sigma_3){\sf g}_{\,{\rm e}\,s_2}^{\hskip 5pt\alpha_2}\!(\sigma_2)\big|_{\rm NS}.
\end{eqnarray*}
It the present case the braiding relation of  $\sf g$ fields we need reads
\begin{eqnarray*}
{\sf g}_{\,{\rm e}\,s_3}^{\hskip 5pt\alpha_3}\!(\sigma_3){\sf g}_{\,{\rm e}\,s_2}^{\hskip 5pt\alpha_2}\!(\sigma_2)\big|_{\rm NS}
& = &
\int\limits_{\mathbb S}\!\frac{d\alpha_u}{2i} \sum\limits_{\vec{\scriptscriptstyle \rm\bf g}}
\mathbb{B}^{\epsilon,\,\rm NS}_{\alpha_s\alpha_u}\!\!\left[^{\alpha_3\:\alpha_2}_{\alpha_4\:\alpha_1}\right]^{\scriptscriptstyle\rm\bf g_2g_3}_{\scriptstyle\rm\, e\hskip 3pt e}\
{\sf g}_{\,{\scriptscriptstyle\rm\bf g_2}\,t_2}^{\hskip 8pt\alpha_2}\!(\sigma_2){\sf g}_{\,{\scriptscriptstyle\rm\bf g_3}\,t_3}^{\hskip 8pt\alpha_3}\!(\sigma_3)\big|_{\rm NS}.
\end{eqnarray*}
Using braiding relations (\ref{def_sigma}) and representation (\ref{chiral:matrix:elements:RR}) one gets:
\begin{eqnarray*}
&&
\hskip -2cm
\sigma^-(\sigma_2)\sigma^+(\sigma_3)
{\sf g}_{\,{\rm e}\,t_2}^{\hskip 5pt\alpha_2}\!(\sigma_2){\sf g}_{\,{\rm e}\,t_3}^{\hskip 5pt\alpha_3}\!(\sigma_3)\big|_{\rm NS}
\; = \;
{\sf W}_{{\rm o}\hskip 4.5pt t_2}^{-\, \beta_2}(\sigma_2){\sf W}_{{\rm e}\hskip 4.5pt t_3}^{+\, \beta_3}(\sigma_3)\big|_{\rm NS}
\\[6pt]
& = &
\bra{\nu_4}{\sf W}_{{\rm e}\hskip 4.5pt t_2}^{+\, \beta_2}(1)\ket{w_u^+}
\bra{w_u^+}{\sf W}_{{\rm e}\hskip 4.5pt t_3}^{+\, \beta_3}(1)\ket{\nu_1}\,
V_{\rm o}^+\chiral{\beta_2}{\Delta_4}{\beta_u}(z_2)
V_{\rm e}^+\chiral{\beta_3}{\beta_u}{\Delta_1}(z_3),
\\[6pt]
&&
\hskip -2cm
\sigma^+(\sigma_2)\sigma^-(\sigma_3)
{\sf g}_{\,{\rm e}\,t_2}^{\hskip 5pt\alpha_2}\!(\sigma_2){\sf g}_{\,{\rm e}\,t_3}^{\hskip 5pt\alpha_3}\!(\sigma_3)\big|_{\rm NS}
\; = \;
{\sf W}_{{\rm e}\hskip 4.5pt t_2}^{+\, \beta_2}(\sigma_2){\sf W}_{{\rm o}\hskip 4.5pt t_3}^{-\, \beta_3}(\sigma_3)\big|_{\rm NS}
\\[6pt]
& = &
\bra{\nu_4}{\sf W}_{{\rm e}\hskip 4.5pt t_2}^{+\, \beta_2}(1)\ket{w_u^+}
\bra{w_u^+}{\sf W}_{{\rm e}\hskip 4.5pt t_3}^{+\, \beta_3}(1)\ket{\nu_1}\,
V_{\rm e}^+\chiral{\beta_2}{\Delta_4}{\beta_u}(z_2)
V_{\rm o}^+\chiral{\beta_3}{\beta_u}{\Delta_1}(z_3),
\end{eqnarray*}
\begin{eqnarray*}
&&
\hskip -1cm
\sigma^-(\sigma_2)\sigma^+(\sigma_3)
{\sf g}_{\,{\rm o}\,t_2}^{\hskip 5pt\alpha_2}\!(\sigma_2){\sf g}_{\,{\rm o}\,t_3}^{\hskip 5pt\alpha_3}\!(\sigma_3)\big|_{\rm NS}
\; = \;
{\rm e}^{\frac{i\pi}{2}(\epsilon-1)}\,{\sf W}_{{\rm e}\hskip 4.5pt t_2}^{-\, \beta_2}(\sigma_2){\sf W}_{{\rm o}\hskip 4.5pt t_3}^{+\, \beta_3}(\sigma_3)\big|_{\rm NS}
\\[6pt]
& = &
-i{\rm e}^{\frac{i\pi}{2}(\epsilon-1)}\,\bra{\nu_4}{\sf W}_{{\rm e}\hskip 4.5pt t_2}^{-\, \beta_2}(1)\ket{w_u^+}
\bra{w_u^+}{\sf W}_{{\rm e}\hskip 4.5pt t_3}^{-\, \beta_3}(1)\ket{\nu_1}\,
V_{\rm e}^-\chiral{\beta_2}{\Delta_4}{\beta_u}(z_2)
V_{\rm o}^-\chiral{\beta_3}{\beta_u}{\Delta_1}(z_3),
\\[6pt]
&&
\hskip -1cm
\sigma^+(\sigma_2)\sigma^-(\sigma_3)
{\sf g}_{\,{\rm o}\,t_2}^{\hskip 5pt\alpha_2}\!(\sigma_2){\sf g}_{\,{\rm o}\,t_3}^{\hskip 5pt\alpha_3}\!(\sigma_3)\big|_{\rm NS}
\; = \;
{\rm e}^{\frac{i\pi}{2}(\epsilon+1)}\,{\sf W}_{{\rm o}\hskip 4.5pt t_2}^{+\, \beta_2}(\sigma_2){\sf W}_{{\rm e}\hskip 4.5pt t_3}^{-\, \beta_3}(\sigma_3)\big|_{\rm NS}
\\[6pt]
& = &
i{\rm e}^{\frac{i\pi}{2}(\epsilon-1)}\,\bra{\nu_4}{\sf W}_{{\rm e}\hskip 4.5pt t_2}^{-\, \beta_2}(1)\ket{w_u^+}
\bra{w_u^+}{\sf W}_{{\rm e}\hskip 4.5pt t_3}^{-\, \beta_3}(1)\ket{\nu_1}\,
V_{\rm o}^-\chiral{\beta_2}{\Delta_4}{\beta_u}(z_2)
V_{\rm e}^-\chiral{\beta_3}{\beta_u}{\Delta_1}(z_3).
\end{eqnarray*}
We are thus lead to following form of the braiding relation
\begin{eqnarray*}
&&
\hskip -0.5cm
V_{\rm e}^+\chiral{\beta_3}{\Delta_4}{\beta_s}(z_3)
V_{\rm o}^+\chiral{\beta_2}{\beta_s}{\Delta_1}(z_2)\; =
\\[4pt]
& = &
\int\limits_{\mathbb S}\!\frac{d\alpha_u}{2i}
\left[
{\sf B}^\epsilon_{\alpha_s\alpha_u}\!\left[^{+\beta_3\:+\beta_2}_{\hskip 4pt \Delta_4\hskip 6pt \Delta_1}\right]^{++}_{\rm\, e\,o}
\left(
V_{\rm e}^+\chiral{\beta_2}{\Delta_4}{\beta_u}(z_2)
V_{\rm o}^+\chiral{\beta_3}{\beta_u}{\Delta_1}(z_3)
+
V_{\rm o}^+\chiral{\beta_2}{\Delta_4}{\beta_u}(z_2)
V_{\rm e}^+\chiral{\beta_3}{\beta_u}{\Delta_1}(z_3)
\right)
\right.
\\[4pt]
&&
\hskip .9cm
\left.
+\;
{\sf B}^\epsilon_{\alpha_s\alpha_u}\!\left[^{+\beta_3\:+\beta_2}_{\hskip 4pt \Delta_4\hskip 6pt \Delta_1}\right]^{--}_{\rm\, e\,o}
\left(
V_{\rm e}^-\chiral{\beta_2}{\Delta_4}{\beta_u}(z_2)
V_{\rm o}^-\chiral{\beta_3}{\beta_u}{\Delta_1}(z_3)
-
V_{\rm o}^-\chiral{\beta_2}{\Delta_4}{\beta_u}(z_2)
V_{\rm e}^-\chiral{\beta_3}{\beta_u}{\Delta_1}(z_3)
\right)
\right],
\end{eqnarray*}
where
\begin{equation}
\begin{array}{rrrl}
\nonumber
{\sf B}^\epsilon_{\alpha_s\alpha_u}\!\left[^{+\beta_3\:+\beta_2}_{\hskip 4pt \Delta_4\hskip 6pt \Delta_1}\right]^{++}_{\rm\, e\,o}
& = & \displaystyle
\frac{1}{\sqrt 2}\,{\rm e}^{\frac{i\pi}{8}}\,
\frac{
{\cal M}^{\rm R\,R\,N}_{\alpha_4,\alpha_2,\alpha_u}\,                        
{\cal M}^{\rm R\,N\,N}_{\alpha_u,\alpha_3,\alpha_1}                          
}{
{\cal M}^{\rm R\,R\,N}_{\alpha_4,\alpha_3,\alpha_s}\,                        
{\cal M}^{\rm R\,N\,N}_{\alpha_s,\alpha_2,\alpha_1}                          
}\
\mathbb{B}^{\epsilon,\,\rm NS}_{\alpha_s\alpha_u}\!\!\left[^{\alpha_3\:\alpha_2}_{\alpha_4\:\alpha_1}\right]^{\rm e\, e}_{\rm e\, e}
& ,
\\[18pt]
{\sf B}^\epsilon_{\alpha_s\alpha_u}\!\left[^{+\beta_3\:+\beta_2}_{\hskip 4pt \Delta_4\hskip 6pt \Delta_1}\right]^{--}_{\rm\, e\,o}
& = & \displaystyle
\frac{1}{\sqrt 2}\,{\rm e}^{\frac{i\pi}{8}+\frac{i\pi\epsilon}{2}}\,
\frac{
{\cal M}^{\rm N\,N\,R}_{\alpha_4,\alpha_2,\alpha_u}\,                         
{\cal M}^{\rm N\,R\,R}_{\alpha_u,\alpha_3,\alpha_1}                           
}{
{\cal M}^{\rm R\,R\,N}_{\alpha_4,\alpha_3,\alpha_s}\,                        
{\cal M}^{\rm R\,N\,N}_{\alpha_s,\alpha_2,\alpha_1}                          
}\
\mathbb{B}^{\epsilon,\,\rm NS}_{\alpha_s\alpha_u}\!\!\left[^{\alpha_3\:\alpha_2}_{\alpha_4\:\alpha_1}\right]^{\rm o\, o}_{\rm e\, e}
& .
\end{array}
\end{equation}
As our last  example we present the result for a composition of two Ramond vertex operators in the Ramond sector,
\begin{eqnarray*}
&&
\hskip -0.5cm
V_{\rm e}^+\chiral{\beta_3}{\beta_4}{\Delta_s}(z_3)
V_{\rm o}^-\chiral{\beta_2}{\Delta_s}{\beta_1}(z_2)\; =
\\[4pt]
& = &
\int\limits_{\mathbb S}\!\frac{d\alpha_u}{2i}
\left[
{\sf B}^\epsilon_{\alpha_s\alpha_u}\!\left[^{+\beta_3\:-\beta_2}_{\hskip 6pt \beta_4\hskip 8pt \beta_1}\right]^{+-}_{\rm\, e\,o}
\left(
V_{\rm e}^+\chiral{\beta_2}{\beta_4}{\Delta_u}(z_2)
V_{\rm o}^-\chiral{\beta_3}{\Delta_u}{\beta_1}(z_3)
+
V_{\rm o}^+\chiral{\beta_2}{\beta_4}{\Delta_u}(z_2)
V_{\rm e}^-\chiral{\beta_3}{\Delta_u}{\beta_1}(z_3)
\right)
\right.
\\[4pt]
&&
\hskip .9cm
\left.
+\;
{\sf B}^\epsilon_{\alpha_s\alpha_u}\!\left[^{+\beta_3\:-\beta_2}_{\hskip 6pt \beta_4\hskip 8pt \beta_1}\right]^{-+}_{\rm\, e\,o}
\left(
V_{\rm e}^-\chiral{\beta_2}{\beta_4}{\Delta_u}(z_2)
V_{\rm o}^+\chiral{\beta_3}{\Delta_u}{\beta_1}(z_3)
-
V_{\rm o}^-\chiral{\beta_2}{\beta_4}{\Delta_u}(z_2)
V_{\rm e}^+\chiral{\beta_3}{\Delta_u}{\beta_1}(z_3)
\right)
\right],
\end{eqnarray*}
where
$$
\begin{array}{rcr}
\nonumber
{\sf B}^\epsilon_{\alpha_s\alpha_u}\!\left[^{+\beta_3\:-\beta_2}_{\hskip 6pt \beta_4\hskip 8pt \beta_1}\right]^{+-}_{\rm\, e\,o}
& = &   \displaystyle
-\frac{i}{\sqrt 2}{\rm e}^{\frac{i\pi}{8}}\,
\frac{
{\cal M}^{\rm R\,N\,N}_{\alpha_4,\alpha_2,\alpha_u}\,                        
{\cal M}^{\rm N\,N\,R}_{\alpha_u,\alpha_3,\alpha_1}                          
}{
{\cal M}^{\rm R\,N\,N}_{\alpha_4,\alpha_3,\alpha_s}\,                        
{\cal M}^{\rm N\,N\,R}_{\alpha_s,\alpha_2,\alpha_1}                          
}\
\mathbb{B}^{\epsilon,\,\rm R}_{\alpha_s\alpha_u}\!\!\left[^{\alpha_3\:\alpha_2}_{\alpha_4\:\alpha_1}\right]^{\rm e\, o}_{\rm e\, o},
\\[18pt]
\nonumber
{\sf B}^\epsilon_{\alpha_s\alpha_u}\!\left[^{+\beta_3\:-\beta_2}_{\hskip 6pt \beta_4\hskip 8pt \beta_1}\right]^{-+}_{\rm\, e\,o}
& = &    \displaystyle
\frac{1}{\sqrt 2}{\rm e}^{\frac{i\pi}{8}+\frac{i\pi\epsilon}{2}}\,
\frac{
{\cal M}^{\rm N\,R\,R}_{\alpha_4,\alpha_2,\alpha_u}\,                        
{\cal M}^{\rm R\,R\,N}_{\alpha_u,\alpha_3,\alpha_1}                          
}{
{\cal M}^{\rm R\,N\,N}_{\alpha_4,\alpha_3,\alpha_s}\,                        
{\cal M}^{\rm N\,N\,R}_{\alpha_s,\alpha_2,\alpha_1}                          
}\
\mathbb{B}^{\epsilon,\,\rm R}_{\alpha_s\alpha_u}\!\!\left[^{\alpha_3\:\alpha_2}_{\alpha_4\:\alpha_1}\right]^{\rm o\, e}_{\rm e\, o}.
\end{array}
$$

\section*{Acknowledgements}
We are grateful to Paulina Suchanek  for numerous discussions
and in particular for sharing with us her understanding  of chiral vertices involving Ramond weights
and to Micha{\l} Pawe{\l}kiewicz for sharing with us some of his results on supersymmetric Saalshutz formulae.

\newpage
\appendix
\section{Properties of Ising chiral vertex operators}
\label{Appendix:Ising}

Properties of the chiral vertex operators in the Ising model can be easily derived from  well known
 analytic
properties of the 4-point conformal blocks \cite{Belavin:1984vu,AlvarezGaume:1989vk}.
The fusion  matrix yields the coefficients of the chiral OPE:
\begin{equation*}
\begin{array}{rrrrrrrrrrrrrrrr}
	\V{1}{\varepsilon}{\varepsilon}{z_3}\V{\varepsilon}{\sigma}{\sigma}{z_2}
    &\sim&
    \tfrac{1}{\sqrt{z_{32}}} \V{1}{\sigma}{\sigma}{z_2}
    ,&\hspace{10pt}&
	\V{\varepsilon}{\varepsilon}{1}{z_3}\V{1}{\sigma}{\sigma}{z_2}
    &\sim& \tfrac{1}{2\sqrt{z_{32}}} \V{\varepsilon}{\sigma}{\sigma}{z_2}
    ,
\\[7pt]
	\V{\sigma}{\varepsilon}{\sigma}{z_3}\V{\sigma}{\sigma}{\varepsilon}{z_2}
    &\sim&\!-\tfrac{1}{\sqrt{z_{32}}} \V{\sigma}{\sigma}{\varepsilon}{z_2}
    ,&&
	\V{\sigma}{\varepsilon}{\sigma}{z_3}\V{\sigma}{\sigma}{1}{z_2}
    &\sim&
    \tfrac{1}{\sqrt{z_{32}}} \V{\sigma}{\sigma}{1}{z_2}
    ,
\\[7pt]
      \V{\varepsilon}{\sigma}{\sigma}{z_3}\V{\sigma}{\sigma}{1}{z_2}
      &\sim&
      z_{32}^{3 \over 8} \V{\varepsilon}{\varepsilon}{1}{z_2}
      ,&&
      \V{1}{\sigma}{\sigma}{z_3}\V{\sigma}{\sigma}{\varepsilon}{z_2}
      &\sim&
      z_{32}^{3  \over 8} \V{1}{\varepsilon}{\varepsilon}{z_2}
      ,
\\[7pt]
	\V{1}{\sigma}{\sigma}{z_3}\V{\sigma}{\sigma}{1}{z_2}
    &\sim&
    z_{32}^{-{1\over 8}}  \V{1}{1}{1}{z_2}
    ,&&
	\V{\varepsilon}{\sigma}{\sigma}{z_3}\V{\sigma}{\sigma}{\varepsilon}{z_2}
    &\sim &
    \!
    2 z_{32}^{-{1\over 8}} \V{\varepsilon}{1}{\varepsilon}{z_2}
    ,
\end{array}
\end{equation*}
      \vspace{-8pt}
\begin{equation}
\label{fuzje:set}
\begin{array}{rrrrrrr}
	\V{\sigma}{\sigma}{1}{z_3}\V{1}{\sigma}{\sigma}{z_2}
    &\sim&     \!
    \tfrac{1}{\sqrt{2}}z_{32}^{-{1\over 8}} \,\V{\sigma}{1}{\sigma}{z_2}
    \!&+& \!
	\tfrac{1}{2\sqrt{2}}z_{32}^{3 \over 8}\, \V{\sigma}{\varepsilon}{\sigma}{z_2}
    ,
\\[7pt]
	\V{\sigma}{\sigma}{\varepsilon}{z_3}\V{\varepsilon}{\sigma}{\sigma}{z_2}
    &\sim&   \!
    \sqrt{2}z_{32}^{-{1\over 8}} \,  \V{\sigma}{1}{\sigma}{z_2}
    \!&-& \!
	\tfrac{1}{\sqrt{2}}  z_{32}^{3 \over 8} \, \V{\sigma}{\varepsilon}{\sigma}{z_2}
    .
\end{array}
\end{equation}
The braiding matrix gives the braiding relations:
\begin{equation*}
\begin{array}{rrrrrrrrrr}
 	\V{1}{\varepsilon}{\varepsilon}{z_3}\V{\varepsilon}{\sigma}{\sigma}{z_2}
    \!&=&  \!
    {\rm e}^{- {i \over 2}\pi\epsilon} \V{1}{\sigma}{\sigma}{z_2}\V{\sigma}{\varepsilon}{\sigma}{z_3}
    ,& \hspace{10pt}&
	\V{\varepsilon}{\varepsilon}{1}{z_3}\V{1}{\sigma}{\sigma}{z_2}
    \!&=&  \!
    \tfrac12 {\rm e}^{{i\over 2} \pi\epsilon} \V{\varepsilon}{\sigma}{\sigma}{z_2}\V{\sigma}{\varepsilon}{\sigma}{z_3}
    ,
\\[7pt]
	\V{\sigma}{\varepsilon}{\sigma}{z_3}\V{\sigma}{\sigma}{\varepsilon}{z_2}
    \!&=&  \!
    2{\rm e}^{ {i\over 2} \pi\epsilon}\V{\sigma}{\sigma}{1}{z_2}\V{1}{\varepsilon}{\varepsilon}{z_3}
    ,&&
	\V{\sigma}{\varepsilon}{\sigma}{z_3}\V{\sigma}{\sigma}{1}{z_2}
    \!&=&  \!
    {\rm e}^{-{i\over 2} \pi\epsilon}\V{\sigma}{\sigma}{\varepsilon}{z_2}\V{\varepsilon}{\varepsilon}{1}{z_3}
    ,
\\[7pt]
	\V{1}{\sigma}{\sigma}{z_3}\V{\sigma}{\sigma}{1}{z_2}
    \!&=&  \!
    {\rm e}^{-{i\over 8} \pi\epsilon} \V{1}{\sigma}{\sigma}{z_2}\V{\sigma}{\sigma}{1}{z_3}
    ,&&
	\V{\varepsilon}{\sigma}{\sigma}{z_3}\V{\sigma}{\sigma}{\varepsilon}{z_2}
    \!&=&  \!
    {\rm e}^{-{i\over 8} \pi\epsilon}\V{\varepsilon}{\sigma}{\sigma}{z_2}\V{\sigma}{\sigma}{\varepsilon}{z_3}
    ,
\\[7pt]
      \V{\varepsilon}{\sigma}{\sigma}{z_3}\V{\sigma}{\sigma}{1}{z_2}
      \!&=&  \!
      {\rm e}^{{i3\over 8} \pi\epsilon}\V{\varepsilon}{\sigma}{\sigma}{z_2}\V{\sigma}{\sigma}{1}{z_3}
      ,&&
      \V{1}{\sigma}{\sigma}{z_3}\V{\sigma}{\sigma}{\varepsilon}{z_2}
      \!&=&  \!
      {\rm e}^{{i3\over 8} \pi\epsilon}\V{1}{\sigma}{\sigma}{z_2}\V{\sigma}{\sigma}{\varepsilon}{z_3}
      ,
\end{array}
\end{equation*}
\vspace{-6pt}
\begin{equation}
\label{braidingi:set}
\begin{array}{rrrrrrrrrrr}
	\V{\sigma}{\sigma}{1}{z_3}\V{1}{\sigma}{\sigma}{z_2}
    \!&=&  \!
    \tfrac{1}{\sqrt{2}}{\rm e}^{{i\over 8} \pi\epsilon}\; \V{\sigma}{\sigma}{1}{z_2}\V{1}{\sigma}{\sigma}{z_3}
    \!&+&  \!
    \tfrac{1}{2\sqrt{2}}{\rm e}^{-{i3\over 8} \pi\epsilon}\; \V{\sigma}{\sigma}{\varepsilon}{z_2}\V{\varepsilon}{\sigma}{\sigma}{z_3}
    &,
\\[7pt]
	\V{\sigma}{\sigma}{\varepsilon}{z_3}\V{\varepsilon}{\sigma}{\sigma}{z_2}
    \!&=&  \!
    \sqrt{2}{\rm e}^{-{i3\over 8} \pi\epsilon}\;\V{\sigma}{\sigma}{1}{z_2}\V{1}{\sigma}{\sigma}{z_3}
    \!&+&  \!
    \tfrac{1}{\sqrt{2}}{\rm e}^{{i\over 8} \pi\epsilon}\;\V{\sigma}{\sigma}{\varepsilon}{z_2}\V{\varepsilon}{\sigma}{\sigma}{z_3}
    &.
\end{array}
\end{equation}

\section{Conformal Ward Identities for the fermionic current $S(\xi)$}
\label{Appendix:CWI}
Correlation functions of the fermionic current $S(\xi)$ in the presence of the Ramond fields are no longer single-valued
and derivation of their form, even if standard, is subtle. As as example we shall discuss the  correlator
\[
\bra{\nu_4}
  {\sf W}_{\; {\rm o}\;s_3}^{-\,\beta_3}(1)
  S(\xi)
  {\sf W}_{\; { \rm e}\; }^{+\,\beta_+}(z)
\ket{\nu_1}.
\]
Due to  OPE (\ref{def:R:fields}) it has square root branch
cuts at $\xi = 1$ and $\xi = z.$ The function
\begin{equation}
\label{Appendix:correlator:s:1}
s(\xi) = \sqrt{1-\xi}\sqrt{\xi-z}
\bra{\nu_4}
  {\sf W}_{\; {\rm o}\;s_3}^{-\,\beta_3}(1)
  S(\xi)
  {\sf W}_{\; { \rm e}\; }^{+\,\beta_+}(z)
\ket{\nu_1}
\end{equation}
should then be single-valued and analytic on the complex $\xi$ plane save the simple poles at $\xi = 0,z,1.$
With the principal argument of a complex number $\xi$ in the range $-\pi \leq {\rm Arg}\,\xi < \pi$
one has:
\begin{equation}
\label{Appendix:square:cuts}
\sqrt{1-\xi} = {\rm e}^{-\frac{i\pi}{2}{\rm sgn\,Arg}(\xi-1)}\sqrt{\xi-1},
\hskip 1cm
\sqrt{\xi-z} = {\rm e}^{\frac{i\pi}{2}{\rm sgn\,Arg}(\xi-z)}\sqrt{z-\xi}.
\end{equation}
Braiding relations (\ref{braiding:W:fields})  yield
\begin{eqnarray}
\nonumber
s(\xi)
& = &
i\sqrt{\xi-1}\sqrt{\xi-z}
\bra{\nu_4}
  S(\xi)
  {\sf W}_{\; {\rm o}\;s_3}^{-\,\beta_3}(1)
  {\sf W}_{\; { \rm e}\; }^{+\,\beta_+}(z)
\ket{\nu_1}
\\[-6pt]
\label{Appendix:correlator:s:2}
\\[-6pt]
\nonumber
& = &
-i\sqrt{1-\xi}\sqrt{z-\xi}
\bra{\nu_4}
  {\sf W}_{\; {\rm o}\;s_3}^{-\,\beta_3}(1)
  {\sf W}_{\; { \rm e}\; }^{+\,\beta_+}(z)
  S(\xi)
\ket{\nu_1}.
\end{eqnarray}
Note that in (\ref{Appendix:correlator:s:2}) the multi-valuednes of  braiding relation  (\ref{braiding:W:fields})
is compensated by the signs coming from (\ref{Appendix:square:cuts}). It then follows from  OPE-s (\ref{def:R:fields})
and $S(\xi)\ket{\nu_1} \sim \frac{1}{\xi}S_{-\frac12}\ket{\nu_1}$ that
\[
s(\xi)\; \sim  \;
\left\{
\begin{array}{lcl}
\displaystyle
-\frac{i\sqrt{z}}{\xi}
\bra{\nu_4}
  {\sf W}_{\; {\rm o}\;s_3}^{-\,\beta_3}(1)
  {\sf W}_{\; { \rm e}\; }^{+\,\beta_+}(z)
  S_{-\frac12}
\ket{\nu_1}
& \;\; {\rm for}\;\; & \xi \to 0,
\\[12pt]
\displaystyle
\frac{\sqrt{1-z}}{\xi-z}\,{\rm e}^{\frac{i\pi}{4}}\beta_+
\bra{\nu_4}
  {\sf W}_{\; {\rm o}\;s_3}^{-\,\beta_3}(1)
  {\sf W}_{\; { \rm o}\; }^{-\,\beta_+}(z)
\ket{\nu_1}
& \;\; {\rm for}\;\; & \xi \to z,
\\[12pt]
\displaystyle
-\frac{\sqrt{1-z}}{\xi-1}\,{\rm e}^{\frac{i\pi}{4}}\beta_3
\bra{\nu_4}
  {\sf W}_{\; {\rm e}\;s_3}^{+\,\beta_3}(1)
  {\sf W}_{\; { \rm e}\; }^{+\,\beta_+}(z)
\ket{\nu_1}
& \;\; {\rm for}\;\; & \xi \to 1.
\end{array}
\right.
\]
This, together with the condition $s(\xi) \to 0$ for $\xi\to\infty,$ completely determines the form of the function
$s(\xi)$ and yields
\begin{eqnarray}
&&
\nonumber
\hskip -0.5cm
\bra{\nu_4}
  {\sf W}_{\; {\rm o}\;s_3}^{-\,\beta_3}(1)
  S(\xi)
  {\sf W}_{\; { \rm e}\; }^{+\,\beta_+}(z)
\ket{\nu_1}
\; = \;
\frac{1}{\sqrt{1-\xi}\sqrt{\xi-z}}
\left[
-\frac{i\sqrt{z}}{\xi}
\bra{\nu_4}
  {\sf W}_{\; {\rm o}\;s_3}^{-\,\beta_3}(1)
  {\sf W}_{\; { \rm e}\; }^{+\,\beta_+}(z)
  S_{-\frac12}
\ket{\nu_1}
\right.
\\[-1pt]
\\[-1pt]
\nonumber
& + &
\left.
\frac{\sqrt{1-z}}{\xi-z}\,{\rm e}^{\frac{i\pi}{4}}\beta_+
\bra{\nu_4}
  {\sf W}_{\; {\rm o}\;s_3}^{-\,\beta_3}(1)
  {\sf W}_{\; { \rm o}\; }^{-\,\beta_+}(z)
\ket{\nu_1}
-\frac{\sqrt{1-z}}{\xi-1}\,{\rm e}^{\frac{i\pi}{4}}\beta_3
\bra{\nu_4}
  {\sf W}_{\; {\rm e}\;s_3}^{+\,\beta_3}(1)
  {\sf W}_{\; { \rm e}\; }^{+\,\beta_+}(z)
\ket{\nu_1}
\right].
\end{eqnarray}
Expanding both sides of this equation around $\xi = z$ and equating the coefficients at $(\xi-z)^{-\frac12}$ we get
\begin{eqnarray}
\label{Appendix:CWIS:1}
&&
\hskip -1.5cm
\bra{\nu_4}
  {\sf W}_{\; {\rm o}\;s_3}^{-\,\beta_3}(1)
  S_{-1}
  {\sf W}_{\; { \rm e}\; }^{+\,\beta_+}(z)
\ket{\nu_1}
\; = \;
\frac{-i}{\sqrt{1-z}\sqrt{z}}\bra{\nu_4}
  {\sf W}_{\; {\rm o}\;s_3}^{-\,\beta_3}(1)
  {\sf W}_{\; { \rm e}\; }^{+\,\beta_+}(z)
  S_{-\frac12}
\ket{\nu_1}
\\
\nonumber
& + &
\frac{{\rm e}^{\frac{i\pi}{4}}}{1-z}
\left[
\frac12\beta_+ \bra{\nu_4}
  {\sf W}_{\; {\rm o}\;s_3}^{-\,\beta_3}(1)
  {\sf W}_{\; { \rm o}\; }^{-\,\beta_+}(z)
\ket{\nu_1}
+
\beta_3 \bra{\nu_4}
  {\sf W}_{\; {\rm e}\;s_3}^{+\,\beta_3}(1)
  {\sf W}_{\; { \rm e}\; }^{+\,\beta_+}(z)
\ket{\nu_1}
\right].
\end{eqnarray}
Formula (\ref{Appendix:CWIS:1}) is used in a derivation of differential equation (\ref{diff:eq:1}).
\section{Some properties of special functions related to the Barnes double gamma}
\label{Appendix:Barnes}
For $\Re\,x > 0$ the Barnes double function $\Gamma_b(x)$ has an integral
representation of the form:
\[
\log\,\Gamma_b(x)
\; = \;
\int\limits_{0}^{\infty}\frac{dt}{t}
\left[
\frac{{\rm e}^{- x t} - {\rm e}^{- Qt/2}}
{\left(1-{\rm e}^{- tb}\right)\left(1-{\rm e}^{- t/b}\right)}
-
\frac{\left(Q/2-x\right)^2}{2{\rm e}^{t}}
-
\frac{Q/2-x}{t}
\right].
\]
With a help of relations
\begin{equation}
\label{Gamma:b:shift}
\Gamma_b(x+b)
=
\frac{\sqrt{2\pi}\,b^{bx-\frac12}}{\Gamma(bx)}\Gamma_b(x),
\hskip 1cm
\Gamma_b\left(x + b^{-1}\right)
=
\frac{\sqrt{2\pi}\,b^{-\frac{x}{b}+\frac12}}{\Gamma(\frac{x}{b})}\Gamma_b(x),
\end{equation}
one can   continue $\Gamma_b(x)$ analytically to a meromorphic function of $x \in \mathbb C$ with
no zeroes and with simple poles located at $ x = -m b - nb^{-1},\; m, n \in
{\mathbb N}.$

\noindent
Borrowing the notation from \cite{Belavin:2007gz} we define
\begin{eqnarray}
\label{otherspecial:b}
S_b(x) = \frac{\Gamma_b(x)}{\Gamma_b(Q-x)},
\hskip 10mm
G_b(x) = {\rm e}^{-\frac{i\pi}{2}x(Q-x)} S_b(x),
\end{eqnarray}
and
\begin{equation}
\label{susyspecial:defs}
\begin{array}{lllclll}
\Gamma_{\rm NS}(x)
& = &
\Gamma_b\left(\frac{x}{2}\right)\Gamma_b\left(\frac{x+Q}{2}\right),
& \hskip 1.1cm &
\Gamma_{\rm R}(x)
& = &
\Gamma_b\left(\frac{x+b}{2}\right)\Gamma_b\left(\frac{x+b^{-1}}{2}\right),
\\[6pt]
S_{\rm NS}(x)
& = &
S_b\left(\frac{x}{2}\right)S_b\left(\frac{x+Q}{2}\right),
& \hskip 1cm &
S_{\rm R}(x)
& = &
S_b\left(\frac{x+b}{2}\right)S_b\left(\frac{x+b^{-1}}{2}\right),
\\[6pt]
G_{\rm NS}(x)
& = &
G_b\left(\frac{x}{2}\right)G_b\left(\frac{x+Q}{2}\right),
& \hskip 1cm &
G_{\rm R}(x)
& = &
G_b\left(\frac{x+b}{2}\right)G_b\left(\frac{x+b^{-1}}{2}\right).
\end{array}
\end{equation}
Using relations (\ref{Gamma:b:shift}) and definitions
(\ref{otherspecial:b}),\ (\ref{susyspecial:defs}) one can easily establish
some basic properties of these functions. 
\begin{itemize}
\item
Relations between $S$ and $G$ functions:
\begin{equation}
\label{G:and:S}
G_{\rm NS}(x) =
\zeta_0\,
{\rm e}^{- \frac{i\pi}{4}x(Q-x)}S_{\rm NS}(x),
\hskip 1cm
G_{\rm R}(x) =
{\rm e}^{- \frac{i\pi}{4}}\zeta_0\,
{\rm e}^{- \frac{i\pi}{4}x(Q-x)}S_{\rm R}(x),
\end{equation}
where
\(
\zeta_0 \;\ = \;\ {\rm e}^{- \frac{i\pi Q^2}{8}}.
\)
\item Shift relations:
\begin{equation}
\label{G:shift}
G_{\rm NS}(x+b^{\pm 1}) = \left(1+{\rm e}^{i\pi b^{\pm 1} x}\right)G_{\rm R}(x),
\hskip .6cm
G_{\rm R}(x+b^{\pm 1}) = \left(1- {\rm e}^{i\pi b^{\pm 1} x}\right)G_{\rm NS}(x).
\end{equation}
\item Reflection properties:
\begin{eqnarray}
\label{reflection:properties}
S_{\rm NS}(x)S_{\rm NS}(Q-x)
& = &
S_{\rm R}(x)S_{\rm R}(Q-x)
\; =  \;
1.
\end{eqnarray}
\item Locations of zeroes and poles:
\begin{eqnarray*}
S_{\rm NS}(x) & = & 0
\hskip 5mm
\Leftrightarrow
\hskip 5mm
x = Q+ mb+nb^{-1},
\hskip 5mm
m,n \in {\mathbb Z}_{\geq 0}, \;
m+n \in 2{\mathbb Z},
\\
S_{\rm R}(x) & = & 0
\hskip 5mm
\Leftrightarrow
\hskip 5mm
x = Q+ mb+nb^{-1},
\hskip 5mm
m,n \in {\mathbb Z}_{\geq 0}, \;
m+n \in 2{\mathbb Z} +1,
\\
S_{\rm NS}(x)^{-1} & = & 0
\hskip 5mm
\Leftrightarrow
\hskip 5mm
x = -mb-nb^{-1},
\hskip 10mm
m,n \in {\mathbb Z}_{\geq 0}, \;
m+n \in 2{\mathbb Z},
\\
S_{\rm R}(x)^{-1} & = & 0
\hskip 5mm
\Leftrightarrow
\hskip 5mm
x = -mb-nb^{-1},
\hskip 10mm
m,n \in {\mathbb Z}_{\geq 0}, \;
m+n \in 2{\mathbb Z} +1.
\end{eqnarray*}
\item Basic residue:
\begin{eqnarray}
\label{residues}
\lim_{x\to 0}\ x\,S_{\rm NS}(x) & = & \frac{1}{\pi}.
\end{eqnarray}
\end{itemize}

\section{Orthogonality relations}
\label{Appendix:Orthogonality}

For $\xi \in i{\mathbb R}_+$ we define
\begin{eqnarray*}
\left\langle \tau\left|^{\rm\scriptscriptstyle N}_{\rm\scriptscriptstyle N}\right|\xi\right\rangle
& = &
\frac{1}{S_{\rm NS}(Q+\tau+\xi-0^+)S_{\rm NS}(Q+\tau-\xi-0^+)}
\;\ = \;\
\frac{S_{\rm NS}(\xi-\tau)}{S_{\rm NS}(Q+\tau+\xi-0^+)},
\\[6pt]
\left\langle \tau\left|^{\rm\scriptscriptstyle R}_{\rm\scriptscriptstyle N}\right|\xi\right\rangle
& = &
\frac{1}{S_{\rm NS}(Q+\tau+\xi-0^+)S_{\rm R}(Q+\tau-\xi)}
\;\ = \;\
\frac{S_{\rm R}(\xi-\tau)}{S_{\rm NS}(Q+\xi+\tau-0^+)},
\end{eqnarray*}
etc.

\noindent
The orthogonality relations
\begin{eqnarray}
\nonumber
\int\limits_{i\mathbb R}\!\frac{d\tau}{i}\
\left[
\left\langle \tau\left|^{\rm\scriptscriptstyle N}_{\rm\scriptscriptstyle N}\right|ip_2\right\rangle^*
\left\langle \tau\left|^{\rm\scriptscriptstyle N}_{\rm\scriptscriptstyle N}\right|ip_1\right\rangle
+
\left\langle \tau\left|^{\rm\scriptscriptstyle R}_{\rm\scriptscriptstyle R}\right|ip_2\right\rangle^*
\left\langle \tau\left|^{\rm\scriptscriptstyle R}_{\rm\scriptscriptstyle R}\right|ip_1\right\rangle
\right]
=
{\cal N}_{\rm NS}^{-1}(p_2)
\,\delta(p_2-p_1),
&&
\\[-10pt]
\label{orthogonality:1}
\\[-4pt]
\nonumber
\int\limits_{i\mathbb R}\!\frac{d\tau}{i}\
\left[
\left\langle \tau\left|^{\rm\scriptscriptstyle N}_{\rm\scriptscriptstyle N}\right|ip_2\right\rangle^*
\left\langle \tau\left|^{\rm\scriptscriptstyle R}_{\rm\scriptscriptstyle R}\right|ip_1\right\rangle
-
\left\langle \tau\left|^{\rm\scriptscriptstyle R}_{\rm\scriptscriptstyle R}\right|ip_2\right\rangle^*
\left\langle \tau\left|^{\rm\scriptscriptstyle N}_{\rm\scriptscriptstyle N}\right|ip_1\right\rangle
\right]
= 0,
&&
\end{eqnarray}
where $p_1,p_2 \in {\mathbb R}_+,$ were derived in \cite{Hadasz:2007wi}. They followed from the ``Saalschutz summation formulae''
\begin{eqnarray}
\label{Saalschutz:1a}
\nonumber
&&
\hskip -2cm
\int\limits_{-i\infty}^{i\infty}\!\! \frac{d\tau}{i}\
{\rm e}^{i\pi\tau Q}
\left[
\frac{G_{\rm NS}(\tau + a) G_{\rm NS}(\tau + b)}
{G_{\rm NS}(\tau +d)G_{\rm NS}(\tau + Q)}
+
\frac{G_{\rm R}(\tau + a) G_{\rm R}(\tau + b)}
{G_{\rm R}(\tau +d)G_{\rm R}(\tau + Q)}
\right]
\\[10pt]
& = &
2\zeta_0^{-3}\,{\rm e}^{\frac{i\pi}{2}d(Q-d)}
\frac{G_{\rm NS}(a)G_{\rm NS}(b)G_{\rm NS}(Q+a-d)G_{\rm NS}(Q+b-d)}
{G_{\rm NS}(Q+a+b-d)},
\end{eqnarray}
and
\begin{eqnarray}
 \label{Saalschutz:1b}
\nonumber
&&
\hskip -2cm
\int\limits_{-i\infty}^{i\infty}\!\! \frac{d\tau}{i}\
{\rm e}^{i\pi\tau Q}
\left[
\frac{G_{\rm NS}(\tau + a) G_{\rm NS}(\tau + b)}
{G_{\rm R}(\tau +d) G_{\rm NS}(\tau + Q)}
+
\frac{G_{\rm R}(\tau + a) G_{\rm R}(\tau + b)}
{G_{\rm NS}(\tau +d)G_{\rm R}(\tau + Q)}
\right]
\\[10pt]
& = &
2i\zeta_0^{-3}\,{\rm e}^{\frac{i\pi}{2}d(Q-d)}
\frac{G_{\rm NS}(a)G_{\rm NS}(b)G_{\rm R}(Q+a-d)G_{\rm R}(Q+b-d)}
{G_{\rm R}(Q+a+b-d)},
\end{eqnarray}
which were also derived in that paper.
For our present purposes we need another two Saalschutz summation formulae, which may be derived
following the steps that lead to (\ref{Saalschutz:1a}), (\ref{Saalschutz:1b}). They read
\begin{eqnarray}
\label{Saalschutz:2a}
\nonumber
&&
\hskip -2cm
\int\limits_{-i\infty}^{i\infty}\!\! \frac{d\tau}{i}\
{\rm e}^{i\pi\tau Q}
\left[
\frac{G_{\rm NS}(\tau + a) G_{\rm R}(\tau + b)}
{G_{\rm R}(\tau +d)G_{\rm NS}(\tau + Q)}
+
\frac{G_{\rm R}(\tau + a) G_{\rm NS}(\tau + b)}
{G_{\rm NS}(\tau +d)G_{\rm R}(\tau + Q)}
\right]
\\[10pt]
& = &
2i\zeta_0^{-3}\,{\rm e}^{\frac{i\pi}{2}d(Q-d)}
\frac{G_{\rm NS}(a)G_{\rm R}(b)G_{\rm NS}(Q+a-d)G_{\rm R}(Q+b-d)}
{G_{\rm NS}(Q+a+b-d)},
\end{eqnarray}
\begin{eqnarray}
\label{Saalschutz:2b}
\nonumber
&&
\hskip -2cm
\int\limits_{-i\infty}^{i\infty}\!\! \frac{d\tau}{i}\
{\rm e}^{i\pi\tau Q}
\left[
\frac{G_{\rm R}(\tau + a) G_{\rm NS}(\tau + b)}
{G_{\rm R}(\tau +d)G_{\rm NS}(\tau + Q)}
+
\frac{G_{\rm NS}(\tau + a) G_{\rm R}(\tau + b)}
{G_{\rm NS}(\tau +d)G_{\rm R}(\tau + Q)}
\right]
\\[10pt]
& = &
2i\zeta_0^{-3}\,{\rm e}^{\frac{i\pi}{2}d(Q-d)}
\frac{G_{\rm R}(a)G_{\rm NS}(b)G_{\rm NS}(Q+a-d)G_{\rm R}(Q+b-d)}
{G_{\rm NS}(Q+a+b-d)}.
\end{eqnarray}
Using the relations between the $S_\natural(x)$ and $G_\natural(x)$ functions
and substituting in (\ref{Saalschutz:2a})
\[
a = 2\varepsilon +\xi_2 - \xi_1,\hskip 1cm
b = 2\varepsilon -\xi_2 - \xi_1,\hskip 1cm
d = Q-2\xi_1
\]
we get
\begin{eqnarray*}
&&
\hskip -1cm
\int\limits_{i\mathbb R}\!\frac{d\tau}{i}
\left[
\left\langle\tau-\epsilon\left|^{\rm\scriptscriptstyle R}_{\rm\scriptscriptstyle N}\right|\xi_2\right\rangle^{*}
\left\langle\tau-\epsilon\left|^{\rm\scriptscriptstyle R}_{\rm\scriptscriptstyle N}\right|\xi_1\right\rangle
+
\left\langle\tau-\epsilon\left|^{\rm\scriptscriptstyle N}_{\rm\scriptscriptstyle R}\right|\xi_2\right\rangle^{*}
\left\langle\tau-\epsilon\left|^{\rm\scriptscriptstyle N}_{\rm\scriptscriptstyle R}\right|\xi_1\right\rangle
\right] =
\\[6pt]
&& = \;
2i\zeta_0^{-3}{\rm e}^{\frac{i\pi}{2}\left(\xi_1^2-\xi_2^2\right)-2i\pi\xi_1^2}\
\frac{G_{\rm NS}(2\epsilon+\xi_-)G_{\rm NS}(2\epsilon-\xi_-)G_{\rm R}(2\epsilon-\xi_+)G_{\rm R}(2\epsilon+\xi_+)}{G_{\rm NS}(4\epsilon)},
\end{eqnarray*}
where $\xi_-=\xi_2-\xi_1,\ \xi_+ = \xi_2+\xi_1.$

For $x\to 0$ we have from (\ref{G:and:S}) and (\ref{residues}):
\[
G_{\rm NS}(x) = \frac{\zeta_0}{\pi x} +{\cal O}(1)
\]
so that
\begin{eqnarray*}
\lim_{\epsilon\to 0}
\frac{G_{\rm NS}(2\epsilon+ip_2-ip_1)G_{\rm NS}(2\epsilon-ip_2+ip_1)}{G_{\rm NS}(4\epsilon)}
& = &
\frac{\zeta_0}{\pi}\,\lim_{\epsilon\to 0}\frac{4\epsilon}{(2\epsilon)^2+(p_2-p_1)^2}
\; = \;
2\zeta_0\delta(p_2-p_1).
\end{eqnarray*}
This finally gives
\begin{eqnarray}
\label{orthogonality:nonzero}
&&
\hskip -2cm
\int\limits_{i\mathbb R}\!\frac{d\tau}{i}
\left[
\left\langle\tau\left|^{\rm\scriptscriptstyle R}_{\rm\scriptscriptstyle N}\right|ip_2\right\rangle^{*}
\left\langle\tau\left|^{\rm\scriptscriptstyle R}_{\rm\scriptscriptstyle N}\right|ip_1\right\rangle
+
\left\langle\tau\left|^{\rm\scriptscriptstyle N}_{\rm\scriptscriptstyle R}\right|ip_2\right\rangle^{*}
\left\langle\tau\left|^{\rm\scriptscriptstyle N}_{\rm\scriptscriptstyle R}\right|ip_1\right\rangle
\right]
\\[6pt]
\nonumber
& = &
4i\zeta_0^{-2}{\rm e}^{2i\pi p_1^2}G_{\rm R}(-2ip_1)G_{\rm R}(2ip_1)\delta(p_2-p_1)
=
{\cal N}_{\rm R}^{-1}\,\delta(p_2-p_1).
\end{eqnarray}
Similarly, with a help of  formula (\ref{Saalschutz:2b}), on gets
\begin{eqnarray*}
&&
\hskip -1cm
\int\limits_{i\mathbb R}\!\frac{d\tau}{i}
\left[
\left\langle\tau-\epsilon\left|^{\rm\scriptscriptstyle N}_{\rm\scriptscriptstyle R}\right|\xi_2\right\rangle^{*}
\left\langle\tau-\epsilon\left|^{\rm\scriptscriptstyle R}_{\rm\scriptscriptstyle N}\right|\xi_1\right\rangle
+
\left\langle\tau-\epsilon\left|^{\rm\scriptscriptstyle R}_{\rm\scriptscriptstyle N}\right|\xi_2\right\rangle^{*}
\left\langle\tau-\epsilon\left|^{\rm\scriptscriptstyle N}_{\rm\scriptscriptstyle R}\right|\xi_1\right\rangle
\right] =
\\
&& = \;
2i\zeta_0^{-3}{\rm e}^{\frac{i\pi}{2}\left(\xi_1^2-\xi_2^2\right)-2i\pi\xi_1^2}
\frac{G_{\rm NS}(2\epsilon+\xi_+)G_{\rm NS}(2\epsilon-\xi_+)G_{\rm R}(2\epsilon-\xi_-)G_{\rm R}(2\epsilon+\xi_-)}{G_{\rm NS}(4\epsilon)}.
\end{eqnarray*}
Now, since $\xi_+= i(p_1+p_2)$ does not vanish for $p_i\in {\mathbb R}_+,$ we do not have a singular contribution from
\[
G_{\rm NS}(2\epsilon+\xi_+)G_{\rm NS}(2\epsilon-\xi_+)
\]
in the limit $\epsilon\to 0.$
Moreover, $G_{\rm R}(x)$ is regular in the vicinity of the imaginary $x$ axis and
\[
\lim_{\epsilon\to 0\,}G_{\rm R}(2\epsilon+\xi_-)G_{\rm R}(2\epsilon-\xi_-)
\]
is finite. This gives
\begin{equation}
\label{orthogonality:zero}
\int\limits_{i\mathbb R}\!\frac{d\tau}{i}
\left[
\left\langle\tau\left|^{\rm\scriptscriptstyle N}_{\rm\scriptscriptstyle R}\right|ip_2\right\rangle^{*}
\left\langle\tau\left|^{\rm\scriptscriptstyle R}_{\rm\scriptscriptstyle N}\right|ip_1\right\rangle
+
\left\langle\tau\left|^{\rm\scriptscriptstyle R}_{\rm\scriptscriptstyle N}\right|ip_2\right\rangle^{*}
\left\langle\tau\left|^{\rm\scriptscriptstyle N}_{\rm\scriptscriptstyle R}\right|ip_1\right\rangle
\right]
=
0.
\end{equation}
Using the reflection formula for the Barnes $S-$functions, Eq.\ (\ref{reflection:properties}),
the fact that $S_{\natural}(x)$ are real analytic and that $p_i\in {\mathbb R},\ \tau \in i{\mathbb R},$ we can rewrite
(\ref{orthogonality:1}), (\ref{orthogonality:nonzero}) and (\ref{orthogonality:zero}) in the form
\begin{eqnarray}
\label{ortho:final:nonzero}
\nonumber
&&
\hskip -1cm
\int\limits_{i\mathbb R}\!\frac{d\tau}{i}\
\left\{
\frac{S_{\rm NS}(ip_2+\tau)}{S_{\rm NS}(Q+ip_2-\tau)}\frac {S_{\rm NS}(ip_1-\tau)}{S_{\rm NS}(Q+ip_1+\tau)}
+
\frac{S_{\rm R}(ip_2+\tau)}{S_{\rm R}(Q+ip_2-\tau)}\frac{S_{\rm R}(ip_1-\tau)}{S_{\rm R}(Q+ip_1+\tau)}
\right\}
\\
&&
\hskip 9cm = \;
{\cal N}_{\rm NS}^{-1}\,\delta(p_2-p_1),
\\[10pt]
\nonumber
&&
\hskip -1cm
\int\limits_{i\mathbb R}\!\frac{d\tau}{i}\
\left\{
\frac{S_{\rm R}(ip_2+\tau)}{S_{\rm NS}(Q+ip_2-\tau)}\frac {S_{\rm NS}(ip_1-\tau)}{S_{\rm R}(Q+ip_1+\tau)}
+
\frac{S_{\rm NS}(ip_2+\tau)}{S_{\rm R}(Q+ip_2-\tau)}\frac {S_{\rm R}(ip_1-\tau)}{S_{\rm NS}(Q+ip_1+\tau)}
\right\}
\\
\nonumber
&&
\hskip 9cm = \;
{\cal N}_{\rm R}^{-1}\,\delta(p_2-p_1),
\end{eqnarray}
and
\begin{eqnarray}
\label{ortho:final:zero}
\nonumber
\int\limits_{i\mathbb R}\!\frac{d\tau}{i}\
\left\{
\frac{S_{\rm NS}(ip_2+\tau)}{S_{\rm NS}(Q+ip_2-\tau)}\frac{S_{\rm R}(ip_1-\tau)}{S_{\rm R}(Q+ip_1+\tau)}
-
\frac{S_{\rm R}(ip_2+\tau)}{S_{\rm R}(Q+ip_2-\tau)}\frac {S_{\rm NS}(ip_1-\tau)}{S_{\rm NS}(Q+ip_1+\tau)}
\right\}
& = & 0,
\\[-6pt]
\\[-4pt]
\nonumber
\int\limits_{i\mathbb R}\!\frac{d\tau}{i}\
\left\{
\frac{S_{\rm R}(ip_2+\tau)}{S_{\rm NS}(Q+ip_2-\tau)}\frac {S_{\rm R}(ip_1-\tau)}{S_{\rm NS}(Q+ip_1+\tau)}
+
\frac{S_{\rm NS}(ip_2+\tau)}{S_{\rm R}(Q+ip_2-\tau)}\frac {S_{\rm NS}(ip_1-\tau)}{S_{\rm R}(Q+ip_1+\tau)}
\right\}
& = & 0,
\end{eqnarray}
were for brevity we have not written the ``$-0^+$'' prescription explicitly. It is now easy  to check
that for every choice of the parity indices ${\scriptstyle \rm\bf h}_2,{\scriptstyle \rm\bf h}_3,$
and ${\scriptstyle \rm\bf g}_2,{\scriptstyle \rm\bf g}_3,$ the orthogonality formula
(\ref{orthogonality:matrix:form}) reduces to one of the equations (\ref{ortho:final:nonzero}), (\ref{ortho:final:zero}).

\end{document}